\documentclass[preprint,12pt]{elsarticle}
\usepackage{amssymb}
\usepackage{graphicx}
\usepackage{multirow}%
\usepackage{amsmath,amssymb,amsfonts,bm}%
\usepackage{amsthm}%
\usepackage{multirow}
\usepackage{mathrsfs}%
\usepackage[title]{appendix}%
\usepackage{float}
\usepackage{placeins}
\usepackage{xcolor}
\usepackage{epsfig}
\usepackage[caption=false]{subfig}
\usepackage{textcomp}%
\usepackage{manyfoot}%
\usepackage{booktabs}%
\usepackage{enumitem}
\usepackage{array}
\usepackage{geometry}
\usepackage{float}
\usepackage{algorithm}%
\usepackage{algorithmicx}%
\usepackage{algpseudocode}%
\usepackage{listings}
\usepackage{xcolor}
\usepackage{caption}
\usepackage{dashrule}
\usepackage{hyperref}
\usepackage{subfig} 
\hypersetup{
    colorlinks=true,
    linkcolor=blue,
    citecolor=blue,
    urlcolor=blue
}

\journal{Applied Mathematical Modelling}

\begin{document}

\begin{frontmatter}

\title{Inequality Constrained Minimum Density Power Divergence Estimation in Panel Count Data}

\author[inst1]{Udita Goswami}

\affiliation[inst1]{organization={Department of Mathematics and Computing},
            addressline={IIT(Indian School of Mines) Dhanbad}, 
            city={},
            postcode={826004}, 
            state={Jharkhand},
            country={India}}

\author[inst2]{Shuvashree Mondal*}

\affiliation[inst2]{organization={Department of Mathematics and Computing},
            addressline={IIT(Indian School of Mines) Dhanbad}, 
            city={},
            postcode={826004}, 
            state={Jharkhand},
            country={India}}

\begin{abstract}
The analysis of panel count data has garnered considerable attention in the literature, leading to the development of multiple statistical techniques.  In inferential analysis, most works focus on leveraging estimating equation-based techniques or conventional maximum likelihood estimation.  However, the robustness of these methods is largely questionable.  In this paper, we present a robust density power divergence estimation method for panel count data arising from non-homogeneous Poisson processes correlated through a latent frailty variable.  To cope with real-world incidents, it is often desirable to impose certain inequality constraints on the parameter space, leading to the constrained minimum density power divergence estimator.  Being incorporated with inequality restrictions, coupled with the inherent complexity of our objective function, standard computational algorithms are inadequate for estimation purposes.  To overcome this, we adopt sequential convex programming, which approximates the original problem through a series of subproblems.  Further, we study the asymptotic properties of the resultant estimator, making a significant contribution to this work.  The proposed method ensures high efficiency in the model estimation while providing reliable inference despite data contamination.  Moreover, the density power divergence measure is governed by a tuning parameter \(\gamma\), which controls the trade-off between robustness and efficiency.  To effectively determine the optimal value of \(\gamma\), this study employs a generalized score-matching technique, marking considerable progress in the data analysis.  Simulation studies and real data examples are provided to illustrate the performance of the estimator and to substantiate the theory developed.
\end{abstract}

\begin{highlights}
    \item Proposes a robust estimator for panel count data with latent frailty.
    \item Incorporates inequality constraints into density power divergence estimation.
    \item Establishes asymptotic properties under parameter restrictions.
    \item Finds optimal tuning parameter through generalized score-matching technique.
    \item Validates performance through simulations and real applications.
\end{highlights}


\end{frontmatter}

\section{Introduction}\label{sec1}

In the study of event history, the types of events that occur repetitively are usually referred to as recurrent events.  Some examples of recurrent events refer to hospitalizations of intravenous drug users, multiple instances of specific tumors \citep{wang2001analyzing}, repeated pyogenic infections in patients with inherited disorders \citep{lin2000semiparametric}, and claims for vehicle warranties \citep{kalbfleisch1991methods}.  Panel count data are common when subjects in a study encounter recurrent events observed only at discrete time points instead of being observed continuously in time.  They might occur when continuous monitoring of the subjects under study is too expensive or infeasible.  As a result, panel count data only provide the number of occurrences of the recurrent events of interest between successive observation times, while the exact occurrence times remain unknown; refer to Huang et al. \citep{huang2006analysing},  Sun and Zhao \citep{sun2013statistical}.

In the literature, it is a common practice to characterize the occurrences of recurrent events by leveraging counting processes.  Further, due to the incomplete nature of the panel count data, it is more convenient to model the mean function of the counting processes.  Readers may refer to Sun and Kalbfleisch \citep{sun1995estimation}, Wellner and Zhang \citep{wellner2000two}, Hu et al. \citep{hu2009generalized}, and references therein.  In this work, we focus on the analysis of panel count data where each subject under study is exposed to multiple recurrent events characterized by nonhomogeneous Poisson processes; refer to Tang et al. \citep{tang2017bayesian}, Li and Pham \citep{li2017nhpp}.  It is a non-stationary counting process with independent increments.  In many instances, as it proves challenging to assess all the covariates, a latent frailty variable can be considered that reflects individual-specific heterogeneity typically influencing the frequency of these recurrent incidents; follow Slimacek and Lindqvist \citep{slimacek2016nonhomogeneous}.  In this work, we develop the inferential analysis of panel count data where each subject under study is exposed to multiple recurrent events characterized by nonhomogeneous Poisson processes correlated through a frailty variable.  Based on the counting process, there exists a substantial body of literature concerning inferential studies for panel count data.  Amongst them, He et al. \citep{he2009semiparametric} exploited a semi-parametric approach for the analysis of panel count data.  Moreover, Sun and Wei \citep{sun2000regression} and He et al. \citep{he2008regression} investigated regression analysis based on counting processes concerning univariate and multivariate panel count data, respectively.  

Throughout the years, most research on estimation methods for panel count data has been devoted to maximum likelihood estimation (MLE).  The major contributions consist of the works by Wellner and Zhang \citep{wellner2007two}, Zhu et al. \citep{zhu2018semiparametric}, Li et al. \citep{li2021regression}, Zeng and Lin \citep{zeng2021maximum}.  Maximum likelihood estimation is a popular method because of its recognized characteristics, such as asymptotic efficiency, consistency, sufficiency, and invariance.  However, with small deviations from the assumed model conditions, those methods often produce biased and inefficient estimates, highlighting the necessity for robust estimation techniques.  In this paper, our objective is to perform statistical inference based on multivariate panel count data by applying robust density power divergence estimation methods.  The minimum density power divergence estimation (MDPDE) was first proposed by Basu et al. \citep{basu1998robust} and subsequently applied in numerous studies. Readers can track the publications of Ghosh and Basu \citep{ghosh2013robust}, Basu et al. \citep{basu2022robust}, Mandal et al. \citep{mandal2023robust}, Baghel and Mondal \citep{baghel2024analysis}.  Recently, Balakrishnan et al. \citep{balakrishnan2019robust} developed weighted minimum density power divergence estimators to analyze one-shot device testing data, similar to current status data, which is a specific type of panel count data. Xiong and Zhu \citep{xiong2022minimum} developed a robust MDPDE-based method for negative binomial integer-valued GARCH models, effectively handling overdispersed count time series data and outliers.  Their research motivates our study on panel count data, where robustness and heterogeneity are key concerns.  Upon reviewing the literature, it is found that none of the existing studies on multivariate panel count data generated from the counting process address robust methods for parameter estimation.  Hence, we suggest this novel approach to serve the need for a robust estimation technique.  

Further, in real-life scenarios, it is often observed that one recurring event takes place more frequently than the other event.  Such instances can be accurately characterized by imposing restrictions on the parameter space.  Moreover, research on restricted minimum density power divergence estimators has been sparse, with all current studies concentrating solely on equality constraints.  For example, Basu et al. \citep{basu2018testing}, Felipe et al. \citep{felipe2023restricted}, Baghel and Mondal \citep{baghel2024robust} suggested equality constraints in the parameter space of the model parameters and derived the restricted minimum density power divergence estimators.  In this study, we broaden the scope of restricted minimum density power divergence estimators by introducing inequality constraints into the parameter space.  However, the addition of these constraints, along with the complex nature of the objective function, limits the effectiveness of standard computational algorithms in finding optimal solutions.  This challenge has motivated our exploration of Sequential Convex Programming (SCP), a technique initially developed by Dinh and Diehl \citep{dinh2010local}, which transforms a non-convex problem into a sequence of convex subproblems.  A key component of SCP is the trust region, which constrains each subproblem to a neighborhood around the current iterate, ensuring both convergence and local validity of convex approximations.  SCP has found successful applications in trajectory optimization, model predictive control, and various engineering design problems.  For additional insights and applications, readers are encouraged to consult Debrouwere et al. \citep{debrouwere2013time}, Morgan et al. \citep{morgan2014model}, and Wang and Lu \citep{wang2020improved}, among others.  The principal novelty in this paper lies in deriving the asymptotic distribution of the minimum density power divergence estimators while incorporating these inequality restrictions. 

Moreover, the MDPDE is characterized by a tuning parameter \( \gamma \), which controls the trade-off between robustness and efficiency.  In the literature, Warwick and Jones \citep{warwick2005choosing} were the first to introduce a data-driven algorithm for determining the optimal value of the tuning parameter, which has since been studied and refined by various authors.  Recently, Sugasawa and Yonekura \citep{sugasawa2021selection}, \citep{yonekura2023adaptation} delved into this problem by exploiting the Hyvärinen score-matching method with unnormalized models based on robust divergence.  Hyvärinen and Dayan \citep{hyvarinen2005estimation} introduced the Hyvärinen score, a score-matching method for continuous-valued data.  This approach ensures a consistent estimator by focusing on the gradients of the log of the model and observed data densities, eliminating the need to compute the normalization constant.  The principle of Hyvärinen score matching (Hyvärinen \citep{hyvarinen2007some}) involves minimizing the expected squared difference between the score functions of the parameterized density and the true density.  However, this approach is restricted to continuous-valued data, where the densities have to be differentiable in the space of $\mathbb{R}^n$.  For discrete data, Lyu \citep{lyu2012interpretation} demonstrated a generalized score-matching approach by incorporating the marginalization operator $\mathcal{M}$.  This study was further extended by Xu et al. \citep{xu2022generalized} for discrete independent multivariate data, where he used a general linear operator $\mathcal{L}$, in place of the gradient operator.  In our study, we leverage this generalized score-matching method proposed by Xu et al. \citep{xu2022generalized} to find the optimal value of the tuning parameter $\gamma.$  Finally, a comparative study is executed to evaluate the performance of an existing method and the generalized score-matching approach. 

The rest of the article goes as follows.  In Section \ref{sec2}, the model description is provided along with the study of the likelihood function.  A review of the minimum density power divergence estimator is presented in Section \ref{sec3}.  We derive the restricted minimum density power divergence estimator in Section \ref{sec4}.  The determination of optimal tuning parameter $\gamma$ is discussed in Section \ref{sec5}.  Section \ref{sec6} presents an extensive simulation study and real data analysis in order to evaluate the performance of the developed methods.

\section{Model Description}\label{sec2}

Consider a study involving \( m \) independent subjects who may experience \( w \) different types of recurrent events.  Let \( N_{ij}(t) \) denote the number of occurrences of the $j^{th}$ recurrent event up to time \( t \) for the $i^{th}$ subject, where \( j = 1, 2, \ldots, w \) and \( i = 1, \ldots, m \).   We assume that for any \( i^{th} \) subject, the unobserved heterogeneity is represented by a frailty variable \( Z_{i} \), where $Z_{i}'s$, for $i = 1, \ldots, m $ are independently distributed according to an exponential distribution with a common rate parameter \( \zeta (\zeta > 0) \).  Given \( Z_{i} \),  \( N_{ij}(t) \) are assumed to follow independent nonhomogeneous Poisson process with intensity functions \( \lambda_{j}(t \mid Z_{i}) = Z_{i} a_{j} t^{b_{j}} \), where \( a_{j} > 0 \), and \( b_{j} > 0 \) for \( j = 1, \ldots, w \).  Further, each  \( N_{ij}(t) \) is being observed only at finite time points say, \( \tau_{1}, \tau_{2}, \ldots, \tau_{l} \) for \( j = 1, 2, \ldots, w \) and \( i = 1, \ldots, m \).  Therefore, define \( N_{ijl} \) as the number of occurrences of the $j^{th}$ recurrent event within the interval \( (\tau_{l-1}, \tau_{l}] \) for the $i^{th}$ subject, where \( l = 1, 2, \ldots, k \); \( j = 1, 2, \ldots, w \); and \( i = 1, \ldots, m \).  For simplicity, we focus on two recurrent events (\( w = 2 \)), although the framework and results can be extended to \( w > 2 \) in a similar manner.  Given \( Z_{i} \), the probability mass function of $\boldsymbol{N_{i}} = (N_{i11}, N_{i21}, \ldots, N_{i1k}, N_{i2k}$ can be derived as follows
\begin{align*}
f(\boldsymbol{n_{i}} \vert\ z_{i}) = P(\boldsymbol{N_{i}} = \boldsymbol{n_{i}}\vert\ z_{i}) = \prod_{j=1}^{2}\prod_{l=1}^{k} P(N_{ijl} = n_{ijl}\vert\ z_{i}) \notag\\
= \frac{e^{-z_i \sum_{j=1}^{2} \sum_{l=1}^{k}{a_{j}(\tau_{l}^{b_{j}} - \tau_{l-1}^{b_{j}})} z_i^{\sum_{j=1}^{2} \sum_{l=1}^{k} n_{ijl}} \prod_{j=1}^{2}\prod_{l=1}^{k} (a_{j} (\tau_{l}^{b_{j}} - \tau_{l-1}^{b_{j}}))^{n_{ijl}}}}{\prod_{j=1}^{2}\prod_{l=1}^{k} n_{ijl}!}, \tag{1}
\label{eq:1}
\end{align*}
where $\boldsymbol{n_{i}} = (n_{i11}, n_{i21}, \ldots, n_{i1k},n_{i2k}).$

\noindent The marginal probability mass function of $\boldsymbol{N_{i}}$  can be obtained as
\begin{align*}
f(\boldsymbol{n_{i}}) &= \int_{0}^{\infty} P(\boldsymbol{N_{i}} = \boldsymbol{n_{i}}\vert\ z_{i}) h(z_{i})~dz_{i} \notag\\
&= \int_{0}^{\infty}\prod_{j=1}^{2}\prod_{l=1}^{k} P(N_{ijl} = n_{ijl}\vert\ z_{i}) h(z_{i})~dz_{i} \notag\\
&= \int_{0}^{\infty} \frac{e^{-z_i \sum_{j=1}^{2} \sum_{l=1}^{k}{a_{j}(\tau_{l}^{b_{j}} - \tau_{l-1}^{b_{j}})} z_i^{\sum_{j=1}^{2} \sum_{l=1}^{k} n_{ijl}} \prod_{j=1}^{2}\prod_{l=1}^{k} (a_{j} (\tau_{l}^{b_{j}} - \tau_{l-1}^{b_{j}}))^{n_{ijl}}}}{\prod_{j=1}^{2}\prod_{l=1}^{k} n_{ijl}!} \zeta e^{-\zeta z_{i}}~dz_{i} \notag\\
&=   \frac {\zeta \prod_{j=1}^{2}\prod_{l=1}^{k}(a_{j} (\tau_{l}^{b_{j}} - \tau_{l-1}^{b_{j}}))^{n_{ijl}} (\sum_{j=1}^{2} \sum_{l=1}^{k} n_{ijl})!} {\prod_{j=1}^{2}\prod_{l=1}^{k} n_{ijl}! (\zeta + a_{j}(\tau_{l}^{b_{j}} - \tau_{l-1}^{b_{j}}))^{\sum_{j=1}^{2} \sum_{l=1}^{k} n_{ijl}}} \notag\\
&= \frac {\zeta \prod_{l=1}^{k}(a_{1} (\tau_{l}^{b_{1}} - \tau_{l-1}^{b_{1}}))^{n_{i1l}} (a_{2} (\tau_{l}^{b_{2}} - \tau_{l-1}^{b_{2}}))^{n_{i2l}} \sum_{l=1}^{k} (n_{i1l} + n_{i2l})!} {\prod_{l=1}^{k} n_{i1l}! n_{i2l}! (\zeta + \sum_{l=1}^{k} (a_{1}(\tau_{l}^{b_{1}} - \tau_{l-1}^{b_{1}}) + a_{2}(\tau_{l}^{b_{2}} - \tau_{l-1}^{b_{2}})))^{\sum_{l=1}^{k} (n_{i1l} + n_{i2l})}}.  \tag{2}
\label{eq:2}
\end{align*}

\noindent Here, we denote ${ \boldsymbol{\theta} } = (\zeta, a_1, b_1, a_2, b_2).$  Based on the count data across the intervals, the likelihood function can be written as
\begin{eqnarray*}
L({ \boldsymbol{\theta} })&\propto& \prod_{i=1}^{m} f(\boldsymbol{n_{i}}) \nonumber\\
&=& \frac {\zeta^m \prod_{l=1}^{k}(a_{1} (\tau_{l}^{b_{1}} - \tau_{l-1}^{b_{1}}))^{m n_{i1l}} (a_{2} (\tau_{l}^{b_{2}} - \tau_{l-1}^{b_{2}}))^{m n_{i2l}} (\sum_{l=1}^{k} (n_{i1l} + n_{i2l})!)^m} {\prod_{l=1}^{k} n_{i1l}^m! n_{i2l}^m! (\zeta + \sum_{l=1}^{k} (a_{1}(\tau_{l}^{b_{1}} - \tau_{l-1}^{b_{1}}) + a_{2}(\tau_{l}^{b_{2}} - \tau_{l-1}^{b_{2}})))^{m \sum_{l=1}^{k} (n_{i1l} + n_{i2l})}}. \nonumber \\
\end{eqnarray*}

\noindent Therefore, the log-likelihood can be obtained as
\begin{eqnarray*}
l({ \boldsymbol{\theta} } )
&=& m\log{\zeta}+m\sum_{l=1}^{k}(a_{1} n_{i1l} \log(\tau_{l}^{b_{1}} - \tau_{l-1}^{b_{1}}) + a_{2} n_{i2l} \log(\tau_{l}^{b_{2}} - \tau_{l-1}^{b_{2}}))\nonumber\\&&+ 
m\sum_{l=1}^{k} \log(n_{i1l} + n_{i2l})! - 
m\sum_{l=1}^{k}(\log(n_{i1l}!)+\log(n_{i2l}!)) -  \nonumber\\&&-
m\sum_{l=1}^{k} (n_{i1l}+n_{i2l})\log(\zeta + \sum_{l=1}^{k} (a_{1}(\tau_{l}^{b_{1}} - \tau_{l-1}^{b_{1}}) + a_{2}(\tau_{l}^{b_{2}} - \tau_{l-1}^{b_{2}}))).
\end{eqnarray*}

\noindent  Hence, the maximum likelihood estimator (MLE) of ${ \boldsymbol{\theta} }$, say ${ \hat{\boldsymbol{\theta} }}$ would be derived as
\begin{equation*}
{ \hat{\boldsymbol{\theta} }} = \underset{\boldsymbol{\theta}}{\arg\min}\ l({ \boldsymbol{\theta} } ).
\end{equation*}

\noindent The maximum likelihood estimator (MLE) is widely regarded due to its several properties like efficiency and consistency.  
However, with small deviations from the assumed model conditions, those methods tend to lead to biased and inefficient estimates, raising the need for robust estimation methods.  In the following section, a robust estimation approach will be discussed.

\section{ Minimum Density Power Divergence Estimator}\label{sec3}

In a study of robust estimation, Basu et al. \citep{basu1998robust} were the first to propose the density power divergence method.  They considered the parametric model of densities $\{f_t\}$ with respect to the Lebesgue measure where the unknown parameter $t\in \Theta$ where $\Theta$, the parameter space.  Let $G$ be the class of all distributions with densities $g$ for the same measure.  Under this assumption, the density power divergence measure between density functions $g$ and $f_{t}$ is defined as
\begin{equation*}
d_{\gamma}(g, f_t)=\int \left\{f_t^{1+\gamma}(x)-\left(1+\frac{1}{\gamma}\right) g(x)\,f_t^{\gamma}(x) +\frac{1}{\gamma}g^{1+\gamma}(x)\right\} dx,\quad \gamma>0. \tag{3}
\label{eq:3}
\end{equation*}

The hyperparameter, $\gamma,$ termed as a tuning parameter, plays an essential role in estimating the unknown model parameters. It governs the trade-off between model efficiency and robustness in estimation based on the density power divergence measure.  Further, $d_{\gamma}(g, f_t)$ tends to become the Kullback-Leibler divergence between $g$ and $f_t$ when $\gamma$ tends to 0.  For estimation purposes, $f$ is set as the assumed theoretical model where true density $g$ is replaced by the empirical one, say $\widehat{g}$.  In the current model set-up, the density power divergence between the theoretical probability mass function $f$ and the empirical probability mass function $\widehat{g}$ can be found as follows
\begin{align*}
d_{\gamma}(\widehat{g}, f) &= \sum_{\substack{n_{jl}=0 \\ j=1,2 \\ l=1,2,\dots,k}}^{\infty}
f^{1+\gamma}(\boldsymbol{n}) - \left(1+\frac{1}{\gamma}\right)\frac{1}{m}\sum_{i=1}^{m}f^{\gamma}(\boldsymbol{N_{i}}),\quad \gamma>0 \tag{4}
\label{eq:4}
\end{align*}
where $ \boldsymbol{n}= (n_{11}, n_{21}, \ldots, n_{1k}, n_{2k}).$   For a given tuning parameter ${\gamma}$, minimizing $d_{\gamma}(\widehat{g}, f)$ with respect to the parameter ${ \boldsymbol{\theta} }$ is the same as minimizing $\frac{1}{m} \sum_{i=1}^{m} V_{\boldsymbol{\theta}}(\boldsymbol{N_{i}})$ where
\begin{equation*}
V_{\boldsymbol{\theta}}(\boldsymbol{N_{i}}) = \sum_{\substack{n_{jl}=0 \\ j=1,2 \\ l=1,2,\dots,k}}^{\infty}
f^{1+\gamma}(\boldsymbol{n}) - \left(1+\frac{1}{\gamma}\right) f^{\gamma}(\boldsymbol{N_{i}}).\tag{5}
\label{eq:5}
\end{equation*}
In the particular instance where \(\gamma \rightarrow 0\), the objective function simplifies to

\[
-\frac{1}{m} \sum_{i=1}^{m} \log f(\boldsymbol{n});
\]

\noindent the relevant minimizer is identified as the maximum likelihood estimator (MLE) of ${ \boldsymbol{\theta} }$.  Based on \eqref{eq:4}, and \eqref{eq:5}, the minimum density power divergence estimator of $\boldsymbol{\theta}$
can be derived as
\begin{equation*}
{\hat{\boldsymbol{\theta}} }_{\gamma} = \underset{\boldsymbol{\theta}}{\arg\min} \ \frac{1}{m} \sum_{i=1}^{m} V_{\boldsymbol{\theta}}(\boldsymbol{N_{i}})
 , \qquad \gamma>0. \tag{6}
\label{eq:6}
\end{equation*}

\section{Restricted Minimum Density Power Divergence Estimator}\label{sec4}

In this section, we establish an estimation framework utilizing the density power divergence method under certain constraints applied to the parameter space.  In many practical situations, it is common to observe that the frequency of one recurring event surpasses that of the other event.  To incorporate such scenarios in estimation, restrictions based on inequalities are imposed on the parameter space.  Let $p$ denote the number of model parameters, and consider a set of $r<p$ constraints that define the restricted parameter space in the form
\begin{align*}
\boldsymbol {h}\left(\boldsymbol{\theta}\right) \geq \mathbf{0}_{r} \tag{7}
\label{eq:7}
\end{align*} on $\Theta \subset \mathbb{R}^{p} $, where $\boldsymbol{h}: \mathbb{R}^{p} \rightarrow \mathbb{R}^{r}$ 
is a vector-valued function.  The associated $p \times r$ matrix
\begin{equation*}
\mathbf{H}(\boldsymbol{\theta})=\frac{\partial \boldsymbol{h}^{T}(\boldsymbol{\theta})}{\partial \boldsymbol{\theta}} \tag{8}
\label{eq:8}
\end{equation*} 
exists, is continuous in $\boldsymbol{\theta}$, and satisfies $\operatorname{rank}(\mathbf{H}(\boldsymbol{\theta}))=r$.  Under the model assumption above, the restricted MDPD (RMDPD) estimator can be derived as
\begin{equation*}
{\tilde{\boldsymbol{\theta}} }_{\gamma} = \underset{\boldsymbol{h(\theta) \geq 0_{r}}}{\arg\min} \ \frac{1}{m} \sum_{i=1}^{m} V_{\boldsymbol{\theta}}(\boldsymbol{N_{i}}) , \qquad \gamma>0 .  \tag{9}
\label{eq:9}
\end{equation*}

\noindent Further, leveraging the application of the Karush–Kuhn–Tucker (KKT) conditions, any value of $\boldsymbol{\theta}$ satisfying the following system of equations and inequalities will be the minimizer of \eqref{eq:5}  with the restriction $\boldsymbol {h}\left(\boldsymbol{\theta}\right) \geq \mathbf{0}_{r},$

\
\begin{equation*}
 \frac{\partial}{\partial \boldsymbol{\theta}} \frac{1}{m} \sum_{i=1}^{m} V_{\boldsymbol{\theta}}(\boldsymbol{N_{i}})  + \boldsymbol{\lambda} \boldsymbol{H}\left(\boldsymbol{\theta}\right) =\mathbf{0}_{p},  \tag{10}
\label{eq:10}\\
\end{equation*}

\begin{equation*}
\boldsymbol {h}\left(\boldsymbol{\theta}\right) \geq \mathbf{0}_{r}, \tag{11}
\label{eq:11}\\
\end{equation*}

\begin{equation*}
\boldsymbol{\lambda}^{\prime} \boldsymbol
{h}\left(\boldsymbol{\theta}\right)=\mathbf{0}, \tag{12}
\label{eq:12}\\
\end{equation*}

\begin{equation*}
\boldsymbol{\lambda} \geq \mathbf{0}, \tag{13}
\label{eq:13}\\
\end{equation*}
where $\boldsymbol{\lambda}$ denotes the vector of Lagrange multipliers. 

\noindent In numerical experiments, we have studied a specific form of $\boldsymbol {h}\left(\boldsymbol{\theta}\right)$ defined as 
\begin{equation*}
    \boldsymbol {A}\boldsymbol{\theta} \geq \mathbf{0}_{r},
\end{equation*}
where
\begin{equation*}
\boldsymbol {A} = \left(
\begin{array}{ccccc}
0  & 1 & 0 & -1 & 0 \\
0  & 0 & 1 & 0 & -1
\end{array}
\right). \\
\end{equation*}

\noindent The performance of MLE and MDPDEs has been extensively examined, taking into account this restricted parameter space, in simulation experiments and real data analysis. 

\subsection*{4.1. Computational Algorithm}

The objective function derived in this work may exhibit either convexity or non-convexity.  To effectively handle such cases, we exploit the computationally efficient framework of Sequential Convex Programming (SCP), as introduced by Dinh and Diehl \citep{dinh2010local}.  SCP is an iterative optimization methodology designed for non-convex problems, wherein each iteration involves constructing a convex surrogate--typically via linearization or convexification--of the original objective and constraints around the current estimate of the solution.  This enables the application of convex optimization techniques to refine the solution progressively. Owing to its robustness and real-time feasibility, SCP has gained widespread use in domains such as trajectory optimization, control systems, and robotics.

SCP is widely known for its capability to tackle intricate, constrained problems where other straightforward non-convex optimization techniques would be too computationally intensive.  Here, to find the optimal values of $\boldsymbol{\theta}$, our strategy is to consider the classical gradient; see Duchi et al. \citep{duchi2018scp}.  Let us assume 
\begin{equation*}
    f(\boldsymbol{\theta}) = \frac{1}{m} \sum_{i=1}^{m} V_{\boldsymbol{\theta}}(\boldsymbol{N_{i}}).
\end{equation*}

\noindent Therefore, the MDPD estimation is reduced to a general non-convex optimization problem as follows,
\begin{equation}
\begin{aligned}
\min_{\boldsymbol{\theta} \in \mathbb{R}^p_{+}} \quad & f(\boldsymbol{\theta}) \\
\text{subject to} \quad & h^{*}_j(\boldsymbol{\theta}) \leq 0, \quad j = 1, \ldots, r,
\end{aligned} \tag{14}
\label{eq:14} \\
\end{equation}

\noindent where $h^{*}_j(\boldsymbol{\theta}) = -h_j(\boldsymbol{\theta})$.  At any $k^{th}$ iteration, the SCP approach maintains an estimate of the solution $\boldsymbol{\theta}^{(k)}$, and a trust region $\mathcal{T}^{(k)} \subset \mathbb{R}^p_+$ over which the approximations are deemed valid.  The trust region method is a classical tool within SCP and typically involves an $\ell_2$-norm ball
\[
    \mathcal{T}^{(k)} = \left\{ \boldsymbol{\theta} \in \mathbb{R}^p \,\big|\, \|\boldsymbol{\theta} - \boldsymbol{\theta}^{(k)}\|_2 \leq \rho \right\}, \quad \rho > 0
\]
or an $\ell_1$-norm box
\[
    \mathcal{T}^{(k)} = \left\{ \boldsymbol{\theta} \in \mathbb{R}^p \,\big|\, |\boldsymbol{\theta_i} - \boldsymbol{\theta_i}^{(k)}| \leq \rho_i, \quad i=1, \ldots\, p \right\}.
\]

\noindent Here, we see that the $\ell_2$-norm ball trust region offers greater flexibility in the development of modeling strategies. In contrast, the $\ell_1$-norm box trust region limits the indices $\boldsymbol{\theta_i}$ to engage solely in convex objectives, convex inequality constraints, and unconstrained linear equality constraints (i.e., $\rho_i = +\infty$).  Thereafter, we form a convex approximation of $f$ applying a first-order Taylor approximation as follows
 
\begin{equation*}
    \widehat{f}(\boldsymbol{\theta}) = f(\boldsymbol{\theta}^{(k)}) + \nabla f(\boldsymbol{\theta}^{(k)})^T (\boldsymbol{\theta} - \boldsymbol{\theta}^{(k)}). \tag{15}
\label{eq:15} \\
\end{equation*}

\noindent Similarly, we can proceed with the convex approximations $\tilde{h}^{*}_j$ of the inequality constraints $h^{*}_j$ over $\mathcal{T}^{(k)}$ for $j = 1, \dots, r.$

\noindent Finally, we can solve the convex subproblem as follows,

\begin{equation}
\begin{aligned}
\min_{\boldsymbol{\theta} \in \mathcal{T}^{(k)}} \quad & \widehat{f}(\boldsymbol{\theta}) \\
\text{subject to} \quad & \tilde{h}^{*}_j(\boldsymbol{\theta}) \leq 0, \quad j = 1, \ldots, r.
\end{aligned} \tag{16}
\label{eq:16} \\
\end{equation}

\noindent The steps for SCP are provided in the following algorithm.

\begin{table}[h!]
\renewcommand{\arraystretch}{1.2}
\setlength{\tabcolsep}{4pt}
\begin{tabular}{p{0.95\textwidth}}
\hline
\textbf{Algorithm: Sequential Convex Programming} \\
\hline
\begin{minipage}[t]{0.95\textwidth}
\footnotesize
\begin{itemize}[leftmargin=1.2em]
    \item Initialize $\boldsymbol{\theta}^{(1)}$ and set the trust region radius $\rho_i > 0$, $i = 1, \ldots, p$.
    \item \textbf{Iteration:}
    \begin{itemize}[leftmargin=1em]
        \item[\textbf{-}] Set $k=1$.
        \item[\textbf{-}] Form the affine (first-order Taylor) approximation around $\boldsymbol{\theta}^{(k)}$:
        \[
        \widehat{f}(\boldsymbol{\theta}) = f(\boldsymbol{\theta}^{(k)}) + \nabla f(\boldsymbol{\theta}^{(k)})^T (\boldsymbol{\theta} - \boldsymbol{\theta}^{(k)}).
        \]
        \item[\textbf{-}] Define the trust region:
        \[
        \mathcal{T}^{(k)} = \left\{ \boldsymbol{\theta} \in \mathbb{R}^p \,\big|\, |\boldsymbol{\theta} - \boldsymbol{\theta}^{(k)}| \leq \rho_i, \quad i=1, \ldots\, p \right\}.
        \]
        \item[\textbf{-}] Construct convex approximations $\tilde{h}^{*}_j(\boldsymbol{\theta})$ of the constraints $h^{*}_j(\boldsymbol{\theta})$, for $j = 1, \ldots, r$.
        \item[\textbf{-}] Solve the convex subproblem:
        \begin{equation*}
        \begin{aligned}
            \min_{\boldsymbol{\theta} \in \mathcal{T}^{(k)}} \quad & \widehat{f}(\boldsymbol{\theta}) \\
            \text{subject to} \quad & \tilde{h}^{*}_j(\boldsymbol{\theta}) \leq 0, \quad j = 1, \ldots, r.
        \end{aligned} \\
        \end{equation*}

    \end{itemize}
    \item \textbf{Update:}
    \begin{itemize}[leftmargin=1em]
        \item[\textbf{-}] Set $k=k+1$ and $\boldsymbol{\theta}^{(k+1)}$ as the optimal solution of the above subproblem.
    \end{itemize}
    \item \textbf{Repeat:} Until convergence
        \item Return $\boldsymbol{\theta}^{(k+1)}$ as the optimal estimate.\\
\end{itemize}
\end{minipage} \\
\hline
\end{tabular}
\end{table}

\subsection*{4.2. Asymptotic Properties}

The RMDPD estimator cannot be derived in explicit form, making the examination of its exact sample properties quite challenging. This leads us to investigate the asymptotic properties of RMDPDE. For further development, we modified the regulatory conditions suggested by Basu et al. \citep{basu2018testing} according to our context.

\begin{itemize}
    \item[\textbf{(R1)}] For every $\boldsymbol{\theta} \in \Theta,$ the model distributions $f(\boldsymbol{n})$  have a common support, ensuring that the set $\mathcal{S}=\left\{\boldsymbol{n} \mid f(\boldsymbol{n})>0\right\}$ remains independent of $\boldsymbol{\theta}$.  Additionally, the true distribution $G$ is supported within $\mathcal{S}$, where the corresponding density ${g}$ remains strictly positive.
    
    \item[\textbf{(R2)}] There exists an open subset $\omega$ of the parameter space $\Theta$, consisting of the best fitting parameter $\boldsymbol{\theta}_{0}$ such that for nearly all $\boldsymbol{n} \in \mathcal{S}$ and every $\boldsymbol{\theta} \in \omega$, the density $f(\boldsymbol{n})$ is three times differentiable with respect to $\boldsymbol{\theta}$, and its third-order partial derivatives are continuous in terms of $\boldsymbol{\theta}$.
    
    \item[\textbf{(R3)}] The summations $\sum_{\substack{n_{jl}=0 \\ j=1,2 \\ l=1,2,\dots,k}}^{\infty} f^{\gamma + 1}(\boldsymbol{n})$ and $\sum_{\substack{n_{jl}=0 \\ j=1,2 \\ l=1,2,\dots,k}}^{\infty}  f^{\gamma}(\boldsymbol{n})g(\boldsymbol{n})$ can be differentiated three times with respect to $\boldsymbol{\theta}$, where $g$ represents the true joint mass function of $\boldsymbol{N}$.  Moreover, these derivatives can be taken directly under the summation sign.
    
    \item[\textbf{(R4)}] The $p \times p$ matrix $\boldsymbol{J}_{\gamma}(\boldsymbol{\theta})$, as specified in equation (\hyperref[eq:12]{12}), is positive definite.

    \item[\textbf{(R5)}] There is a function $M_{t u v}(\boldsymbol{n})$ such that $\left|\nabla_{t u v} V_{\boldsymbol{\theta}}(\boldsymbol{n})\right| \leq M_{t u v}(\boldsymbol{n})$ for all $\boldsymbol{\theta} \in \omega$, where $E_{g}\left[M_{t u v}(\boldsymbol{N})\right]=m_{t u v}<\infty$ for all $t, u$, and $v.$
\end{itemize}

\noindent \textbf{Additional condition: (A)}  For every $\boldsymbol{\theta} \in \omega$, the second-order partial derivatives 
$\frac{\partial^{2} \boldsymbol{h}_{t}(\boldsymbol{\theta})}{\partial \boldsymbol{\theta}_{u} \partial \boldsymbol{\theta}_{v}}$ 
are bounded for all $t, u$, and $v$, where 
$\boldsymbol{h}_{t}(\cdot)$ represents the $t^{th}$ element of $\boldsymbol{h}(\cdot)$. \\
 
\noindent Under these assumptions, the following theorem establishes the existence, consistency, and distribution of the restricted robust estimator within an asymptotic framework.   \\

\noindent \textbf{Theorem.}  Suppose the true distribution $G$ belongs to the assumed model, and let $\boldsymbol{\theta_{0}} \in \Theta_{0}$ represent the true parameter.  Assume that the regulatory conditions \textbf{(R1)} - \textbf{(R5)} and the condition \textbf{(A)} hold.  Then, the Restricted Minimum Density Power Divergence estimator $\tilde{\boldsymbol{\theta}}_{\gamma}$ of $\boldsymbol{\theta}$ derived under the constraints $\boldsymbol {h}\left(\boldsymbol{\theta}\right) \geq \mathbf{0}_{r}$ possesses the following properties.

\begin{itemize}
    \item[\textbf{a)}] The set of equations (\hyperref[eq:10]{10}) - (\hyperref[eq:13]{13}) for estimating restricted minimum density power divergence exhibit a consistent sequence of roots $\tilde{\boldsymbol{\theta}}_{m, \gamma}$ (hereafter referred to as $\tilde{\boldsymbol{\theta}}_{\gamma}$ ), that is, $\tilde{\boldsymbol{\theta}}_{\gamma} \xrightarrow[m \rightarrow \infty]{\mathcal{P}} \boldsymbol{\theta}_{0}$.
    
    \item[\textbf{b)}] The asymptotic distribution of the restricted minimum density power divergence estimator  $\tilde{\boldsymbol{\theta}}_{\gamma}$ is expressed as
    \begin{equation*}
    \sqrt{m}\left(\tilde{\boldsymbol{\theta}}_{\gamma}-\boldsymbol{\theta}_{0}\right) \sim N(\mathbf{0}_{p}, \boldsymbol{\Sigma}_{\gamma}(\bm{\theta_0})),
    \end{equation*} 
    \noindent where $\boldsymbol{\Sigma}_{\gamma}(\bm{\theta_0}) = L^{T}(\bm{\theta_0})K_{\gamma}(\bm{\theta_0})L(\bm{\theta_0})$, \\
    
    \indent $\boldsymbol{L} (\bm{\theta_0}) = -[J^{-1}_{\gamma}(\bm{\theta_0}) + J^{-1}_{\gamma}(\bm{\theta_0})(H_{s}(\bm{\theta_0})P^{-1}H^{T}_{s}(\bm{\theta_0}))J^{-1}_{\gamma}(\bm{\theta_0})]$,
    \begin{equation*}
    \boldsymbol{J}_{\gamma}(\boldsymbol{\theta}) = \sum_{\substack{n_{jl}=0 \\ j=1,2 \\ l=1,2,\dots,k}}^{\infty} \boldsymbol{u}(\boldsymbol{n}) \boldsymbol{u}^{T}(\boldsymbol{n}) f^{\gamma+1}(\boldsymbol{n}),
    \tag{17}
    \label{eq:17}
    \end{equation*}
    \begin{equation*}
    \boldsymbol{K}_{\gamma}(\boldsymbol{\theta}) = \sum_{\substack{n_{jl}=0 \\ j=1,2 \\ l=1,2,\dots,k}}^{\infty} \boldsymbol{u}(\boldsymbol{n}) \boldsymbol{u}^{T}(\boldsymbol{n}) f^{2\gamma+1}(\boldsymbol{n}) - \boldsymbol{\xi}_{\gamma}(\boldsymbol{\theta}) \boldsymbol{\xi}_{\gamma}^{T}(\boldsymbol{\theta}),
    \tag{18}
    \label{eq:18}
    \end{equation*}
    \begin{equation*}
    \boldsymbol{\xi}_{\gamma}(\boldsymbol{\theta}) = \sum_{\substack{n_{jl}=0 \\ j=1,2 \\ l=1,2,\dots,k}}^{\infty} \boldsymbol{u}(\boldsymbol{n}) f^{\gamma+1}(\boldsymbol{n}),\\  \quad
    \boldsymbol{u}(\boldsymbol{n})=\frac{\partial}{\partial \boldsymbol{\theta}} \log f(\boldsymbol{n}),
    \end{equation*}                                                              
    and $\boldsymbol{I}_{\boldsymbol{\theta}}(\boldsymbol{n})=-\frac{\partial}{\partial \boldsymbol{\theta}} \boldsymbol{u}(\boldsymbol{n})$ is the information function of the model. \\
    
    PROOF. See the Appendix.
\end{itemize}

\section{Optimal tuning Parameter}\label{sec5}

In estimation based on the density power divergence measure, the tuning parameter \( \gamma \) is crucial for balancing the trade-off between efficiency and robustness of the RMDPD estimator in the presence of data contamination.  Selecting an optimal value for \( \gamma \) is vital to ensure the robustness of the estimator without compromising its efficiency.  Several studies have been performed in the literature in this direction.  Warwick and Jones \citep{warwick2005choosing} proposed a data-driven procedure to achieve the optimal tuning parameter for the MDPDE by minimizing its asymptotic mean squared error (MSE).  The expression for this minimization is defined as
\[
\widehat{\text{MSE}}(\boldsymbol{\gamma}) = (\boldsymbol{\hat{\theta}_\gamma} - \boldsymbol{\theta_P})^T (\boldsymbol{\hat{\theta}_\gamma} - \boldsymbol{\theta_P}) + \frac{1}{m} \operatorname{Tr} \left( \boldsymbol{\Sigma}_{\gamma}(\bm{\hat{\theta}_\gamma}) \right),
\]

\noindent where $\boldsymbol{\theta_P}$ denotes the pilot estimator, and \(\operatorname{Tr}(\cdot)\) represents the trace of the matrix.  However, the procedure introduced by Warwick and Jones \citep{warwick2005choosing} has a limitation, as the optimal value of $\gamma$  is dependent on the choice of the pilot estimator.  To mitigate this issue, Basak et al. \citep{basak2021optimal} introduced an iterative approach.  This is referred to as the Iterative Warwick and Jones approach, more commonly known as the "IWJ" technique.  In this method, the pilot estimator is substituted at every iteration with the revised estimator calculated using the optimal value of $\gamma$, and this process continues until stabilization is achieved.  

Recently, Sugasawa and Yonekura \citep{sugasawa2021selection} delved into this problem by exploiting the asymptotic approximation of Hyvärinen score with unnormalized models based on robust divergence.  This criterion eliminates the need for the estimator's asymptotic variance formula, which is required in the previous method.  The Hyvärinen score-matching method, proposed by Hyvärinen and Dayan \citep{hyvarinen2005estimation}, is designed to estimate statistical models with an unknown normalization constant.  Rather than relying on traditional approaches such as Markov Chain Monte Carlo (MCMC) or approximating the normalization constant, this method minimizes the expected squared distance between the gradient of the log density derived from the model and that obtained from the observed data.  However, this method is only applicable when all variables have a continuous value and are defined over $\mathbb{R}^n$.  Consequently, Hyvärinen \citep{hyvarinen2007some} expanded this framework to accommodate binary variables and to estimate non-normalized models defined within the non-negative real domain, denoted as \(\mathbb{R}^n_+\).  Later, Lyu \citep{lyu2012interpretation} introduced a generalized version of score matching by incorporating general linear operators into the Fisher divergence.  He also presented a specific application of this generalized score-matching to discrete data, demonstrating it as a more intuitive extension of score-matching for discrete cases.  Recently, Xu et al. \citep{xu2022generalized} introduced a generalized score-matching method tailored for regression-type models with count data.  They also developed a comprehensive theoretical framework for estimating score matching in cases involving independent observations within a fixed design set-up for regression-type models.  This method is referred to as the "GSM" method, for short.

In this work, our novel contribution includes determining the optimal value of the tuning parameter $\gamma$ through a generalized score-matching technique, proposed by Xu et al. \citep{xu2022generalized}.  The following developments are made in this direction.  As discussed in \ref{sec2}, in this study we get \( m \) independent observation vectors $(n_{i11}, n_{i21}, \ldots, n_{i1k}, n_{i2k})$ for $i=1,\ldots,m.$  Define, $p(\boldsymbol{n_{i}} | \boldsymbol{\theta})$ as
\begin{equation*}
p(\boldsymbol{n_{i}} | \boldsymbol{\theta}) = \exp\left(-\left(\sum_{\substack{n_{jl}=0 \\ j=1,2 \\ l=1,2,\dots,k}}^{\infty}
f^{1+\gamma}(\boldsymbol{n}) - \left(1+\frac{1}{\gamma}\right)f^{\gamma}(\boldsymbol{n_{i}})\right)\right)
\tag{19}
\label{eq:19}
\end{equation*} 

\noindent Note that, $p( \boldsymbol{n}| \boldsymbol{\theta})$ poses to be a parametrized model without the normalization constant.  Assume $q( \boldsymbol{n })$ be the true probability model, where $q(\boldsymbol{n}) \propto p(\boldsymbol{n})$, when the parametrized model is correctly specified.  According to Lyu \citep{lyu2012interpretation}, a generalized form of score matching can be achieved by substituting the gradient operator with a general linear operator \( \mathcal{L} \).  As a result, the generalized score matching objective function, denoted by \( D_\mathcal{L}(q, p) \), is defined  as follows  
\begin{equation*}
D_\mathcal{L}(q, p) =  
\mathbb{E} \left[ 
\left\| 
\frac{\mathcal{L}q(\boldsymbol{n})}{q(\boldsymbol{n})} - 
\frac{\mathcal{L}p(\boldsymbol{n}|\boldsymbol{\theta})}{p(\boldsymbol{n}|\boldsymbol{\theta})}\right\|^2 
\right], \tag{20}
\label{eq:20}
\end{equation*} 
where \( \| \cdot \| \) denotes the Euclidean norm.  Following Xu et al. \citep{xu2022generalized}, we define a linear operator \( \mathcal{L} \) as  
\begin{equation*}
\mathcal{L} p\left(\boldsymbol{n} \mid \boldsymbol{\theta}\right) =
\left(
\begin{array}{c}
\vdots  \\
\mathcal{L}_{s} p\left(\boldsymbol{n} \mid \boldsymbol{\theta}\right) \\
\vdots 
\end{array}
\right)
=
\left( 
\begin{array}{c}
\vdots  \\
p\left(\boldsymbol{n}^{(s+)} \mid \boldsymbol{\theta}\right) - p\left(\boldsymbol{n} \mid \boldsymbol{\theta}\right) \\
\vdots 
\end{array}
\right),
\end{equation*}

\noindent where $\boldsymbol{n}^{(s+)} $ is derived from $\boldsymbol{n}$ by adding 1 to its $s^{th}$ element without altering the other elements.  Similarly, $\boldsymbol{n}^{(s-)}$ can be obtained by subtracting 1 from its $s^{th}$ element while leaving the other elements unchanged.  If \( \boldsymbol{n} \) reaches the boundary of its range, we set \( p\left(\boldsymbol{n}^{(s+)} \mid \boldsymbol{\theta}\right) = 0 \), and \( p\left(\boldsymbol{n}^{(s-)} \mid \boldsymbol{\theta}\right) = 0 \).  In the discrete multivariate scenario, the operator $\mathcal{L} p\left(\boldsymbol{n} \mid \boldsymbol{\theta}\right)$ serves as a discrete approximation to the gradient of \( p\left(\boldsymbol{n} \mid \boldsymbol{\theta}\right) \) at  \( \boldsymbol{n} \). 
By rewriting 
\begin{equation*}
     \frac{\mathcal{L} p\left(\boldsymbol{n} \mid \boldsymbol{\theta}\right)}{p\left(\boldsymbol{n} \mid \boldsymbol{\theta}\right)} = \frac{p\left(\boldsymbol{n}^{(s+)} \mid \boldsymbol{\theta}\right)}{p\left(\boldsymbol{n} \mid \boldsymbol{\theta}\right)} - 1,
\end{equation*} 

\noindent we focus on minimizing the difference between the ratio \( \frac{p\left(\boldsymbol{n}^{(s+)} \mid \boldsymbol{\theta}\right)}{p\left(\boldsymbol{n_{i}} \mid \boldsymbol{\theta}\right)} \) and the corresponding empirical ratio \( \frac{q\left(\boldsymbol{n}^{(s+)}\right)}{q\left(\boldsymbol{n_{i}}\right)} \).

\noindent In order to handle potential division by zero in these slope expressions, we adopt the following transformation proposed by Hyvärinen \citep{hyvarinen2007some}, 
\[
t(u) = \frac{1}{1 + u}.
\]  
This ensures that infinite ratios resulting from zero probabilities are transformed to \( t(\infty) = 0 \).  Upon implementing this transformation, we arrive at the subsequent objective function for \( q \) and \( p \), 
\begin{align*}
D_{\mathrm{GSM}}\left(q, p\right) = & \mathbb{E}\left\{ \left[ t\left( \frac{p\left(\boldsymbol{n}^{(s+)} \mid \boldsymbol{\theta}\right)}{p\left(\boldsymbol{n} \mid \boldsymbol{\theta}\right)} \right) - t\left( \frac{q\left(\boldsymbol{n}^{(s+)}\right)}{q\left(\boldsymbol{n}\right)} \right) \right]^2 \right. \\
& \left. + \left[ t\left( \frac{p\left(\boldsymbol{n} \mid \boldsymbol{\theta}\right)}{p\left(\boldsymbol{n}^{(s-)} \mid \boldsymbol{\theta}\right)} \right) - t\left( \frac{q\left(\boldsymbol{n}\right)}{q\left(\boldsymbol{n}^{(s-)} \right)} \right) \right]^2 \right\}. \tag{21}
\label{eq:21}
\end{align*}

\noindent Based on \eqref{eq:21}, the empirical estimator of the objective function is given by
\begin{align*}
\widehat{D}_{\mathrm{GSM}}\left(q, p\right) = & \frac{1}{m} \sum_{i=1}^{m} \Bigg[ 
t\left(\frac{p\left(\boldsymbol{n}^{(s+)} \mid \boldsymbol{\theta}\right)}{p\left(\boldsymbol{n} \mid \boldsymbol{\theta}\right)}\right)^{2} +  t\left(\frac{p\left(\boldsymbol{n} \mid \boldsymbol{\theta}\right)}{p\left(\boldsymbol{n}^{(s-)} \mid \boldsymbol{\theta}\right)}\right)^{2} \\
& \quad - 2 t\left(\frac{p\left(\boldsymbol{n}^{(s+)} \mid \boldsymbol{\theta}\right)}{p\left(\boldsymbol{n} \mid \boldsymbol{\theta}\right)}\right) 
\Bigg].
\tag{22}
\end{align*}

\noindent The optimal tuning parameter \( \gamma \) is then derived as 
\begin{equation*}
{\hat{\boldsymbol{\gamma}}} = \underset{\boldsymbol{\theta}}{\arg\min} \ \widehat{D}_{\mathrm{GSM}}\left(q, p\right). \tag{23}
\end{equation*}

\noindent For comparison, we have used the GSM method and the IWJ method in our numerical experiments to determine which is the most efficient in identifying the optimal tuning parameter.

\section{Numerical Experiments}\label{sec6}

Numerical experiments, including simulations and real-world data analysis, assess the effectiveness of the proposed methods, demonstrating their robustness, accuracy, and practical use while also identifying areas for improvement.

\subsection*{6.1. Simulation Study}

In this section, a simulation environment is developed exploiting 1000 Monte Carlo generations to evaluate the performance of MLE and MDPDEs.  Additionally, a comparative analysis is performed to examine the effectiveness of these estimators in both restricted and unrestricted scenarios.  The number of occurrences of the recurrent events is considered to follow the non-homogeneous Poisson process.  In our formulation, we assume that the recurrent event \( N_{i1l} \) occurs more frequently than \( N_{i2l} \).  To reflect this assumption in the model, we impose the inequality constraints \( a_1 \geq a_2 \) and \( b_1 \geq b_2 \), ensuring that the rate parameters associated with \( N_{i1l} \) are at least as large as those of \( N_{i2l} \).  These constraints help incorporate the inherent ordering of event frequencies into the estimation process.  The true values of the model parameters are assigned as \( \zeta = 4.5 \), \( a_1 = 0.9 \), \( b_1 = 0.5 \), \( a_2 = 0.6 \), \( b_2 = 0.2 \), and the time intervals are defined as \( (0.01, 0.35] \), \( (0.35, 0.69] \), and \( (0.69, 1.12] \).  The experiment is performed with a fixed sample size \( m = 100 \) and tuning parameters \( \gamma = 0.2, 0.5, 0.8 \).

Moreover, the behavior of the estimators has been thoroughly examined for both pure and contaminated data.  Contamination is incorporated in the data set by generating an $\epsilon$ proportional of values from the Inverse-Gaussian distribution.  In this context, we examine four distinct contamination rates, $\epsilon = 0.006, 0.044, 0.085, 0.153$, and analyze the results accordingly.  Here, we have utilized the sequential convex programming (SCP) algorithm to obtain the MLE and MDPDEs.

An examination of the parameter estimates reported in \hyperref[tab1]{Table \ref{tab1}} and \hyperref[tab2]{Table \ref{tab2}} reveals that the inequality constraints \( a_1 \geq a_2 \) and \( b_1 \geq b_2 \) are consistently satisfied under both the unrestricted and restricted estimation schemes.

\begin{figure}[H]
    \centering

    \subfloat[\(\zeta\)]{\includegraphics[width=0.35\textwidth]{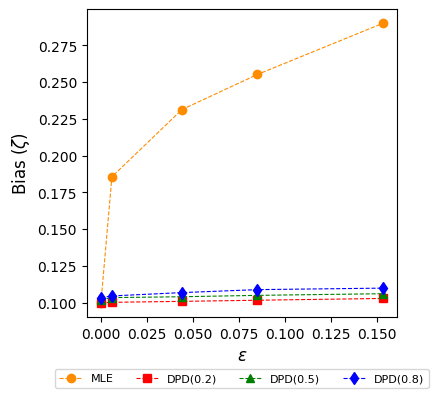}}\hspace{0.5cm}
    \subfloat[\(a_1\)]{\includegraphics[width=0.35\textwidth]{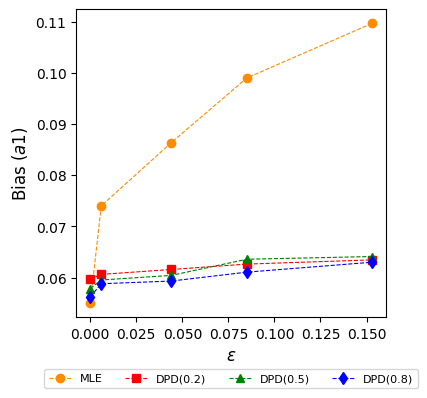}}

    \vskip 0.5cm
    \subfloat[\(b_1\)]{\includegraphics[width=0.35\textwidth]{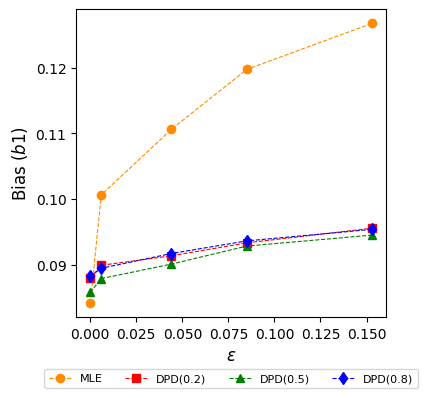}}

    \vskip 0.5cm
    \subfloat[\(a_2\)]{\includegraphics[width=0.35\textwidth]{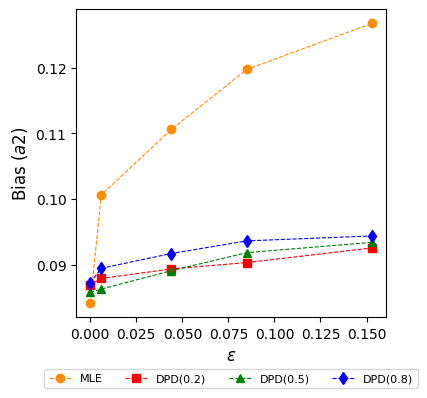}}\hspace{0.5cm}
    \subfloat[\(b_2\)]{\includegraphics[width=0.35\textwidth]{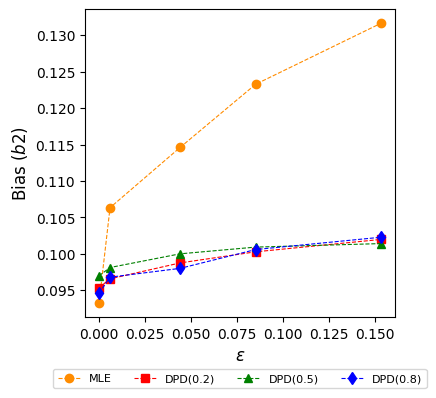}}

    \caption{Bias of estimates of parameters under Unrestricted case}
    \label{fig1}
\end{figure}

\begin{figure}[H]
    \centering

    \subfloat[\(\zeta\)]{\includegraphics[width=0.35\textwidth]{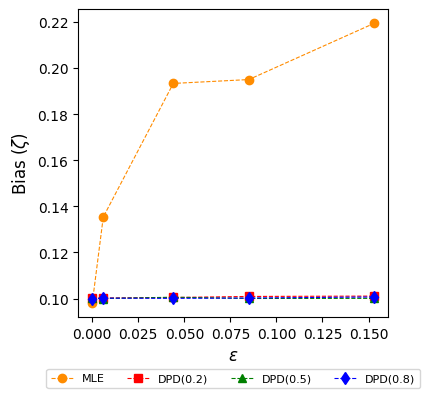}}\hspace{0.5cm}
    \subfloat[\(a_1\)]{\includegraphics[width=0.35\textwidth]{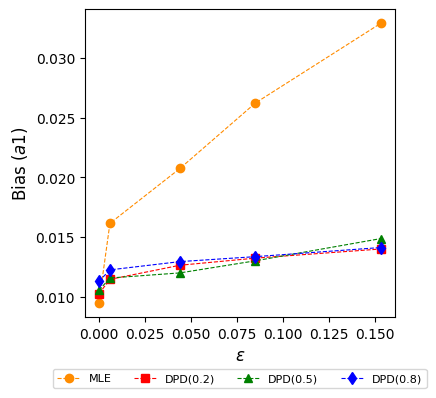}}

    \vskip 0.5cm
    \subfloat[\(b_1\)]{\includegraphics[width=0.35\textwidth]{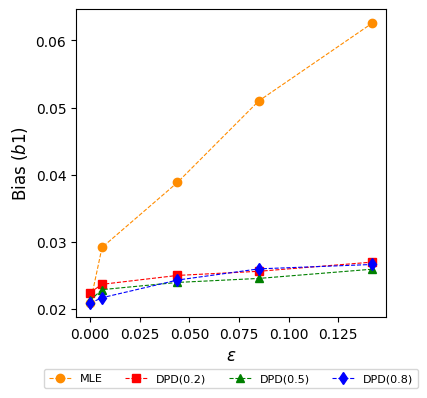}}

    \vskip 0.5cm
    \subfloat[\(a_2\)]{\includegraphics[width=0.35\textwidth]{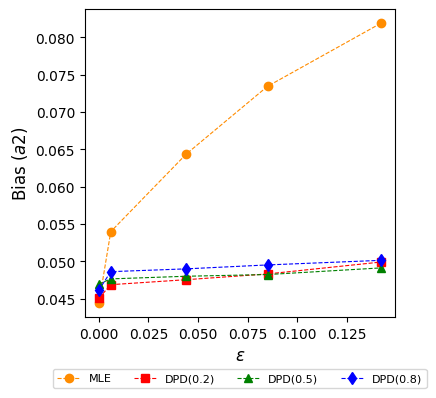}}\hspace{0.5cm}
    \subfloat[\(b_2\)]{\includegraphics[width=0.35\textwidth]{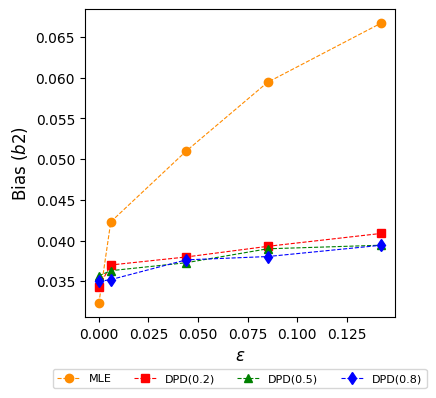}}

    \caption{Bias of estimates of parameters under Restricted case}
    \label{fig2}
\end{figure}

Further, in both unrestricted and restricted scenarios, it has been observed that MLE exhibits lower bias and MSE than MDPDEs when dealing with pure data, making it the superior choice in such cases.  However, under various levels of data contamination, MDPDEs show reduced bias and MSE in comparison to MLE.  Moreover, the increase in bias for MDPDEs from pure to contaminated data is less significant than that for MLE.  This indicates that MDPDE is a robust estimator that does not significantly compromise efficiency.  The graphical representations of bias estimates for model parameters in both pure and contaminated datasets under restricted and unrestricted scenarios are provided in \hyperref[fig1]{Figure \ref{fig1}} and \hyperref[fig2]{Figure \ref{fig2}}, respectively.  Owing to the superiority of MDPDE on the basis of performance demonstrated here, it is a preferable choice over MLE when the data is contaminated.

From \hyperref[tab3]{Table \ref{tab3}} and \hyperref[tab4]{Table \ref{tab4}}, it is evident that, regardless of the contamination levels in the simulated data, there is a significant decrease in the MSE of MLE and the MDPDE when transitioning from the unrestricted case to the restricted case.  Therefore, with prior knowledge of inequality constraints in the parameter space, the restricted MDPDE is the most favorable option compared to the other methods applied here.

\begin{table}[htb!]
    \begin{center}
    \fontsize{10}{12}\selectfont
    \caption{Estimated values of the parameters under Unrestricted case}\label{tab1}
    \vspace{0.2cm}
    \begin{tabular}{lcccccc}
    \hline
    \textbf{m = 100} & $\boldsymbol{\epsilon = 0.000}$ & $\boldsymbol{\epsilon = 0.006}$ & $\boldsymbol{\epsilon = 0.044}$ & $\boldsymbol{\epsilon = 0.085}$ & $\boldsymbol{\epsilon = 0.153}$ \\ 
    \hline
    \hspace{1.61cm} $\boldsymbol{\zeta}$& 4.599616 & 4.685667 & 4.731346 & 4.755229 & 4.789967 \\
    \hspace{1.61cm} \textbf{a\textsubscript{1}} & 0.954962 & 0.974019 & 0.986361 & 0.998989 & 1.009662 \\
    \textbf{MLE} \hspace{17.5pt} \textbf{b\textsubscript{1}} & 0.584146 & 0.600669 & 0.610611 & 0.619753 & 0.626762 \\
    \hspace{1.61cm} \textbf{a\textsubscript{2}} & 0.684146 & 0.700669 & 0.710611 & 0.719753 & 0.726762 \\
    \hspace{1.61cm} \textbf{b\textsubscript{2}} & 0.293205 & 0.306336 & 0.314596 & 0.323264 & 0.331632 \\
    \hline
    \hspace{1.61cm} $\boldsymbol{\zeta}$ & 4.600026 & 4.600325 & 4.600974 & 4.601743 & 4.602945 \\
    \hspace{1.61cm} \textbf{a\textsubscript{1}}  & 0.959689 & 0.960625 & 0.961580 & 0.962635 & 0.963452 \\
    $\boldsymbol{\gamma = 0.2}$ \hspace{7.5pt} \textbf{b\textsubscript{1}} & 0.587931 & 0.589936 & 0.591382 & 0.593357 & 0.595581 \\
    \hspace{1.61cm} \textbf{a\textsubscript{2}} & 0.686931 & 0.687936 & 0.689382 & 0.690357 & 0.692581 \\
    \hspace{1.61cm} \textbf{b\textsubscript{2}} & 0.295359 & 0.296568 & 0.298754 & 0.300279 & 0.301976 \\
    \hline
    \hspace{1.61cm} $\boldsymbol{\zeta}$ & 4.602129 & 4.603525 & 4.604113 & 4.605053 & 4.606175 \\
    \hspace{1.61cm} \textbf{a\textsubscript{1}} & 0.957855 & 0.959529 & 0.960423 & 0.963579 & 0.964110 \\
    $\boldsymbol{\gamma = 0.5}$ \hspace{7.5pt} \textbf{b\textsubscript{1}} & 0.585801 & 0.587922 & 0.590115 & 0.592865 & 0.594531 \\
    \hspace{1.61cm} \textbf{a\textsubscript{2}} & 0.685801 & 0.686322 & 0.689115 & 0.691865 & 0.693431 \\
    \hspace{1.61cm} \textbf{b\textsubscript{2}} & 0.296975 & 0.298123 & 0.300002 & 0.300925 & 0.301407 \\
    \hline
    \hspace{1.61cm} $\boldsymbol{\zeta}$ & 4.603475 & 4.604672 & 4.606924 & 4.608952 & 4.609963 \\
    \hspace{1.61cm} \textbf{a\textsubscript{1}}  & 0.956227 & 0.958796 & 0.959321 & 0.961059 & 0.962998 \\
    $\boldsymbol{\gamma = 0.8}$ \hspace{7.5pt} \textbf{b\textsubscript{1}} & 0.588284 & 0.589483 & 0.591731 & 0.593654 & 0.595411 \\
    \hspace{1.61cm} \textbf{a\textsubscript{2}} & 0.687284 & 0.689483 & 0.691731 & 0.693654 & 0.694411 \\
    \hspace{1.61cm} \textbf{b\textsubscript{2}} & 0.294589 & 0.296798 & 0.298005 & 0.300588 & 0.302278 \\
    \hline
    \end{tabular}
    \end{center}
\end{table}

Further, the optimal behavior of the tuning parameter $\gamma$ is studied for pure data and contaminated data.  In addition to finding the value of $\gamma$, the MSE of the model parameters is calculated under varying contamination levels with a fixed sample size of $m = 100$.  This same analysis has been conducted for both the GSM and IWJ methods for comparison purposes.  \hyperref[tab5]{Table \ref{tab5}} and \hyperref[tab6]{Table \ref{tab6}} show the optimal tuning parameters identified through the GSM and IWJ approaches at $\boldsymbol{\epsilon} = 0.000, 0.006, 0.044, 0.085, 0.153$.  According to these optimal values of $\gamma$, the MSE of the model parameters is calculated to assess the performance of each method.  The results reveal that the MSE obtained through the GSM method is significantly lower than that achieved using the IWJ method, indicating the superior accuracy and effectiveness of the GSM approach in identifying the optimal tuning parameter.  Consequently, it can be inferred that GSM provides a more effective and reliable approach for finding the optimal tuning parameter $\gamma$ compared to the IWJ method.

\begin{table}[htb!]
    \begin{center}
    \fontsize{10}{12}\selectfont
    \caption{Estimated values of the parameters under Restricted case}\label{tab2}
    \vspace{0.2cm}
    \begin{tabular}{lcccccc}
    \hline
    \textbf{m = 100} & $\boldsymbol{\epsilon = 0.000}$ & $\boldsymbol{\epsilon = 0.006}$ & $\boldsymbol{\epsilon = 0.044}$ & $\boldsymbol{\epsilon = 0.085}$ & $\boldsymbol{\epsilon = 0.153}$ \\ 
    \hline
    \hspace{1.61cm} $\boldsymbol{\zeta}$ & 4.597892 & 4.635432 & 4.693444 & 4.695126 & 4.719544 \\
    \hspace{1.61cm} \textbf{a\textsubscript{1}} & 0.909448 & 0.916151 & 0.920746 & 0.926234 & 0.932904 \\
    \textbf{MLE} \hspace{17.5pt} \textbf{b\textsubscript{1}} & 0.520801 & 0.529136 & 0.538825 & 0.550985 & 0.562547 \\
    \hspace{1.61cm} \textbf{a\textsubscript{2}} & 0.644396 & 0.653997 & 0.664375 & 0.673456 & 0.681883 \\
    \hspace{1.61cm} \textbf{b\textsubscript{2}} & 0.232281 & 0.242296 & 0.250991 & 0.259443 & 0.266687 \\
    \hline
    \hspace{1.61cm} $\boldsymbol{\zeta}$ & 4.600052 & 4.600212 & 4.600440 & 4.600891 & 4.601045 \\
    \hspace{1.61cm} \textbf{a\textsubscript{1}} & 0.910248 & 0.911457 & 0.912640 & 0.913237 & 0.914001 \\
    $\boldsymbol{\gamma = 0.2}$ \hspace{7.5pt} \textbf{b\textsubscript{1}} & 0.522408 & 0.523640 & 0.524965 & 0.525582 & 0.526946 \\
    \hspace{1.61cm} \textbf{a\textsubscript{2}} & 0.645081 & 0.646897 & 0.647564 & 0.648336 & 0.649924 \\
    \hspace{1.61cm} \textbf{b\textsubscript{2}} & 0.234308 & 0.236996 & 0.237980 & 0.239287 & 0.240867 \\
    \hline
    \hspace{1.61cm} $\boldsymbol{\zeta}$ & 4.600012 & 4.600000 & 4.600560 & 4.600011 & 4.600069 \\
    \hspace{1.61cm} \textbf{a\textsubscript{1}} & 0.910528 & 0.911572 & 0.911999 & 0.913014 & 0.914872 \\
    $\boldsymbol{\gamma = 0.5}$ \hspace{7.5pt} \textbf{b\textsubscript{1}} & 0.521265 & 0.522847 & 0.523921 & 0.524518 & 0.525889 \\
    \hspace{1.61cm} \textbf{a\textsubscript{2}} & 0.646988 & 0.647671 & 0.648025 & 0.648253 & 0.649140 \\
    \hspace{1.61cm} \textbf{b\textsubscript{2}} & 0.235658 & 0.236296 & 0.237274 & 0.238985 & 0.239423 \\
    \hline
    \hspace{1.61cm} $\boldsymbol{\zeta}$ & 4.600000 & 4.600156 & 4.600063 & 4.600215 & 4.600658 \\
    \hspace{1.61cm} \textbf{a\textsubscript{1}} & 0.911289 & 0.912253 & 0.912945 & 0.913361 & 0.914125 \\
    $\boldsymbol{\gamma = 0.8}$ \hspace{7.5pt} \textbf{b\textsubscript{1}} & 0.520836 & 0.521642 & 0.524275 & 0.525943 & 0.526596 \\
    \hspace{1.61cm} \textbf{a\textsubscript{2}} & 0.646186 & 0.648639 & 0.649015 & 0.649541 & 0.650154 \\
    \hspace{1.61cm} \textbf{b\textsubscript{2}} & 0.234975 & 0.235197 & 0.237625 & 0.238032 & 0.239413 \\
    \hline
    \end{tabular}
    \end{center}
\end{table}

\subsection*{6.2. Real Data Analysis}

This research examines survey data from the National Health and Nutrition Examination Survey (NHANES) carried out from 2017 until March 2020.  As a result of the COVID-19 pandemic, field operations for the 2019–2020 cycle were suspended, resulting in data that is incomplete and not representative of the nation.  To tackle this issue, a nationally representative dataset was formed by merging data from the 2017–2018 cycle with pre-pandemic information from the 2019–March 2020 cycle.  This data is retrieved from the adjacent link, National Health and Nutrition Examination Survey (NHANES);   NHANES 2017-March 2020 Pre-pandemic Questionnaire Data - \href{https://www.cdc.gov/nchs/nhanes}{https://www.cdc.gov/nchs/nhanes}. 
Here, the study centers on 50 subjects with equivalent follow-up durations, exploring \( N_{i1l}\) and \( N_{i2l}\), which represent the number of cold/flu and pneumonia events, correspondingly, in the period \( (\tau_{l-1}, \tau_l] \).  These subjects have been observed every 3 months, over an entire period of 12 months.

\begin{table}[htb!]
\begin{center}
\fontsize{10}{12}\selectfont
\caption{MSE of estimates of parameters under Unrestricted case}\label{tab3}
\vspace{0.2cm}
\begin{tabular}{lcccccc}
\hline
\textbf{m = 100} & $\boldsymbol{\epsilon = 0.000}$ & $\boldsymbol{\epsilon = 0.006}$ & $\boldsymbol{\epsilon = 0.044}$ & $\boldsymbol{\epsilon = 0.085}$ & $\boldsymbol{\epsilon = 0.153}$ \\ 
\hline
\hspace{1.61cm} $\boldsymbol{\zeta}$ & 0.009983 & 0.021282 & 0.039132 & 0.050652 & 0.061431 \\
\hspace{1.61cm} \textbf{a\textsubscript{1}} & 0.002659 & 0.004533 & 0.005967 & 0.007151 & 0.008087 \\
\textbf{MLE} \hspace{0.6cm} \textbf{b\textsubscript{1}} & 0.004597 & 0.005888 & 0.006484 & 0.007597 & 0.008521 \\
\hspace{1.61cm} \textbf{a\textsubscript{2}} & 0.004021 & 0.005782 & 0.006487 & 0.008123 & 0.009353 \\
\hspace{1.61cm} \textbf{b\textsubscript{2}} & 0.004233 & 0.005999 & 0.006650 & 0.007248 & 0.008747 \\
\hline
\hspace{1.61cm} $\boldsymbol{\zeta}$ & 0.010010 & 0.010112 & 0.012147 & 0.013074 & 0.015075 \\
\hspace{1.61cm} \textbf{a\textsubscript{1}}  & 0.003218 & 0.003517 & 0.003742 & 0.004047 & 0.004513 \\
$\boldsymbol{\gamma = 0.2}$ \hspace{0.2cm} \textbf{b\textsubscript{1}} & 0.004864 & 0.004981 & 0.005142 & 0.005535 & 0.005691 \\
\hspace{1.61cm} \textbf{a\textsubscript{2}} & 0.004266 & 0.005071 & 0.005256 & 0.005397 & 0.005788 \\
\hspace{1.61cm} \textbf{b\textsubscript{2}} & 0.004686 & 0.004781 & 0.005042 & 0.005651 & 0.006040 \\
\hline
\hspace{1.61cm} $\boldsymbol{\zeta}$ & 0.010045 & 0.010127 & 0.011147 & 0.012002 & 0.013142 \\
\hspace{1.61cm} \textbf{a\textsubscript{1}}  & 0.003619 & 0.004218 & 0.004315 & 0.005017 & 0.005318 \\
$\boldsymbol{\gamma = 0.5}$ \hspace{0.2cm} \textbf{b\textsubscript{1}} & 0.004081 & 0.004356 & 0.004605 & 0.004864 & 0.005251 \\
\hspace{1.61cm} \textbf{a\textsubscript{2}} & 0.004266 & 0.004923 & 0.005063 & 0.005071 & 0.005366 \\
\hspace{1.61cm} \textbf{b\textsubscript{2}} & 0.004389 & 0.004525 & 0.004887 & 0.005686 & 0.005991 \\
\hline
\hspace{1.61cm} $\boldsymbol{\zeta}$ & 0.010000 & 0.011015 & 0.012002 & 0.012089 & 0.013012 \\
\hspace{1.61cm} \textbf{a\textsubscript{1}}  & 0.003619 & 0.004018 & 0.004123 & 0.004218 & 0.004417 \\
$\boldsymbol{\gamma = 0.8}$ \hspace{0.2cm} \textbf{b\textsubscript{1}} & 0.004081 & 0.008864 & 0.009043 & 0.008864 & 0.008864 \\
\hspace{1.61cm} \textbf{a\textsubscript{2}} & 0.004661 & 0.004961 & 0.005134 & 0.005266 & 0.005371 \\
\hspace{1.61cm} \textbf{b\textsubscript{2}} & 0.004922 & 0.005312 & 0.005563 & 0.005686 & 0.005716 \\
\hline
\end{tabular}
\end{center}
\end{table}

\begin{table}[htb!]
\begin{center}
\fontsize{10}{12}\selectfont
\caption{MSE of estimates of parameters under Restricted case}\label{tab4}
\vspace{0.2cm}
\begin{tabular}{lcccccc}
\hline
\textbf{m = 100} & $\boldsymbol{\epsilon = 0.000}$ & $\boldsymbol{\epsilon = 0.006}$ & $\boldsymbol{\epsilon = 0.044}$ & $\boldsymbol{\epsilon = 0.085}$ & $\boldsymbol{\epsilon = 0.153}$ \\ 
\hline
\hspace{1.61cm} $\boldsymbol{\zeta}$ & 0.009064 & 0.012095 & 0.025129 & 0.035234 & 0.038696 \\
\hspace{1.61cm} \textbf{a\textsubscript{1}} & 0.002824 & 0.004354 & 0.005814 & 0.006404 & 0.007299 \\
\textbf{MLE} \hspace{0.6cm} \textbf{b\textsubscript{1}} & 0.002223 & 0.003026 & 0.003538 & 0.004674 & 0.005206 \\
\hspace{1.61cm} \textbf{a\textsubscript{2}} & 0.002349 & 0.003304 & 0.004723 & 0.005711 & 0.006564 \\
\hspace{1.61cm} \textbf{b\textsubscript{2}} & 0.002691 & 0.004415 & 0.005463 & 0.006706 & 0.007087 \\
\hline
\hspace{1.61cm} $\boldsymbol{\zeta}$ & 0.010001 & 0.010097 & 0.011001 & 0.012142 & 0.013956 \\
\hspace{1.61cm} \textbf{a\textsubscript{1}} & 0.002948 & 0.002993 & 0.003220 & 0.003303 & 0.003802 \\
$\boldsymbol{\gamma = 0.2}$ \hspace{0.2cm} \textbf{b\textsubscript{1}} & 0.002446 & 0.002783 & 0.003115 & 0.003656 & 0.004069 \\
\hspace{1.61cm} \textbf{a\textsubscript{2}} & 0.002637 & 0.002813 & 0.003161 & 0.003567 & 0.003656 \\
\hspace{1.61cm} \textbf{b\textsubscript{2}} & 0.003654 & 0.003799 & 0.004158 & 0.004399 & 0.004668 \\
\hline
\hspace{1.61cm} $\boldsymbol{\zeta}$ & 0.009856 & 0.010000 & 0.010161 & 0.010002 & 0.010443 \\
\hspace{1.61cm} \textbf{a\textsubscript{1}} & 0.003220 & 0.003419 & 0.003604 & 0.003918 & 0.004159 \\
$\boldsymbol{\gamma = 0.5}$ \hspace{0.2cm} \textbf{b\textsubscript{1}} & 0.003115 & 0.003598 & 0.003897 & 0.004081 & 0.004266 \\
\hspace{1.61cm} \textbf{a\textsubscript{2}} & 0.003661 & 0.003768 & 0.003891 & 0.004166 & 0.004260 \\
\hspace{1.61cm} \textbf{b\textsubscript{2}} & 0.003358 & 0.003492 & 0.003716 & 0.003922 & 0.004169 \\
\hline
\hspace{1.61cm} $\boldsymbol{\zeta}$ & 0.009975 & 0.010000 & 0.010981 & 0.010000 & 0.010012 \\
\hspace{1.61cm} \textbf{a\textsubscript{1}} & 0.002619 & 0.002820 & 0.003012 & 0.003218 & 0.003433 \\
$\boldsymbol{\gamma = 0.8}$ \hspace{0.2cm} \textbf{b\textsubscript{1}} & 0.002932 & 0.003198 & 0.003303 & 0.003598 & 0.003781 \\
\hspace{1.61cm} \textbf{a\textsubscript{2}} & 0.002673 & 0.002856 & 0.003017 & 0.003166 & 0.003366 \\
\hspace{1.61cm} \textbf{b\textsubscript{2}} & 0.002830 & 0.003058 & 0.003228 & 0.003458 & 0.003622 \\
\hline
\end{tabular}
\end{center}
\end{table}

\begin{table}[htb!]
\begin{center}
\fontsize{10}{12}\selectfont
\caption{Comparison between tuning parameters of the two methods}\label{tab5}
\vspace{0.2cm}
\begin{tabular}{lcccccc}
\hline
\multirow{2}{*}{\textbf{m = 100}} & \multicolumn{2}{c}{$\boldsymbol{\epsilon = 0.000}$} & \multicolumn{2}{c}{$\boldsymbol{\epsilon = 0.006}$} & \multicolumn{2}{c}{$\boldsymbol{\epsilon = 0.044}$} \\ 
& \textbf{GSM} & \textbf{IWJ} & \textbf{GSM} & \textbf{IWJ} & \textbf{GSM} & \textbf{IWJ}\\ 
\hline
\hspace{1.21cm} $\boldsymbol{\gamma}$ & 0.55 & 0.60 & 0.48 & 0.40 & 0.59 & 0.35 \\
\hline
\hspace{1.21cm} $\boldsymbol{\zeta}$ & 0.000811 & 0.001198 & 0.000530 & 0.001451 & 0.000472 & 0.001878 \\
\hspace{1.21cm} \textbf{a\textsubscript{1}} & 0.001074 & 0.001960 & 0.001918 & 0.002562 & 0.001174 & 0.002535 \\
\textbf{MSE} \hspace{0.2cm} \textbf{b\textsubscript{1}} & 0.000069 & 0.000103 & 0.000604 & 0.000922 & 0.000057 & 0.000631 \\
\hspace{1.21cm} \textbf{a\textsubscript{2}} & 0.000485 & 0.001099 & 0.000478 & 0.001005 & 0.000571 & 0.001327 \\
\hspace{1.21cm} \textbf{b\textsubscript{2}} & 0.000519 & 0.000935 & 0.000700 & 0.001547 & 0.000731 & 0.001378 \\
\hline
\end{tabular}
\end{center}
\end{table}

\begin{table}[htb!]
\begin{center}
\fontsize{10}{12}\selectfont
\caption{Comparison between tuning parameters of the two methods}\label{tab6}
\vspace{0.2cm}
\begin{tabular}{lcccccc}
\hline
\multirow{2}{*}{\textbf{m = 100}} & \multicolumn{2}{c}{$\boldsymbol{\epsilon = 0.085}$} & \multicolumn{2}{c}{$\boldsymbol{\epsilon = 0.153}$} \\ 
& \textbf{GSM} & \textbf{IWJ} & \textbf{GSM} & \textbf{IWJ} \\ 
\hline
\hspace{1.21cm} $\boldsymbol{\gamma}$ & 0.58 & 0.34 & 0.41 & 0.47 \\
\hline
\hspace{1.21cm} $\boldsymbol{\zeta}$ & 0.000591 & 0.001717 & 0.000662 & 0.001447 \\
\hspace{1.21cm} \textbf{a\textsubscript{1}} & 0.001115 & 0.003672 & 0.001458 & 0.002965 \\
\textbf{MSE} \hspace{0.2cm} \textbf{b\textsubscript{1}} & 0.000060 & 0.000908 & 0.000805 & 0.001261 \\
\hspace{1.21cm} \textbf{a\textsubscript{2}} & 0.000443 & 0.001190 & 0.000998 & 0.001330 \\
\hspace{1.21cm} \textbf{b\textsubscript{2}} & 0.000766 & 0.001393 & 0.000967 & 0.001857 \\
\hline
\end{tabular}
\end{center}
\end{table}

\begin{figure}[H]
    \centering
    \includegraphics[width=0.85\textwidth]{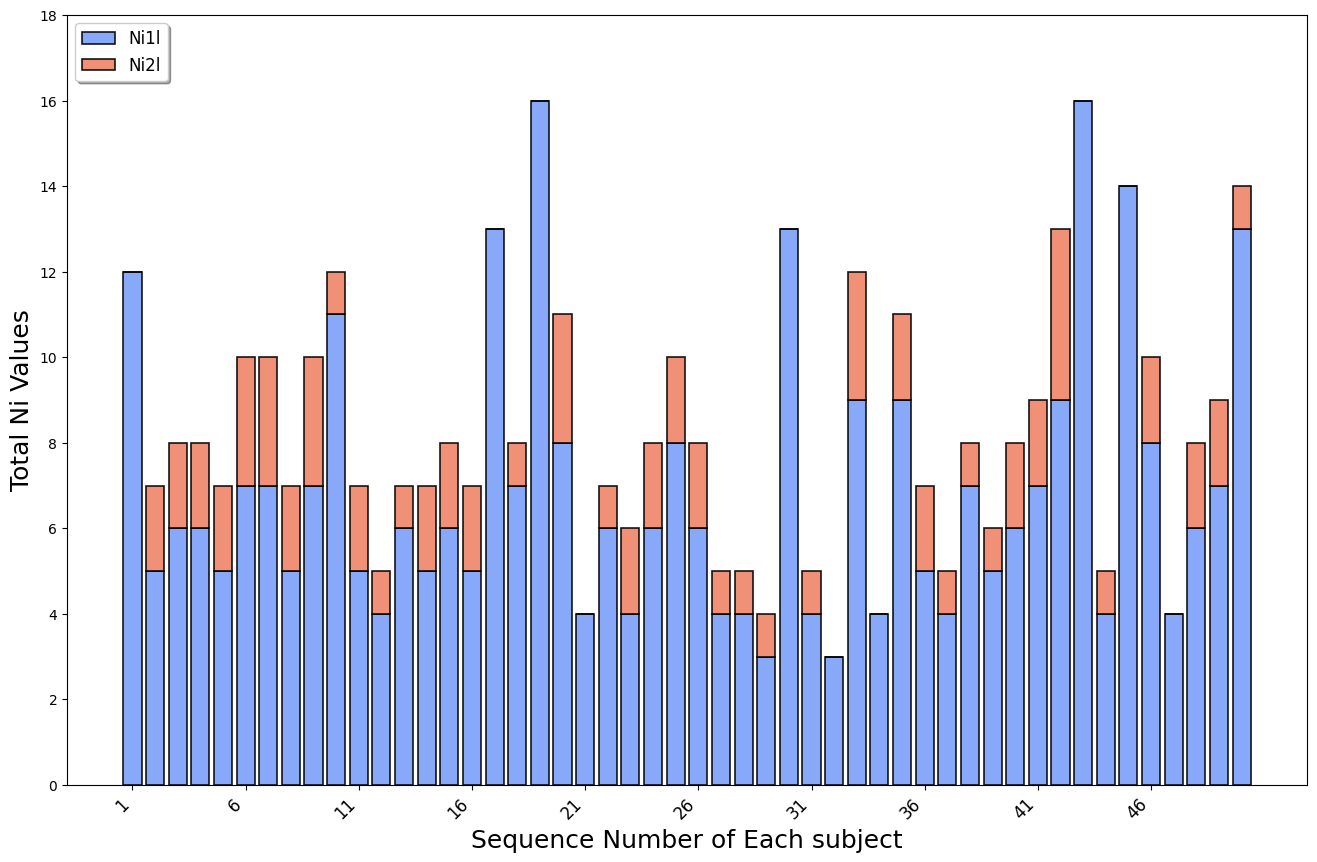} 
    \caption{Instances of cold/flu ($N_{i1l}$) and pneumonia ($N_{i2l}$) in different individuals}
    \label{fig3}
\end{figure}

\begin{figure}[H]
    \centering
    \includegraphics[width=0.85\textwidth]{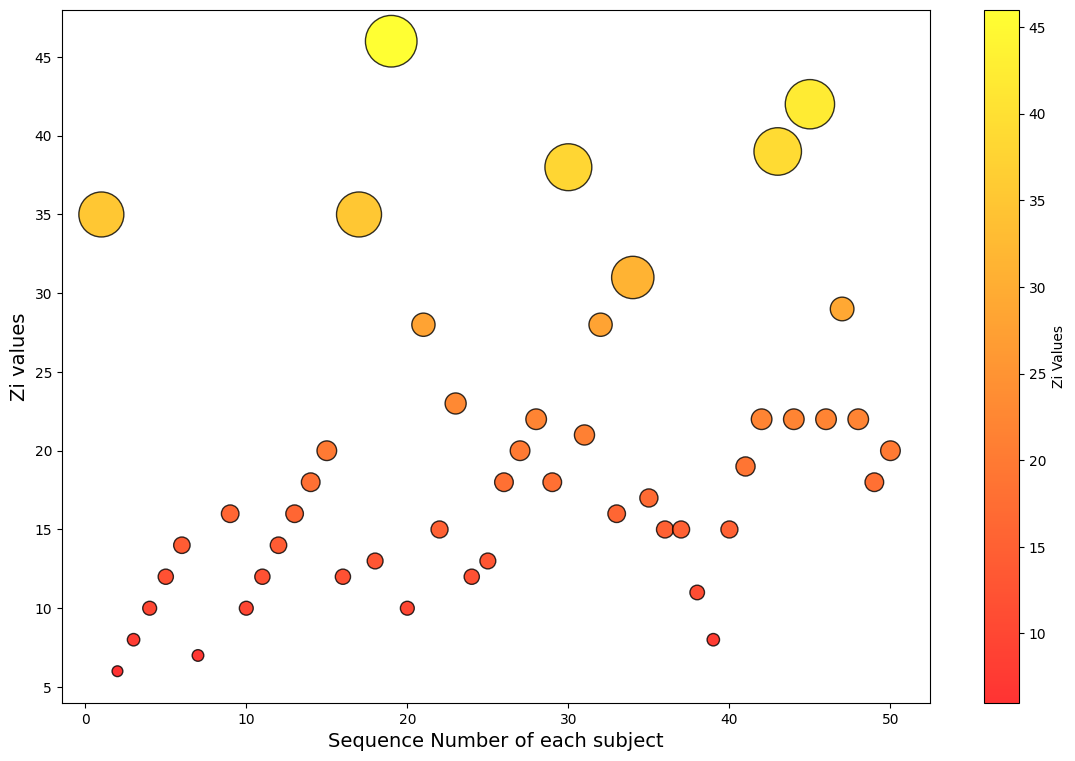}
    \caption{Number of hours each individual does physical activities weekly}
    \label{fig4} 
\end{figure}

The real dataset reveals a few outliers, particularly in the relationship between frequent cold/flu and the likelihood of developing pneumonia.  In most cases, individuals who have experienced cold or flu episodes four to five times within a specific interval tend to develop pneumonia at least once in that period.  However, 6 out of 50 individuals deviate from this pattern.  Despite suffering from cold/flu five to six times in a given timeframe, they did not develop pneumonia at all.  A key observation is that these individuals engage in a higher-than-average duration of physical activity, as depicted in the stacked bar graph (\textcolor{blue}{\hyperref[fig3]{Figure \ref{fig3}}}), which visualizes the recurrence of cold/flu ($N_{i1l}$) and pneumonia ($N_{i2l}$) across individuals.  Moreover, Figure~3 clearly illustrates that cold/flu episodes, represented by \( N_{i1l} \), occur more frequently than pneumonia episodes, denoted by \( N_{i2l} \). This observed pattern justifies the inequality constraints \( a_1 \geq a_2 \) and \( b_1 \geq b_2 \), which are imposed to reflect the higher recurrence rate of cold/flu compared to pneumonia in the model.  The frailty variable \( Z_i \), representing the number of hours a person engages in physical activities weekly, emerges as a crucial determinant in this anomaly.   It appears that higher physical activity levels serve as a protective factor, preventing pneumonia during the follow-up period.  This correlation is further illustrated in the bubble chart (\textcolor{blue}{\hyperref[fig4]{Figure \ref{fig4}}}), where these six individuals are represented by large bubbles, reflecting their high exercise rates.  Keeping in view the current scenario, the robust restricted minimum density power divergence (RMDPD) may be a suitable estimation technique.

Further, a dissimilarity measure-based test statistic to evaluate the suitability of the nonhomogeneous Poisson process for the data is defined as
\begin{equation*}
T = \frac{1}{mk} \sum_{i=1}^m \sum_{l=1}^{k} \left[\left|\frac{n_{i1l} - e_{i1l}} {e_{i1l}} \right| + 
\left| \frac{n_{i2l} - e_{i2l}} {e_{i2l}}\right|\right].
\tag{24}
\end{equation*} 

\noindent where \( n_{i1l} \) and \( e_{i1l} \) represent the observed and expected counts of cold or flu occurrences for the \( i^{\text{th}} \) individual during the interval \( (\tau_{l-1}, \tau_{l}] \), respectively, and \( n_{i2l} \) and \( e_{i2l} \) represent the observed and expected counts of pneumonia occurrences for the same individual within the interval \( (\tau_{l-1}, \tau_{l}] \), with \( i = 1, 2, 3, \ldots, 50 \).  The bootstrap (Bt) technique is employed to estimate the p-value of the test using 10,000 generated samples.  A test statistic is computed for each sample.  The proportion of bootstrap test statistics that exceed the test statistic from the real data, T, divided by the total number of bootstrap samples, provides the approximate p-value.  A smaller test statistic indicates a greater likelihood that the data conform to the nonhomogeneous Poisson process.
The computed values of the test statistic, T along with the associated p-values for various tuning parameters $(\gamma = 0.2, 0.5, 0.8)$ are reported in {Table~\ref{tab7}}.  These p-values confirm the suitability of the nonhomogeneous Poisson process for modeling data.  Therefore, it can be inferred that the instances of cold or flu and pneumonia in people can be adequately represented by the nonhomogeneous Poisson process.

\begin{table}[htb!]
\centering
\fontsize{10}{12}\selectfont
\caption{Test statistic and p-value for different $\gamma$}\label{tab7}
\vspace{0.2cm} 
\begin{tabular}{cccccccccc}
\hline
$\boldsymbol{\gamma}$ & \textbf{T} & \textbf{p-value} \\ \hline
0.2    & 2.052768  & 0.510800 \\
0.5    & 1.933244  & 0.654200  \\
0.8    & 2.126994  & 0.359600   \\ \hline
\end{tabular}
\end{table}

The MLE and RMDPDE of the model parameters are obtained, and the bootstrap estimates of Bias and MSE generated from 10,000 bootstrap samples are computed.  The initial parameter value of the Sequential convex programming algorithm is set as (6.75, 0.82, 0.63, 0.42, 0.23), which is selected through a comprehensive grid search method.  It is noticed that MDPDE shows lower Bootstrap Bias and MSE (Bt Bias and Bt MSE) in comparison to MLE. 

Considering that the dataset contains some outliers, the efficacy of RMDPDE in this context aligns with the aim of the current study.  Consequently, the restricted MDPDE has demonstrated outstanding performance on the real dataset.  \textcolor{blue}{\hyperref[tab8]{Table \ref{tab8}}} displays the parameter estimates under both unrestricted and restricted settings.  These results indicate that the inequality constraints, namely $a_1 \geq a_2$ and $b_1 \geq b_2$, are satisfied.  The Bootstrap Bias and MSE (Bt Bias and Bt MSE) of MLE and RMDPDE for different $\gamma$ values are presented in \textcolor{blue}{\hyperref[tab9]{Table \ref{tab9}}} for the unrestricted case and in \textcolor{blue}{\hyperref[tab10]{Table \ref{tab10}}} for the restricted case.  From these tables, it is evident that both MLE and RMDPDE exhibit a notable reduction in Bootstrap Bias and MSE under the restricted framework, highlighting the advantage of incorporating inequality constraints.  This suggests that the restricted MDPDE offers superior performance among the estimators considered.  The optimal tuning parameter, \(\gamma\), for the model, is determined using real data.  Based on this value of \(\gamma\), the model parameters \(\zeta\), \(a_1\), \(b_1\), \(a_2\), and \(b_2\) are estimated, and the bootstrap bias and mean squared error (MSE) of the parameters are calculated.  The results obtained using GSM method are compared with those from the IWJ method.  \textcolor{blue}{\hyperref[tab11]{Table \ref{tab11}}} provides a detailed summary of the optimal tuning parameter \(\gamma\), the estimated model parameters, and their corresponding Bt Bias and Bt MSE values.  It has been noted that the estimates derived from the GSM method consistently show lower Bt Bias and Bt MSE compared to those obtained from the IWJ method.  This comparison reveals that the GSM method outperforms the IWJ method by achieving lower MSE, underscoring its effectiveness in tuning parameter selection.

\begin{table}[htb!]
\begin{center}
\fontsize{10}{12}\selectfont
\caption{Estimated values of the parameters under Unrestricted and Restricted cases}\label{tab8}
\vspace{0.2cm}
\begin{tabular}{llccccc}
\hline
& & $\boldsymbol{\zeta}$ & \textbf{a\textsubscript{1}} & \textbf{b\textsubscript{1}} & \textbf{a\textsubscript{2}} & \textbf{b\textsubscript{2}}\\ 
\hline
\multirow{4}{*}{\textbf{Unrestricted}} & \textbf{MLE} & 6.872250 & 0.914841 & 0.725169 & 0.330802 & 0.193911 \\
& $\boldsymbol{\gamma = 0.2}$ & 6.766742 & 0.844871 & 0.655538 & 0.406642 & 0.218145 \\
& $\boldsymbol{\gamma = 0.5}$ & 6.761689 & 0.840328 & 0.652417 & 0.404723
& 0.213945 \\
& $\boldsymbol{\gamma = 0.8}$ & 6.739611 & 0.803824 & 0.613709 & 0.409200
& 0.219717 \\
\\
\multirow{4}{*}{\textbf{Restricted}} & \textbf{MLE} & 6.799012 & 0.829115 & 0.638724 & 0.425638 & 0.232427 \\
& $\boldsymbol{\gamma = 0.2}$ & 6.752852 & 0.822441 & 0.632660 & 0.423361 & 0.231178 \\
& $\boldsymbol{\gamma = 0.5}$ & 6.752557 & 0.821889 & 0.632196 & 0.422114 & 0.231008 \\
& $\boldsymbol{\gamma = 0.8}$ & 6.751632 & 0.821745 & 0.631684 & 0.421248 & 0.230181 \\
\hline
\end{tabular}
\end{center}
\end{table}

\begin{table}[htb!]
\begin{center}
\fontsize{10}{12}\selectfont
\caption{Bootstrap Bias and MSE of estimates of parameters under Unrestricted case}\label{tab9}
\vspace{0.2cm}
\begin{tabular}{llccccc}
\hline
& & $\boldsymbol{\zeta}$ & \textbf{a\textsubscript{1}} & \textbf{b\textsubscript{1}} & \textbf{a\textsubscript{2}} & \textbf{b\textsubscript{2}}\\ 
\hline
\multirow{4}{*}{\textbf{Bt Bias}} & \textbf{MLE} & 0.122140 & 0.094960 & 0.095150 & -0.089300 & -0.036238 \\
& $\boldsymbol{\gamma = 0.2}$ & 0.016694 & 0.024955 & 0.025646 & -0.013667 & -0.011995 \\
& $\boldsymbol{\gamma = 0.5}$ & 0.011575 & 0.020198 & 0.022350 & -0.015346
& -0.016190 \\
& $\boldsymbol{\gamma = 0.8}$ & -0.010576 & -0.016552 & -0.016488 & -0.010609
& -0.010283 \\
\\
\multirow{4}{*}{\textbf{Bt MSE}} & \textbf{MLE} & 0.049346 & 0.009017 & 0.008864 & 0.005825 & 0.002493 \\
& $\boldsymbol{\gamma = 0.2}$ & 0.002993 & 0.002345 & 0.002416 & 0.003212 & 0.001096 \\
& $\boldsymbol{\gamma = 0.5}$ & 0.002491 & 0.001907 & 0.002012 & 0.002017 & 0.001008 \\
& $\boldsymbol{\gamma = 0.8}$ & 0.001715 & 0.001808 & 0.001894 & 0.001591 & 0.000945 \\
\hline
\end{tabular}
\end{center}
\end{table}

\begin{table}[htb!]
\begin{center}
\fontsize{10}{12}\selectfont
\caption{Bootstrap Bias and MSE of estimates of parameters under Restricted case}\label{tab10}
\vspace{0.2cm}
\begin{tabular}{llccccc}
\hline
& & $\boldsymbol{\zeta}$ & \textbf{a\textsubscript{1}} & \textbf{b\textsubscript{1}} & \textbf{a\textsubscript{2}} & \textbf{b\textsubscript{2}}\\ 
\hline
\multirow{4}{*}{\textbf{Bt Bias}} & \textbf{MLE} & -0.024464 & -0.035629 & -0.021732 & -0.040737 & -0.009072 \\
& $\boldsymbol{\gamma = 0.2}$ & 0.001589 & 0.002138 & 0.002899 & -0.003837 & -0.002089 \\
& $\boldsymbol{\gamma = 0.5}$ & 0.001289 & 0.001744 & 0.002390 & -0.003081 & -0.001932 \\
& $\boldsymbol{\gamma = 0.8}$ & -0.001943 & -0.001389 & -0.001938 & -0.003128 & -0.001867 \\
\\
\multirow{4}{*}{\textbf{Bt MSE}} & \textbf{MLE} & 0.002436 & 0.003273 & 0.004123 & 0.002287 & 0.000931 \\
& $\boldsymbol{\gamma = 0.2}$ & 0.000199 & 0.000136 & 0.000243 & 0.000172 & 0.000087 \\
& $\boldsymbol{\gamma = 0.5}$ & 0.000161 & 0.000112 & 0.000204 & 0.000141 & 0.000085 \\
& $\boldsymbol{\gamma = 0.8}$ & 0.000163 & 0.000148 & 0.000433 & 0.000158 & 0.000035 \\
\hline
\end{tabular}
\end{center}
\end{table}

\begin{table}[htb!]
\centering
\fontsize{10}{12}\selectfont
\caption{Comparison between tuning parameters of the two methods}\label{tab11}
\vspace{0.2cm} 
\begin{tabular}{cccccccccc}
\hline
\rule{0pt}{12pt} & $\boldsymbol{\widehat{\gamma}}$ & $\boldsymbol{\widehat{\zeta}}$ & $\widehat{\mathbf{a}}_{\mathbf{1}}$
 & $\widehat{\mathbf{b}}_{\mathbf{1}}$
 & $\widehat{\mathbf{a}}_{\mathbf{2}}$
 & $\widehat{\mathbf{b}}_{\mathbf{2}}$
 \\  
\hline
\textbf{GSM} & 0.36   & 6.768821 & 0.884960 & 0.71415 & 0.419214 & 0.229217 \\
\textbf{Bt Bias} &  & 0.002271 & 0.004258 & 0.006042 & -0.000523 & -0.000289 \\
\textbf{Bt MSE} & & 0.000223 & 0.000266 & 0.000508 & 0.000016 & 0.000008 \\
\\
\textbf{IWJ} & 0.20 &   6.942140 & 0.994960 & 0.774150 & 0.411938 & 0.221970 \\
\textbf{Bt Bias} &  & 0.003883 & 0.007909 & 0.011158 & -0.001042 & -0.000591 \\
\textbf{Bt MSE} & & 0.000377 & 0.000496 & 0.000940 & 0.000036 & 0.000019 \\
\hline
\end{tabular}
\end{table}

\section{Conclusion}\label{sec7}

In this work, we have explored a robust estimation method utilizing the density power divergence measure based on panel count data.  In order to make our model more realistic, we incorporate a set of inequality constraints within the parameter space, leading to the restricted minimum density power divergence estimation (RMDPDE) method.  The large sample properties of the RMDPD estimator are studied analytically, while the finite sample properties of the proposed estimator are investigated through rigorous simulation studies.  Numerical experiments reveal that the resulting estimators are more robust than the conventional maximum likelihood estimators (MLE) when handling contaminated data.  Additionally, it is noted that in the presence of inequality restrictions, the restricted robust estimator surpasses the performance of restricted MLE in terms of both robustness and efficiency.  Further, we employ a generalized score-matching (GSM) method to determine the optimal value of the tuning parameter, $\gamma$.  Through comprehensive numerical analyses, we can deduce that the GSM approach is more effective at determining the optimal tuning parameter than the existing IWJ method.  Finally, our analysis of the recent Questionnaire data from the National Health and Nutrition Examination Survey (NHANES) illustrates the ability of the RMDPD estimation method to handle outliers. 

In the future, a different form of divergence measure may be used to estimate the parameters of the panel count data model.  We can additionally incorporate multiple observable covariates into the model. Works in those areas are underway, and we are hopeful about sharing those results shortly.

\section*{Data availability}

National Health and Nutrition Examination Survey (NHANES);   NHANES 2017-March 2020 Pre-pandemic Questionnaire Data -
\href{https://www.cdc.gov/nchs/nhanes}{https://www.cdc.gov/nchs/nhanes}.

\section*{Acknowledgements}

The authors extend their gratitude to the anonymous reviewers, the administrative editor, the associate editor, and the editor for their valuable and constructive suggestions, which have greatly improved the quality of this paper.

\section*{Declaration of competing interest}

The authors confirm that they have no recognized financial conflicts of interest or personal relationships that could have influenced the work presented in this paper.

\section*{Funding}

This research was not supported by any specific grant from public, commercial, or non-profit funding organizations.

\appendix
\section{Proof of Theorem (a)}

\noindent According to Hanson \citep{hanson1965inequality}, the demonstration of the theorem unfolds as described below. 

\noindent Define, 
\begin{align}
g_{m}(\boldsymbol{\theta}) = & \frac{\sum_{i=1}^{m} V_{\boldsymbol{\theta}}(\boldsymbol{N_{i}})}{m(1+\gamma)}.  \tag{A.1}
\label{eq:22}
\end{align}
\noindent We adhere to the proof outline in Aitchison and Silvey \citep{aitchison1958maximum}.  For \( \boldsymbol{\theta}_0 \in  \Theta_0 \), the true parameter value, define \( U_{\alpha} \) as the set \(\left\{ \boldsymbol{\theta} : \left\| \boldsymbol{\theta}  - \boldsymbol{\theta}_0 \right\| \leq \alpha  \right\} \).  There exists a constant \( \delta > 0 \) such that \(  U_{\delta} \in \omega \), where \( \omega \) is specified in regulatory condition, (\textbf{R2}). 
We aim to show that, under the constraint $\boldsymbol {h}\left(\boldsymbol{\theta}\right) \geq \mathbf{0}_{r}$, the estimating equation for the density power divergence possesses a root in \(  U_{\delta} \) as \( m \to \infty \). \\

\noindent \textbf{Conditions on $h$}

\begin{itemize}
    \item[$\boldsymbol{\mathscr{H}_{1}}$.] For every $\boldsymbol{\theta} \in U_{a}$, the partial derivatives $\frac{\partial \boldsymbol{h}_{k}(\boldsymbol{\theta})}{\partial \boldsymbol{\theta}_{i}}$, for $i = 1, 2, \dots, p$ and $k = 1, 2, \dots, r$, exist and are continuous functions of $\boldsymbol{\theta}$.
    
    \item[$\boldsymbol{\mathscr{H}_{2}}$.] For every $\boldsymbol{\theta} \in U_{\alpha}$, the second-order partial derivatives $\frac{\partial^{2} \boldsymbol{h}_{k}(\boldsymbol{\theta})}{\partial \boldsymbol{\theta}_{i} \partial \boldsymbol{\theta}_{j}}$, for $i, j = 1, 2, \dots, p$ and $k = 1, 2, \dots, r$, exist and satisfy $\left|\frac{\partial^{2} \boldsymbol{h}_{k}(\boldsymbol{\theta})}{\partial \boldsymbol{\theta}_{i} \partial \boldsymbol{\theta}_{j}}\right| < 2\kappa_{2}$, where $\kappa_{2}$ is a given constant.

    \item[$\boldsymbol{\mathscr{H}_{3}}$.] The $p \times r$ matrix $\boldsymbol{H}_{\boldsymbol{\theta}} = \left[\frac{\partial \boldsymbol{h}_{j}(\boldsymbol{\theta})}{\partial \boldsymbol{\theta}_{i}}\right]$, evaluated at $\boldsymbol{\theta} = \boldsymbol{\theta}_{0}$, has maximum rank. If $r > p$, then at most $p$ components of $\boldsymbol{h}$ can vanish at each point of $U_{\alpha}$. For the components $\boldsymbol{h}_{k}$, say for $k = 1, 2, \dots, s$, that vanish at some point in $U_{\alpha}$, the $p \times s$ matrix $\left[\frac{\partial \boldsymbol{h}_{k}(\boldsymbol{\theta})}{\partial \boldsymbol{\theta}_{i}}\right]$ also has maximum rank.
\end{itemize}

\noindent The restricted minimum density power divergence (RMDPD) estimator of $\boldsymbol{\theta}$, i.e., $\widetilde{\boldsymbol{\theta}}_{\gamma}$, satisfies the following system of estimating equations and inequalities,

\begin{equation*}
m \frac{\partial}{\partial \boldsymbol{\theta}} g_{m}(\boldsymbol{\theta}) + \boldsymbol{\lambda} \boldsymbol{H}\left(\boldsymbol{\theta}\right) =\mathbf{0}_{p} ,\tag{A.2}
\label{eq:23}\\
\end{equation*}

\begin{equation*}
\boldsymbol {h}\left(\boldsymbol{\theta}\right) \geq \mathbf{0}_{r}, \tag{A.3}
\label{eq:24}\\
\end{equation*}

\begin{equation*}
\boldsymbol{\lambda}^{\prime} \boldsymbol
{h}\left(\boldsymbol{\theta}\right)=\mathbf{0}, \tag{A.4}
\label{eq:25}\\
\end{equation*}

\begin{equation*}
\boldsymbol{\lambda} \geq \mathbf{0}, \tag{A.5}
\label{eq:26}\\
\end{equation*}
where $\boldsymbol{\lambda}$ denotes the vector of Lagrange multipliers.

\noindent Now, Taylor series expansions of (\hyperref[eq:23]{A.2}) around $\boldsymbol{\theta}_{0}$ gives

\begin{align*}
\left.m \frac{\partial}{\partial \boldsymbol{\theta}} g_{m}(\boldsymbol{\theta})\right|_{\boldsymbol{\theta}=\boldsymbol{\theta}_{0}} & +\left.m \frac{\partial^{2}}{\partial \boldsymbol{\theta}^{T} \partial \boldsymbol{\theta}} g_{m}(\boldsymbol{\theta})\right|_{\boldsymbol{\theta}=\boldsymbol{\theta}_{0}}\left(\boldsymbol{\theta}-\boldsymbol{\theta}_{0}\right)+\boldsymbol{\nu}_{m}^{(1)}(\boldsymbol{\theta}) \\
& + \boldsymbol{\lambda}\boldsymbol{H}\left(\boldsymbol{\theta}_{0}\right) +\boldsymbol{\nu}_{m}^{(2)}(\boldsymbol{\theta})=\mathbf{0}_{p},  \tag{A.6}
\label{eq:27}
\end{align*}
\noindent where\\

\begin{itemize}
    \item[$\boldsymbol{(i)}$] The vector $\boldsymbol{\nu}_{m}^{(1)}(\boldsymbol{\theta})$ has its $c^{th}$ element given by $\frac{1}{2}\left(\boldsymbol{\theta}-\boldsymbol{\theta}_{0}\right)^{T} \boldsymbol{G}_{c}\left(\boldsymbol{\theta}-\boldsymbol{\theta}_{0}\right)$, where $\boldsymbol{G}_{c}$ is a matrix.  The $(u, v)^{th}$ component of $\boldsymbol{G}_{c}$ is defined as \\
    $\partial^{3} g_{m}(\boldsymbol{\theta}) /\left.\partial \boldsymbol{\theta}_{c} \partial \boldsymbol{\theta}_{u} \partial \boldsymbol{\theta}_{v}\right|_{\boldsymbol{\theta}=\boldsymbol{\theta}_{0}^{(c, 1)}}$, with indices $u, v, c=1,2, \dots, p$.  Additionally, the norm condition $\left\|\boldsymbol{\theta}^{(c, 1)}-\boldsymbol{\theta}_{0}\right\| \leq\left\|\boldsymbol{\theta}-\boldsymbol{\theta}_{0}\right\| $ holds.
    
    \item[$\boldsymbol{(ii)}$] The vector $\boldsymbol{\nu}_{m}^{(2)}(\boldsymbol{\theta})$ has its $c^{th}$ element given by $\frac{1}{2}\left(\boldsymbol{\theta}-\boldsymbol{\theta}_{0}\right)^{T} \boldsymbol{W}_{c}\left(\boldsymbol{\theta}-\boldsymbol{\theta}_{0}\right)$, where $\boldsymbol{W}_{c}$ is a matrix.  The $(u, v)^{th}$ component of $\boldsymbol{W}_{c}$ is defined as \\
    $\partial^{2} \boldsymbol{h}_{m}(\boldsymbol{\theta}) /\left.\partial \boldsymbol{\theta}_{i} \partial \boldsymbol{\theta}_{j}\right|_{\boldsymbol{\theta}=\boldsymbol{\theta}_{0}^{(c, 2)}}$, with indices $u, v= 1,2, \dots, p$, and $c=1,2, \dots, r$.  Further, the norm condition $\left\|\boldsymbol{\theta}^{(c, 2)}-\boldsymbol{\theta}_{0}\right\| \leq\left\|\boldsymbol{\theta}-\boldsymbol{\theta}_{0}\right\|$ holds.
\end{itemize}

\noindent By utilizing the regulatory conditions and implementing the weak law of large numbers (WLLN), we obtain
\begin{itemize}
    \item[$\boldsymbol{(i)}$] $\left.\frac{\partial}{\partial \boldsymbol{\theta}} g_{m}(\boldsymbol{\theta})\right|_{\boldsymbol{\theta}=\boldsymbol{\theta}_{0}} \rightarrow \mathbf{0}_{p}$.
    
    \item[$\boldsymbol{(ii)}$] $\left.\frac{\partial^{2}}{\partial \boldsymbol{\theta}^{T} \partial \boldsymbol{\theta}} g_{m}(\boldsymbol{\theta})\right|_{\boldsymbol{\theta}=\boldsymbol{\theta}_{0}} \rightarrow \boldsymbol{J}_{\gamma}\left(\boldsymbol{\theta}_{0}\right)$.

    \item[$\boldsymbol{(ii)}$] The elements of $(1 / m) \boldsymbol{G}_{c}$ for $c=1,2, \dots, r$, remain bounded in $\boldsymbol{\theta} \in U_{\delta}$ as $m \rightarrow \infty$.

    \item[$\boldsymbol{(ii)}$] The elements of $(1/m) \boldsymbol{W}_{c}$ for $c=1,2, \dots, r$, remain bounded in $\boldsymbol{\theta} \in U_{\delta}$ as $m \rightarrow \infty$.
\end{itemize}

\noindent Under these conditions, there exists an integer $m_{1}$, such that $\forall$ $m>m_{1}$ equation (\hyperref[eq:27]{A.6}) divided by $m$ can be written as
\begin{equation*}
\boldsymbol{J}_{\gamma}(\boldsymbol{\theta_0})\left(\boldsymbol{\theta}-\boldsymbol{\theta}_{0}\right)+\frac{\boldsymbol{\lambda}}{m} \boldsymbol{H}\left(\boldsymbol{\theta}_{0}\right) +\delta^{2} \boldsymbol{\nu}_{m}^{(3)}(\boldsymbol{\theta})=\mathbf{0}_{p}, \tag{A.7}
\label{eq:28}
\end{equation*}
where $\left\|\boldsymbol{\nu}_{m}^{(3)}(\boldsymbol{\theta})\right\|$ is bounded for $\boldsymbol{\theta} \in U_{\delta}$. \\
Next, it will be shown that the system (\hyperref[eq:23]{A.2}) - (\hyperref[eq:26]{A.5}) has a solution ${\widehat {\boldsymbol{\theta}}}$, ${\widehat{\boldsymbol{\lambda}}}$, where ${\widehat{\boldsymbol{\theta}}} \in U_{\delta}$. \\

\noindent Since $\boldsymbol{\theta}_{0}$ satisfies the constraint, (\hyperref[eq:24]{A.3}) we may write
\begin{equation*}
\boldsymbol{h_{i}}\left(\boldsymbol{\theta_{0}}\right)=\mathbf{0}, \quad i=1,2, \dots, s,\tag{A.8}
\label{eq:29}
\end{equation*}
and
\begin{equation*}
\boldsymbol{h_{j}}\left(\boldsymbol{\theta_{0}}\right)-\boldsymbol{u_{j, 0}}=\mathbf{0}, \quad j=s+1, \dots, r . \tag{A.9}
\label{eq:30}
\end{equation*}
for some value of $s \leq r$, where $\boldsymbol{u_{j, 0}} > \mathbf{0},$ $j=s+1, \dots, r$. Let $\boldsymbol{u_{0}}=\left(\boldsymbol{u_{s+1,0}}\right., \left.\boldsymbol{u_{s+2,0}}, \dots, \boldsymbol{u_{r, 0}}\right)$. If $\boldsymbol{\lambda_{0}}=\left( \boldsymbol{\lambda_{1,0}},  \boldsymbol{\lambda_{2,0}}, \dots,  \boldsymbol{\lambda_{r, 0}}\right)$ is accordingly defined in such a way that
\begin{equation*}
 \boldsymbol{\lambda_{k, 0}} \geq \mathbf{0}, \quad k=1,2, \dots, s, \tag{A.10}
 \label{eq:31}
\end{equation*}
and
\begin{equation*}
\boldsymbol{\lambda_{l, 0}}=\mathbf{0}, \quad l=s+1, \dots, r , \tag{A.11}
\label{eq:32}
\end{equation*}
then conditions (\hyperref[eq:24]{A.3}) and (\hyperref[eq:25]{A.4}) are also satisfied by $\boldsymbol{\lambda_{0}}$ and $\boldsymbol{\theta_{0}}$.\\

\noindent Thus, (\hyperref[eq:23]{A.2}) - (\hyperref[eq:5]{A.5}) will be satisfied by a solution, if it exists, of the system
\begin{equation*}
m \frac{\partial}{\partial \boldsymbol{\theta}} g_{m}(\boldsymbol{\theta}) + \boldsymbol{\lambda^{\textbf{(s)}}} \boldsymbol{H_{s, \theta}} =\mathbf{0} , \tag{A.12}
\label{eq:33}
\end{equation*}
\begin{equation*}
\boldsymbol{\lambda_{k}} \geq \mathbf{0}, \quad k=1,2, \dots, s , \tag{A.13}
\label{eq:34}
\end{equation*}
\begin{equation*}
\boldsymbol{\lambda_{l}} = \mathbf{0}, \quad k=s+1,s+2, \dots, r , \tag{A.14}
\label{eq:35}
\end{equation*}
\begin{equation*}
\boldsymbol{u_{j}} > \mathbf{0}, \quad j=s+1, \dots, r ,
\tag{A.15}
\label{eq:36}
\end{equation*}
\noindent where $\boldsymbol{H_{s ; \boldsymbol{\theta}}}$ is the $p \times s$ matrix $\left[\frac{\partial \boldsymbol{h}_{i}(\boldsymbol{\theta})}{\partial \boldsymbol{\theta}_{k}}\right], i=1,2, \dots, s$; \quad $k=1,2, \dots, p$ and $\boldsymbol{\lambda}^{\textbf{(s)}}$ is the vector of dimensions $s$ whose elements are the first $s$ elements of $\boldsymbol{\lambda}$.  Let $\boldsymbol{u}=\left(\boldsymbol{u}_{s+1}, \boldsymbol{u}_{s+2}, \dots, \boldsymbol{u}_{r}\right)$.  Let $\boldsymbol{h}^{\textbf{(s)}}(\boldsymbol{\theta})=\left(\boldsymbol{h}_{1}(\boldsymbol{\theta}), \boldsymbol{h}_{2}(\boldsymbol{\theta}), \dots, \boldsymbol{h}_{s}(\boldsymbol{\theta})\right)$.  Expanding $\boldsymbol{h}^{\textbf{(s)}}(\boldsymbol{\theta})$ about $\boldsymbol{\theta}_{0}$, we have
\begin{equation*}
\boldsymbol{h^{\textbf{(s)}}}(\boldsymbol{\theta})=\boldsymbol{h^{\textbf{(s)}}}\left(\boldsymbol{\theta_{0}}\right)+\boldsymbol{H_{s ; \boldsymbol{\theta_{0}}}}^{T}\left(\boldsymbol{\theta}-\boldsymbol{\theta_{0}}\right)+\boldsymbol{\delta^{2}} \boldsymbol{\nu}_{m}^{(4)}(\boldsymbol{\theta}) , \tag{A.16}
\label{eq:37}
\end{equation*}
where $\left\|\boldsymbol{\nu}_{m}^{(4)}(\boldsymbol{\theta})\right\|$ is bounded for $\boldsymbol{\theta} \in U_{\delta}$ because of $\boldsymbol{\mathscr{H}_{1}}$ and $\boldsymbol{\mathscr{H}_{2}}$.  Similarly, if $\boldsymbol{h}^{\left(s^{*}\right)}(\boldsymbol{\theta})=\left(\boldsymbol{h}_{s+1}(\boldsymbol{\theta}), \dots, \boldsymbol{h}_{r}(\boldsymbol{\theta})\right)$,
\begin{equation*}
\boldsymbol{h^{\left(\textbf{s}^{*}\right)}}(\boldsymbol{\theta})=\boldsymbol{h^{\left(\textbf{s}^{*}\right)}}\left(\boldsymbol{\theta_{0}}\right)+\boldsymbol{H_{s^{*} ; \boldsymbol{\theta_{0}}}}^{T}\left(\boldsymbol{\theta}-\boldsymbol{\theta_{0}}\right)+\boldsymbol{\delta^{2}} \boldsymbol{\nu}_{m}^{(5)}(\boldsymbol{\theta}), \tag{A.17} 
\label{eq:38}\\
\end{equation*}
where $\boldsymbol{H_{s^{*} ; \boldsymbol{\theta_{0}}}}$ is the $p \times(r-s)$ matrix $\left[\frac{\partial \boldsymbol{h}_{j}(\boldsymbol{\theta})}{\partial \boldsymbol{\theta}_{k}}\right], j=s+1, \dots, r; \quad k=1,2, \dots, p$ and $\left\|\boldsymbol{\nu}_{m}^{(5)}(\boldsymbol{\theta})\right\|$ remains bounded for $\boldsymbol{\theta} \in U_{\delta}$. \\

\noindent Equation (\hyperref[eq:28]{A.7}) takes the form 

\begin{equation*}
\boldsymbol{J}_{\gamma}\left(\boldsymbol{\theta}-\boldsymbol{\theta}_{0}\right)+\frac{\boldsymbol{\lambda^{(s)}}}{m} \boldsymbol{H_{s ; \boldsymbol{\theta_{0}}}} +\delta^{2} \boldsymbol{\nu}_{m}^{(4)}(\boldsymbol{\theta})=\mathbf{0}_{p}. \tag{A.18}
\label{eq:39} \\
\end{equation*}

\noindent Pre-multiplying $\boldsymbol{H_{s ; \boldsymbol{\theta_{0}}}}^{T}\boldsymbol{J^{-1}_{\gamma}},$ we get

\begin{equation*}
\boldsymbol{H_{s ; \boldsymbol{\theta_{0}}}}^{T}\left(\boldsymbol{\theta}-\boldsymbol{\theta}_{0}\right)+\left(\frac{\boldsymbol{\lambda^{\textbf{(s)}}}}{m}\right)\boldsymbol{H_{s ; \boldsymbol{\theta_{0}}}}^{T}\boldsymbol{J^{-1}_{\gamma}}(\boldsymbol{\theta}_0) \boldsymbol{H_{s ; \boldsymbol{\theta_{0}}}} +\delta^{2} \boldsymbol{H_{s ; \boldsymbol{\theta_{0}}}}^{T}\boldsymbol{J^{-1}_{\gamma}}(\boldsymbol{\theta}_0)\boldsymbol{\nu}_{m}^{(4)}(\boldsymbol{\theta})=\mathbf{0}_{p}. \tag{A.19} 
\label{eq:40}\\
\end{equation*}

\noindent By equations (\hyperref[eq:29]{A.8}), (\hyperref[eq:32]{A.11}) and (\hyperref[eq:37]{A.16})

\begin{align*}
\left(\frac{\boldsymbol{\lambda^{\textbf{(s)}}}}{m}\right) \boldsymbol{H_{s ; \boldsymbol{\theta_{0}}}}^{T}\boldsymbol{J^{-1}_{\gamma}}(\boldsymbol{\theta}_0) \boldsymbol{H_{s ; \boldsymbol{\theta_{0}}}} + \delta^{2}[\boldsymbol{H_{s ; \boldsymbol{\theta_{0}}}}^{T}\boldsymbol{J^{-1}_{\gamma}}(\boldsymbol{\theta}_0)\boldsymbol{\nu}_{m}^{(4)}(\boldsymbol{\theta}) + \boldsymbol{\nu}_{m}^{(5)}(\boldsymbol{\theta})]=\mathbf{0}_{p}.
\tag{A.20}
\label{eq:41}
\end{align*}

\noindent As $(\boldsymbol{H_{s ;\boldsymbol{\theta_{0}}}}^{T}\boldsymbol{J^{-1}_{\gamma}}(\boldsymbol{\theta}_0) \boldsymbol{H_{s ; \boldsymbol{\theta_{0}}}})$ is non-singular, its inverse remains bounded for $\boldsymbol{\theta} \in U_{\delta}$.  Consequently, equation (\hyperref[eq:41]{A.20}) can be solved for $\boldsymbol{\lambda^{\textbf{(s)}}}$ in terms of $\boldsymbol{\theta}.$  Let it be $\boldsymbol{\hat{\lambda}^{\textbf{(s)}}}$, then

\begin{align*}
\frac{\boldsymbol{\hat{\lambda}^{\textbf{(s)}}}}{m} =  [\boldsymbol{H_{s ; \boldsymbol{\theta_{0}}}}^{T}\boldsymbol{J^{-1}_{\gamma}}(\boldsymbol{\theta}_0) \boldsymbol{H_{s ; \boldsymbol{\theta_{0}}}}]^{-1}\delta^{2}[\boldsymbol{H_{s ; \boldsymbol{\theta_{0}}}}^{T}\boldsymbol{J^{-1}_{\gamma}}(\boldsymbol{\theta}_0)\boldsymbol{\nu}_{m}^{(4)}(\boldsymbol{\theta}) + \boldsymbol{\nu}_{m}^{(5)}(\boldsymbol{\theta})].
\tag{A.21}
\label{eq:42}
\end{align*}

\noindent Substituting (\hyperref[eq:42]{A.21}) in, (\hyperref[eq:39]{A.18}), we have
\begin{align*}
\boldsymbol{J}_{\gamma}\left(\boldsymbol{\theta}-\boldsymbol{\theta}_{0}\right) + \delta^{2} \boldsymbol{\nu}^{*}(\boldsymbol{\theta})  =\mathbf{0}, \tag{A.22}
\label{eq:43}
\end{align*}
where $\boldsymbol{\nu}^{*}(\boldsymbol{\theta}) = - \boldsymbol{H_{s ; \boldsymbol{\theta_{0}}}}[\boldsymbol{H_{s ; \boldsymbol{\theta_{0}}}}^{T}\boldsymbol{J^{-1}_{\gamma}}(\boldsymbol{\theta}_0) \boldsymbol{H_{s ; \boldsymbol{\theta_{0}}}}]^{-1}[\boldsymbol{H_{s ; \boldsymbol{\theta_{0}}}}^{T}\boldsymbol{J^{-1}_{\gamma}}(\boldsymbol{\theta}_0)\boldsymbol{\nu}_{m}^{(4)}(\boldsymbol{\theta}) + \boldsymbol{\nu}_{m}^{(5)}(\boldsymbol{\theta})] + \delta^{2} \boldsymbol{\nu}_{m}^{(4)}(\boldsymbol{\theta})$.  The norm $\left\|\boldsymbol{\nu}^{*}(\boldsymbol{\theta})\right\|$ remains bounded for $\boldsymbol{\theta} \in U_{\delta}$ and $m > m_1$.  If $\boldsymbol{\delta}$ is small enough and $m > m_1$, then equation (\hyperref[eq:43]{A.22}) admits a solution $\boldsymbol{\hat{\theta}} \in U_{\delta}.$

\noindent Consequently, $\boldsymbol{\hat{\theta}}$ and $\boldsymbol{\hat{\lambda}}$ constitute a solution to the estimating equation.  Thus, there is a solution to the equations (\hyperref[eq:23]{A.2}) - (\hyperref[eq:26]{A.5})  which converges to the true value $\boldsymbol{\theta_{0}}$ in probability. \\ 

\noindent Thus, $\tilde{\boldsymbol{\theta}}_{\gamma} \xrightarrow[m \rightarrow \infty]{\mathcal{P}} \boldsymbol{\theta}_{0}.$ \\

\section{Proof of Theorem (b)}

\noindent This proof closely follows the approach taken by Sen et al. \citep{sen2010finite}.  Differentiating both sides of equation (\hyperref[eq:22]{A.1}) with respect to $\boldsymbol{\theta}$, we obtain

\begin{equation*}
\frac{\partial}{\partial \boldsymbol{\theta}} g_{m}(\boldsymbol{\theta}) =   \left. \sum_{\substack{n_{jl}=0 \\ j=1,2 \\ l=1,2,\dots,k}}^{\infty}  \boldsymbol{u}(\boldsymbol{n}) f^{\gamma + 1}(\boldsymbol{n}) \right. - \left. \frac {1} {m} \sum_{i=1}^{m} \boldsymbol{u}(\boldsymbol{N_{i}}) f^{\gamma}(\boldsymbol{N_{i}}) \right. ,\tag{B.1}
\label{eq:44}
\end{equation*}
and differentiating again with respect to $\boldsymbol{\theta}$, we get

\begin{align*}
\frac{\partial}{\partial \boldsymbol{\theta}^{T}} \frac{\partial}{\partial \boldsymbol{\theta}} g_{m}(\boldsymbol{\theta}) = & (1 + \gamma) \sum_{\substack{n_{jl}=0 \\ j=1,2 \\ l=1,2,\dots,k}}^{\infty}  \boldsymbol{u}(\boldsymbol{n}) \boldsymbol{u}^{T}(\boldsymbol{n})  f^{\gamma + 1}(\boldsymbol{n}) - \sum_{\substack{n_{jl}=0 \\ j=1,2 \\ l=1,2,\dots,k}}^{\infty}  \boldsymbol{I}(\boldsymbol{n}) f^{\gamma + 1}(\boldsymbol{n}) \\
&- \frac{\gamma}{m} \sum_{i=1}^{m} \boldsymbol{u}(\boldsymbol{N_{i}}) \boldsymbol{u}^{T}(\boldsymbol{N_{i}}) f^{\gamma}(\boldsymbol{N_{i}}) + \frac{1}{m} \sum_{i=1}^{m} \boldsymbol{I}(\boldsymbol{N_{i}})f^{\gamma}(\boldsymbol{N_{i}}). \\ 
\end{align*}

\noindent Therefore, $\left.\frac{\partial}{\partial \boldsymbol{\theta}^{T}} \frac{\partial}{\partial \boldsymbol{\theta}} g_{m}(\boldsymbol{\theta})\right|_{\boldsymbol{\theta}=\boldsymbol{\theta}_{0}}$ converges in probability to

\begin{align*}
\sum_{\substack{n_{jl}=0 \\ j=1,2 \\ l=1,2,\dots,k}}^{\infty} \boldsymbol{u}(\boldsymbol{n}) \boldsymbol{u}^{T}(\boldsymbol{n}) f^{\gamma+1}(\boldsymbol{n})
= \boldsymbol{J}_{\gamma}(\boldsymbol{\theta_0}). \tag{B.2} 
\label{eq:45}
\end{align*}

\noindent Given that $f_{\boldsymbol{\theta}_{0}}$ denotes the true distribution, some simple algebra establishes that

$$
\mathrm{E}\left[\left.\sqrt{m} \frac{\partial}{\partial \boldsymbol{\theta}} g_{m}(\boldsymbol{\theta})\right|_{\boldsymbol{\theta}=\boldsymbol{\theta}_{0}}\right]=\mathbf{0}_{p}, \quad \text { and } \operatorname{Var}\left[\left.\sqrt{m} \frac{\partial}{\partial \boldsymbol{\theta}} g_{m}(\boldsymbol{\theta})\right|_{\boldsymbol{\theta}=\boldsymbol{\theta}_{0}}\right]=\boldsymbol{K}_{\gamma}\left(\boldsymbol{\theta}_{0}\right),
$$
where \begin{equation*}
\boldsymbol{K}_{\gamma}(\boldsymbol{\theta}) = \sum_{\substack{n_{jl}=0 \\ j=1,2 \\ l=1,2,\dots,k}}^{\infty} \boldsymbol{u}(\boldsymbol{n}) \boldsymbol{u}^{T}(\boldsymbol{n}) f^{2\gamma+1}(\boldsymbol{n}) - \boldsymbol{\xi}_{\gamma}(\boldsymbol{\theta}) \boldsymbol{\xi}_{\gamma}^{T}(\boldsymbol{\theta}). \\
\end{equation*}
\noindent Thus,
\begin{equation*} 
\left.\sqrt{m} \frac{\partial}{\partial \boldsymbol{\theta}} g_{m}(\boldsymbol{\theta})\right|_{\boldsymbol{\theta}=\boldsymbol{\theta}_{0}} \xrightarrow[m \rightarrow \infty]{L} \mathcal{N}\left(\mathbf{0}_{p}, \boldsymbol{K}_{\gamma}\left(\boldsymbol{\theta}_{0}\right)\right).\tag{B.3}
\label{eq:46}\\
\end{equation*}

\noindent Let us consider the following estimating equations satisfied by $\boldsymbol{\theta}$.  Here, combining (\hyperref[eq:27]{A.6}) and (\hyperref[eq:28]{A.7}), we have
\begin{equation*}
\left.\sqrt{m} \frac{\partial}{\partial \boldsymbol{\theta}} g_{m}(\boldsymbol{\theta})\right|_{\boldsymbol{\theta}=\boldsymbol{\theta}_{0}}+\boldsymbol{J}_{\gamma}\left(\boldsymbol{\theta}_{0}\right) \sqrt{m}\left(\widetilde{\boldsymbol{\theta}}_{\gamma}-\boldsymbol{\theta}_{0}\right)+\frac{\boldsymbol{\hat{\lambda}^{\textbf{(s)}}}}{\sqrt{m}} \boldsymbol{H}_{s; \boldsymbol{\theta}_{0}}+o_{p}(1)=\mathbf{0}_{p}, \\ \tag{B.4}
\label{eq:47}
\end{equation*}

\noindent Now, from (\hyperref[eq:37]{A.16}), we derive
\begin{equation*}
\boldsymbol{H^{T}_{s; \boldsymbol{\theta}_{0}}} \sqrt{m}\left(\widetilde{\boldsymbol{\theta}}_{\gamma}-\boldsymbol{\theta}_{0}\right)+o_{p}(1)=\mathbf{0}_{s}, \\ \tag{B.5} 
\label{eq:48}\\
\end{equation*}

\noindent Similarly, using (\hyperref[eq:38]{A.17}), we obtain
\begin{equation*}
\sqrt{m} \left(\boldsymbol{h^{\left(\textbf{s}^{*}\right)}}(\widetilde{\boldsymbol{\theta}}_{\gamma}) - \boldsymbol{h^{\left(\textbf{s}^{*}\right)}}\left(\boldsymbol{\theta_{0}}\right)\right) = \boldsymbol{H^{T}_{s^{*}; \boldsymbol{\theta}_{0}}}\sqrt{m}\left(\widetilde{\boldsymbol{\theta}}_{\gamma}-\boldsymbol{\theta}_{0}\right) +o_{p}(1). \\ \tag{B.6}
\label{eq:49}\\
\end{equation*}

\noindent Now, we can express equations (\hyperref[eq:47]{B.4}), (\hyperref[eq:48]{B.5}) and (\hyperref[eq:49]{B.6}) in the matrix form as 
\begin{align*}
&\left(
\begin{array}{ccc}
\boldsymbol{J}_{\gamma}\left(\boldsymbol{\theta}_{0}\right) & \boldsymbol{H}_{s; \boldsymbol{\theta}_{0}} & \boldsymbol{H}_{s^{*}; \boldsymbol{\theta}_{0}} \\
\boldsymbol{H}^{T}_{s; \boldsymbol{\theta}_{0}} & \mathbf{0}_{s \times s} & \mathbf{0}_{s \times q} \\
\boldsymbol{H}^{T}_{s^{*}; \boldsymbol{\theta}_{0}} & \mathbf{0}_{q \times s} & \mathbf{0}_{q \times q}
\end{array}
\right)
\left(
\begin{array}{c}
\sqrt{m}\left(\widetilde{\boldsymbol{\theta}}_{\gamma}-\boldsymbol{\theta}_{0}\right) \\
\frac{\boldsymbol{\hat{\lambda}^{\textbf{(s)}}}}{\sqrt{m}}  \\
0
\end{array}
\right) \\
&=
\left(
\begin{array}{c}
-\sqrt{m} \left.\frac{\partial}{\partial \boldsymbol{\theta}} g_{m}(\boldsymbol{\theta})\right|_{\boldsymbol{\theta}=\boldsymbol{\theta}_{0}} \\
0 \\
\sqrt{m} \left(\boldsymbol{h^{\left(\textbf{s}^{*}\right)}}(\widetilde{\boldsymbol{\theta}}_{\gamma}) - \boldsymbol{h^{\left(\textbf{s}^{*}\right)}}\left(\boldsymbol{\theta_{0}}\right)\right)
\end{array}
\right) + o_{p}(1) .
\end{align*}

\noindent Therefore, 
\begin{align*}
\left(
\begin{array}{c}
\sqrt{m}\left(\widetilde{\boldsymbol{\theta}}_{\gamma}-\boldsymbol{\theta}_{0}\right) \\
\frac{\boldsymbol{\hat{\lambda}^{\textbf{(s)}}}}{\sqrt{m}} \\
0
\end{array}
\right)=\left(
\begin{array}{ccc}
\boldsymbol{J}_{\gamma}\left(\boldsymbol{\theta}_{0}\right) & \boldsymbol{H}_{s; \boldsymbol{\theta}_{0}} & \boldsymbol{H}_{s^{*}; \boldsymbol{\theta}_{0}} \\
\boldsymbol{H}^{T}_{s; \boldsymbol{\theta}_{0}} & \mathbf{0}_{s \times s} & \mathbf{0}_{s \times q} \\
\boldsymbol{H}^{T}_{s^{*}; \boldsymbol{\theta}_{0}} & \mathbf{0}_{q \times s} & \mathbf{0}_{q \times q}
\end{array}
\right)^{-1} &\times \\
\substack{
\left(
\begin{array}{c}
-\sqrt{m} \left.\frac{\partial}{\partial \boldsymbol{\theta}} g_{m}(\boldsymbol{\theta})\right|_{\boldsymbol{\theta}=\boldsymbol{\theta}_{0}} \\
0 \\
\sqrt{m} \left(\boldsymbol{h^{\left(\textbf{s}^{*}\right)}}(\widetilde{\boldsymbol{\theta}}_{\gamma}) - \boldsymbol{h^{\left(\textbf{s}^{*}\right)}}\left(\boldsymbol{\theta_{0}}\right)\right)
\end{array}
\right) 
}
+ o_{p}(1).
\end{align*}

\noindent This gives 
\begin{align*}
\left(
\begin{array}{c}
\sqrt{m}\left(\widetilde{\boldsymbol{\theta}}_{\gamma}-\boldsymbol{\theta}_{0}\right) \\
\frac{\boldsymbol{\hat{\lambda}^{\textbf{(s)}}}}{\sqrt{m}} \\
0
\end{array}
\right)
=
\left(
\begin{array}{ccc}
\boldsymbol{A}^{-1}+\boldsymbol{A}^{-1}\boldsymbol{B}\boldsymbol{A}^{-1}  & -\boldsymbol{A}^{-1}\boldsymbol{C}\boldsymbol{P} & -\boldsymbol{A}^{-1}\boldsymbol{U}\boldsymbol{S}^{-1} \\
-\boldsymbol{P}^{-1}\boldsymbol{D}\boldsymbol{S}^{-1} & \boldsymbol{P}^{-1}+\boldsymbol{P}^{-1}\boldsymbol{Q}\boldsymbol{S}^{-1}\boldsymbol{R}\boldsymbol{P}^{-1} & -\boldsymbol{P}^{-1}\boldsymbol{Q}\boldsymbol{S}^{-1} \\
-\boldsymbol{S}^{-1}\boldsymbol{V}\boldsymbol{A}^{-1} & -\boldsymbol{S}^{-1}\boldsymbol{R}\boldsymbol{P}^{-1} & \boldsymbol{S}^{-1}
\end{array}
\right) &\times \\
\substack{
\left(
\begin{array}{c}
-\sqrt{m} \left.\frac{\partial}{\partial \boldsymbol{\theta}} g_{m}(\boldsymbol{\theta})\right|_{\boldsymbol{\theta}=\boldsymbol{\theta}_{0}} \\
0 \\
\sqrt{m} \left(\boldsymbol{h^{\left(\textbf{s}^{*}\right)}}(\widetilde{\boldsymbol{\theta}}_{\gamma}) - \boldsymbol{h^{\left(\textbf{s}^{*}\right)}}\left(\boldsymbol{\theta_{0}}\right)\right)
\end{array}
\right) 
}
+ o_{p}(1), \tag{B.7}
\label{eq:50}
\end{align*}
\noindent where  $$\begin{array}{c}
\boldsymbol{A} = J_{\gamma}(\boldsymbol{\theta}_{0}), \quad
\boldsymbol{B} = H_{s}(\bm{\theta_0})P^{-1}H^{T}_{s}(\bm{\theta_0}) + US^{-1}V,\\
\boldsymbol{C} =  H_{s}(\bm{\theta_0}) - US^{-1}R,  \quad
\boldsymbol{D} =  H^{T}_{s}(\bm{\theta_0}) - QS^{-1}V, \\
\boldsymbol{P} =  -H^{T}_{s}(\bm{\theta_0})J^{-1}_{\gamma} (\boldsymbol{\theta}_{0})H_{s}(\bm{\theta_0}), \quad 
\boldsymbol{Q} =  -H^{T}_{s}(\bm{\theta_0})J^{-1}_{\gamma}(\boldsymbol{\theta}_{0})H_{s^{*}}(\bm{\theta_0}), \\
\boldsymbol{R} = -H^{T}_{s^{*}}(\bm{\theta_0})J^{-1}_{\gamma}(\boldsymbol{\theta}_{0})H_{s}(\bm{\theta_0}), \quad 
\boldsymbol{S} = -H^{T}_{s^{*}}(\bm{\theta_0})J^{-1}_{\gamma}(\boldsymbol{\theta}_{0})H_{s^{*}}(\bm{\theta_0}), \\ 
\boldsymbol{U} =  H_{s^{*}}(\bm{\theta_0}) - H_{s}(\bm{\theta_0})P^{-1}Q, \quad 
\boldsymbol{V} =  H^{T}_{s^{*}}(\bm{\theta_0}) - RP^{-1}H^{T}_{s}(\bm{\theta_0}). \\
\end{array}
$$ 

\noindent Hence, we get
\begin{align*}
    &\sqrt{m}\left(\widetilde{\boldsymbol{\theta}}_{\gamma}-\boldsymbol{\theta}_{0}\right) = -[\boldsymbol{A}^{-1}+\boldsymbol{A}^{-1}\boldsymbol{B}\boldsymbol{A}^{-1} ]\sqrt{m} \left.\frac{\partial}{\partial \boldsymbol{\theta}} g_{m}(\boldsymbol{\theta})\right|_{\boldsymbol{\theta}=\boldsymbol{\theta}_{0}} \\
    &-[\boldsymbol{A}^{-1}\boldsymbol{U}\boldsymbol{S}^{-1}] \sqrt{m} \left(\boldsymbol{h^{\left(\textbf{s}^{*}\right)}}(\widetilde{\boldsymbol{\theta}}_{\gamma}) - \boldsymbol{h^{\left(\textbf{s}^{*}\right)}}\left(\boldsymbol{\theta_{0}}\right)\right) + o_{p}(1). \tag{B.8}
\label{eq:51}
\end{align*}
\noindent and
\begin{align*}
    -\boldsymbol{S}^{-1}\sqrt{m} \left(\boldsymbol{h^{\left(\textbf{s}^{*}\right)}}(\widetilde{\boldsymbol{\theta}}_{\gamma}) - \boldsymbol{h^{\left(\textbf{s}^{*}\right)}}\left(\boldsymbol{\theta_{0}}\right)\right) = [\boldsymbol{S}^{-1}\boldsymbol{V}\boldsymbol{A}^{-1}]\sqrt{m} \left.\frac{\partial}{\partial \boldsymbol{\theta}} g_{m}(\boldsymbol{\theta})\right|_{\boldsymbol{\theta}=\boldsymbol{\theta}_{0}} + o_p(1). \tag{B.9}
\label{eq:52}
\end{align*}

\noindent Solving the above equations, we obtain
\begin{align*}
    &\sqrt{m}\left(\widetilde{\boldsymbol{\theta}}_{\gamma}-\boldsymbol{\theta}_{0}\right) = -[\boldsymbol{A}^{-1}+\boldsymbol{A}^{-1}(\boldsymbol{B}-\boldsymbol{U}\boldsymbol{S}^{-1}\boldsymbol{V})\boldsymbol{A}^{-1}]\sqrt{m} \left.\frac{\partial}{\partial \boldsymbol{\theta}} g_{m}(\boldsymbol{\theta})\right|_{\boldsymbol{\theta}=\boldsymbol{\theta}_{0}}. \tag{B.10}
\label{eq:53}
\end{align*}

\noindent Finally, by merging (\hyperref[eq:46]{B.3}) with (\hyperref[eq:53]{B.10}), we arrive at
\begin{equation*}
\sqrt{m}\left(\tilde{\boldsymbol{\theta}}_{\gamma}-\boldsymbol{\theta}_{0}\right) \sim N(\mathbf{0}_{p}, \boldsymbol{\Sigma}_{\gamma}(\bm{\theta_0})). \\
\end{equation*} 

\section{Python code for data generation and point estimation}
\begin{lstlisting}
# Importing necessary libraries
import numpy as np
import pandas as pd
import matplotlib.pyplot as plt
import math
import random
import time
import scipy
import scipy.special
from scipy.special import factorial
from scipy.special import gamma as gamma_function
from scipy.special import factorial2 as factorial2_function
from scipy.stats import invgauss
from scipy.optimize import minimize

# Define constants and parameters
n = 100
tau = [0.01, 0.35, 0.69, 1.12]
zeta, a1, b1, a2, b2 = 4.5, 0.9, 0.5, 0.6, 0.2
true_theta = np.array([zeta, a1, b1, a2, b2])

# MLE
# gamma = 1e-10

# DPD
# gamma = 0.2, 0.5, 0.8

# Define helper functions to calculate intervals and samples
def generate_poisson_samples(zi, a, b, tau_diff):
    samples = []
    for interval in tau_diff:
        samples.append(np.random.poisson(zi * a * interval, n))
    return np.concatenate(samples)

# Generate samples Frailty variable with Exponential distribution
zi = np.random.exponential(zeta, n)

# Generating Poisson samples from a recurrent event with parameter (a1, b1)
int1, int2, int3 = [tau[i+1]**b1 - tau[i]**b1 for i in range(3)]
tau_diff1 = [int1, int2, int3]
A1 = zi*a1

ni11 = np.random.poisson(A1*int1, n)
ni12 = np.random.poisson(A1*int2, n)
ni13 = np.random.poisson(A1*int3, n)

ni1l = np.concatenate(np.array([ni11, ni12, ni13]))
# print("ni1l:", ni1l)

# Generate Poisson samples from a recurrent event with parameter (a2, b2)
int4, int5, int6 = [tau[i+1]**b2 - tau[i]**b2 for i in range(3)]
tau_diff2 = [int4, int5, int6]
A2 = zi*a2

ni24 = np.random.poisson(A2*int4, n)
ni25 = np.random.poisson(A2*int5, n)
ni26 = np.random.poisson(A2*int6, n)
\end{lstlisting}
\begin{lstlisting}
ni2l = np.concatenate(np.array([ni24, ni25, ni26]))
# print("ni2l:", ni2l)

# DPD
def compute_pmf1(tau, zeta, a1, b1, a2, b2):
      product1 = 1
      sum2 = 0
      product3 = 1
      sum4 = 0
      sum5 = 0

      for l in range(1, 3):  # Iterate over l = 1, 2, 3
          term1 = a1 * (tau[l]**b1 - tau[l-1]**b1)
          term2 = a2 * (tau[l]**b2 - tau[l-1]**b2)
          sum4 += (term1 + term2)
          for n1l in range(0, 5):
              for n2l in range(0, 5):
                  product1 *= (term1**n1l) * (term2**n2l)
                  sum2 += scipy.special.factorial(n1l+n2l)
                  product3 *= (scipy.special.factorial(n1l))*(scipy.special.factorial(n2l))
                  sum5 += (n1l+n2l)

      r = sum2/product3
      f1 = ((zeta*product1)/(((zeta+sum4)**sum5)))*r
      return f1

f_ni = compute_pmf1(tau, zeta, a1, b1, a2, b2)
# print("f_ni:", f_ni)

def compute_pmf2(tau, ni1l, ni2l, zeta, a1, b1, a2, b2):
      product1 = 1
      sum2 = 0
      product3 = 1
      sum4 = 0
      sum5 = 0

      for l in range(1, 3):  # Iterate over l = 1, 2, 3
          term1 = a1 * (tau[l]**b1 - tau[l-1]**b1)
          term2 = a2 * (tau[l]**b2 - tau[l-1]**b2)
          product1 *= (term1** ni1l[l]) * (term2**ni2l[l])
          sum2 += scipy.special.factorial(ni1l[l]+ni2l[l])
          product3 *= (scipy.special.factorial(ni1l[l]))*(scipy.special.factorial(ni2l[l]))
          sum4 += (term1 + term2)
          sum5 += (ni1l[l]+ni2l[l])

      r = sum2/product3
      f2 = ((zeta*product1)/(((zeta+sum4)**sum5)))*r
      return f2

f_Ni = compute_pmf2(tau, ni1l, ni2l, zeta, a1, b1, a2, b2)
# print("f_Ni:", f_Ni)

term1 = f_ni**(gamma+1)
term2 = 1+(1/gamma)
term3 = f_Ni**gamma

V_theta = term1 - term2*term3
# print("V_theta:", V_theta)

# Gradient of zeta
def der_zeta1(tau, zeta, a1, b1, a2, b2):
      sum4 = 0
      sum5 = 0

      for l in range(1, 3):  # Iterate over l = 1, 2, 3
          term1 = a1 * (tau[l]**b1 - tau[l-1]**b1)
          term2 = a2 * (tau[l]**b2 - tau[l-1]**b2)
          sum4 += (term1 + term2)
          for n1l in range(0, 5):
              for n2l in range(0, 5):
                  sum5 += (n1l+n2l)

      dz1 = (1/zeta) + (sum5/(zeta+sum4))
      return dz1

dzeta1 = der_zeta1(tau, zeta, a1, b1, a2, b2)
\end{lstlisting}
\begin{lstlisting}
def der_zeta2(tau, ni1l, ni2l, zeta, a1, b1, a2, b2):
      sum4 = 0
      sum5 = 0

      for l in range(1, 3):  # Iterate over l = 1, 2, 3
          term1 = a1 * (tau[l]**b1 - tau[l-1]**b1)
          term2 = a2 * (tau[l]**b2 - tau[l-1]**b2)
          sum4 += (term1 + term2)
          sum5 += (ni1l[l]+ni2l[l])

      dz2 = (1/zeta) + (sum5/(zeta+sum4))
      return dz2

dzeta2 = der_zeta2(tau, ni1l, ni2l, zeta, a1, b1, a2, b2)

grad_zeta = ((gamma+1)*term1*dzeta1) - (gamma*term2*term3*dzeta2)
# print("grad_zeta:", grad_zeta)

# Gradient of a1
def der_a11(tau, zeta, a1, b1, a2, b2):
      sum6 = 0
      sum7 = 0
      sum8 = 0
      sum9 = 0

      for l in range(1, 3):  # Iterate over l = 1, 2, 3
          term1 = a1 * (tau[l]**b1 - tau[l-1]**b1)
          term2 = a2 * (tau[l]**b2 - tau[l-1]**b2)
          sum6 += (term1 + term2)
          for n1l in range(0, 5):
              for n2l in range(0, 5):
                  sum7 += (n1l+n2l)
                  sum8 += (n1l*(tau[l]**b1 - tau[l-1]**b1))
                  sum9 += (tau[l]**b1 - tau[l-1]**b1)

      da11 = sum8 - ((sum7*sum9)/(zeta+sum6))
      return da11

da11 = der_a11(tau, zeta, a1, b1, a2, b2)

def der_a12(tau, ni1l, ni2l, zeta, a1, b1, a2, b2):
      sum6 = 0
      sum7 = 0
      sum8 = 0
      sum9 = 0

      for l in range(1, 3):  # Iterate over l = 1, 2, 3
          term1 = a1 * (tau[l]**b1 - tau[l-1]**b1)
          term2 = a2 * (tau[l]**b2 - tau[l-1]**b2)
          sum6 += (term1 + term2)
          sum7 += (ni1l[l]+ni2l[l])
          sum8 += (ni1l[l]*(tau[l]**b1 - tau[l-1]**b1))
          sum9 += (tau[l]**b1 - tau[l-1]**b1)

      da12 = sum8 - ((sum7*sum9)/(zeta+sum6))
      return da12

da12 = der_a12(tau, ni1l, ni2l, zeta, a1, b1, a2, b2)

grad_a1 = ((gamma+1)*term1*da11) - (term2*gamma*term3*da12)
# print("grad_a1:", grad_a1)

# Gradient of a2
def der_a21(tau, zeta, a1, b1, a2, b2):
      sum6 = 0
      sum7 = 0
      sum8 = 0
      sum9 = 0
      for l in range(1, 3):  # Iterate over l = 1, 2, 3
          term1 = a1 * (tau[l]**b1 - tau[l-1]**b1)
          term2 = a2 * (tau[l]**b2 - tau[l-1]**b2)
          sum6 += (term1 + term2)
          for n1l in range(0, 5):
              for n2l in range(0, 5):
                  sum7 += (n1l+n2l)
                  sum8 += (n2l*(tau[l]**b2 - tau[l-1]**b2))
\end{lstlisting}
\begin{lstlisting}
                  sum9 += (tau[l]**b2 - tau[l-1]**b2)

      da21 = sum8 - ((sum7*sum9)/(zeta+sum6))
      return da21

da21 = der_a21(tau, zeta, a1, b1, a2, b2)

def der_a22(tau, ni1l, ni2l, zeta, a1, b1, a2, b2):
      sum6 = 0
      sum7 = 0
      sum8 = 0
      sum9 = 0

      for l in range(1, 3):  # Iterate over l = 1, 2, 3
          term1 = a1 * (tau[l]**b1 - tau[l-1]**b1)
          term2 = a2 * (tau[l]**b2 - tau[l-1]**b2)
          sum6 += (term1 + term2)
          sum7 += (ni1l[l]+ni2l[l])
          sum8 += (ni2l[l]*(tau[l]**b2 - tau[l-1]**b2))
          sum9 += (tau[l]**b2 - tau[l-1]**b2)

      da22 = sum8 - ((sum7*sum9)/(zeta+sum6))
      return da22

da22 = der_a22(tau, ni1l, ni2l, zeta, a1, b1, a2, b2)

grad_a2 = ((gamma+1)*term1*da21) - (term2*gamma*term3*da22)
# print("grad_a2:", grad_a2)

# Gradient of b1
def der_b11(tau, zeta, a1, b1, a2, b2):
      sum10 = 0
      sum11 = 0
      sum12 = 0
      sum13 = 0
      for l in range(1, 3):  # Iterate over l = 1, 2, 3
          term1 = a1 * (tau[l]**b1 - tau[l-1]**b1)
          term2 = a2 * (tau[l]**b2 - tau[l-1]**b2)
          term3 = tau[l]**b1 - tau[l-1]**b1
          term4 = ((tau[l]**b1)*np.log(tau[l])) - ((tau[l-1]**b1)*np.log(tau[l-1]))
          sum10 += (term1 + term2)
          for n1l in range(0, 5):
              for n2l in range(0, 5):
                  sum11 += n1l*term3*term4
                  sum12 += (n1l+n2l)
                  sum13 += term3*term4

      db11 = (a1*sum11) - ((sum12*a1*sum13)/(zeta+sum10))
      return db11

db11 = der_b11(tau, zeta, a1, b1, a2, b2)

def der_b12(tau, ni1l, ni2l, zeta, a1, b1, a2, b2):
      sum10 = 0
      sum11 = 0
      sum12 = 0
      sum13 = 0
      for l in range(1, 3):  # Iterate over l = 1, 2, 3
          term1 = a1 * (tau[l]**b1 - tau[l-1]**b1)
          term2 = a2 * (tau[l]**b2 - tau[l-1]**b2)
          term3 = tau[l]**b1 - tau[l-1]**b1
          term4 = ((tau[l]**b1)*np.log(tau[l])) - ((tau[l-1]**b1)*np.log(tau[l-1]))
          sum10 += (term1 + term2)
          sum11 += ni1l[l]*term3*term4
          sum12 += (ni1l[l]+ni2l[l])
          sum13 += term3*term4

      db12 = (a1*sum11) - ((sum12*a1*sum13)/(zeta+sum10))
      return db12

db12 = der_b12(tau, ni1l, ni2l, zeta, a1, b1, a2, b2)

grad_b1 = ((gamma+1)*term1*db11) - (term2*gamma*term3*db12)
# print("grad_b1:", grad_b1)

# Gradient of b2
def der_b21(tau, zeta, a1, b1, a2, b2):
\end{lstlisting}
\begin{lstlisting}
      sum10 = 0
      sum11 = 0
      sum12 = 0
      sum13 = 0

      for l in range(1, 3):  # Iterate over l = 1, 2, 3
          term1 = a1 * (tau[l]**b1 - tau[l-1]**b1)
          term2 = a2 * (tau[l]**b2 - tau[l-1]**b2)
          term3 = tau[l]**b2 - tau[l-1]**b2
          term4 = ((tau[l]**b2)*np.log(tau[l])) - ((tau[l-1]**b2)*np.log(tau[l-1]))
          sum10 += (term1 + term2)
          for n1l in range(0, 5):
              for n2l in range(0, 5):
                  sum11 += n1l*term3*term4
                  sum12 += (n1l+n2l)
                  sum13 += term3*term4
      db21 = (a2*sum11) - ((sum12*a2*sum13)/(zeta+sum10))
      return db21

db21 = der_b21(tau, zeta, a1, b1, a2, b2)

def der_b22(tau, ni1l, ni2l, zeta, a1, b1, a2, b2):
      sum10 = 0
      sum11 = 0
      sum12 = 0
      sum13 = 0

      for l in range(1, 3):  # Iterate over l = 1, 2, 3
          term1 = a1 * (tau[l]**b1 - tau[l-1]**b1)
          term2 = a2 * (tau[l]**b2 - tau[l-1]**b2)
          term3 = tau[l]**b2 - tau[l-1]**b2
          term4 = ((tau[l]**b2)*np.log(tau[l])) - ((tau[l-1]**b2)*np.log(tau[l-1]))
          sum10 += (term1 + term2)
          sum11 += ni2l[l]*term3*term4
          sum12 += (ni1l[l]+ni2l[l])
          sum13 += term3*term4

      db22 = (a2*sum11) - ((sum12*a2*sum13)/(zeta+sum10))
      return db22

db22 = der_b22(tau, ni1l, ni2l, zeta, a1, b1, a2, b2)

grad_b2 = ((gamma+1)*term1*db21) - (term2*gamma*term3*db22)
# print("grad_b2:", grad_b2)

# Sequential Convex Programming
def scp(f, grad_f, theta_true, theta_init, rho_vec, k):
    theta = np.array(theta_init)

    for _ in range(k):
        grad = grad_f(theta)

        # Linear approximation (f_cap)
        def f_cap(theta_var):
            return f(theta) + grad @ (theta_var - theta)

        # Constraints (Restricted case)
        constraints = [
            {"type": "ineq", "fun": lambda th: th[1] - th[3]},  # a1 - a2 >= 0
            {"type": "ineq", "fun": lambda th: th[2] - th[4]},  # b1 - b2 >= 0
        ]
        for i in range(len(theta_true)):
            constraints.append({"type": "ineq", "fun": lambda th, i=i: rho_vec[i] - abs(th[i] - theta_true[i])}) # Trust region

        # Ensure positivity of parameters
        bounds = [(1e-5, None)] * len(theta_true)

        res = minimize(f_cap, theta, method="SLSQP", bounds=bounds, constraints=constraints)

        if not res.success:

            print(f"Iteration failed: {res.message}")
            break

        theta = res.x
\end{lstlisting}
\begin{lstlisting}
    return theta

def f(theta):
    zeta, a1, b1, a2, b2 = theta
    f_ni = compute_pmf1(tau, zeta, a1, b1, a2, b2)
    f_Ni = compute_pmf2(tau, ni1l, ni2l, zeta, a1, b1, a2, b2)
    term1 = f_ni**(gamma+1)
    term2 = 1+(1/gamma)
    term3 = f_Ni**gamma

    V_theta = term1 - term2*term3
    return V_theta

def grad_f(theta):
    # Gradient of zeta
    dzeta1 = der_zeta1(tau, zeta, a1, b1, a2, b2)
    dzeta2 = der_zeta2(tau, ni1l, ni2l, zeta, a1, b1, a2, b2)
    grad_zeta = (((gamma+1)*term1*dzeta1) - (gamma*term2*term3*dzeta2))/(n*(1+gamma))

    # Gradient of a1
    da11 = der_a11(tau, zeta, a1, b1, a2, b2)
    da12 = der_a12(tau, ni1l, ni2l, zeta, a1, b1, a2, b2)
    grad_a1 = (((gamma+1)*term1*da11) - (gamma*term2*term3*da12))/(n*(1+gamma))

    # Gradient of b1
    db11 = der_b11(tau, zeta, a1, b1, a2, b2)
    db12 = der_b12(tau, ni1l, ni2l, zeta, a1, b1, a2, b2)
    grad_b1 = (((gamma+1)*term1*db11) - (gamma*term2*term3*db12))/(n*(1+gamma))

    # Gradient of a2
    da21 = der_a21(tau, zeta, a1, b1, a2, b2)
    da22 = der_a22(tau, ni1l, ni2l, zeta, a1, b1, a2, b2)
    grad_a2 = (((gamma+1)*term1*da21) - (gamma*term2*term3*da22))/(n*(1+gamma))

    # Gradient of b2
    db21 = der_b21(tau, zeta, a1, b1, a2, b2)
    db22 = der_b22(tau, ni1l, ni2l, zeta, a1, b1, a2, b2)
    grad_b2 = (((gamma+1)*term1*db21) - (gamma*term2*term3*db22))/(n*(1+gamma))

    return np.array([grad_zeta, grad_a1, grad_b1, grad_a2, grad_b2])

# Parameters
theta_true = np.array([4.5, 0.9, 0.5, 0.6, 0.2])
theta_init = np.array([4.6, 0.8, 0.4, 0.5, 0.1])
rho_vec = np.array([0.11464, 0.05746, 0.07415, 0.06524, 0.07670])
k = 10

opt_theta = scp(f, grad_f, theta_true, theta_init, rho_vec, k)
# print("Optimal theta:", opt_theta)

# Simulation across D datasets and storing optimal theta values
D = 1000
results = np.zeros((D, 5))
initial_theta = [4.6, 0.8, 0.4, 0.5, 0.1]

# Set the seed for reproducibility
seed_value = 42
np.random.seed(seed_value)
random.seed(seed_value)

for i in range(D):
    zi = np.random.exponential(zeta, n)
    tau_diff1 = [tau[j + 1]**b1 - tau[j]**b1 for j in range(3)]
    tau_diff2 = [tau[j + 1]**b2 - tau[j]**b2 for j in range(3)]
    ni1l = generate_poisson_samples(zi, a1, b1, tau_diff1)
    ni2l = generate_poisson_samples(zi, a2, b2, tau_diff2)

    # Extract the best_theta from the returned tuple
    best_theta = scp(f, grad_f, theta_true, theta_init, rho_vec, k)
    results[i] = best_theta # Assign only the best_theta to results

# Store the results in a DataFrame
columns = ['zeta', 'a1', 'b1', 'a2', 'b2']
index = [f'D{i+1}' for i in range(D)]
result_df = pd.DataFrame(results, columns=columns, index=index)
# print(result_df)
\end{lstlisting}
\begin{lstlisting}
initial_theta = [4.6, 0.8, 0.4, 0.5, 0.1]
true_theta = [4.5, 0.9, 0.5, 0.6, 0.2]

# Mean, Bias, MSE
column_means = np.sum(result_df, axis=0)/D
column_bias = column_means - true_theta
column_mse = np.sum((result_df - true_theta)**2, axis=0)/D

print("\nMean of each parameter:")
for name, mean in zip(columns, column_means):
    print(f"{name}: {mean:.6f}")

print("\nBias of each parameter:")
for name, bias in zip(columns, column_bias):
    print(f"{name}: {bias:.6f}")

print("\nMSE of each parameter:")
for name, mse in zip(columns, column_mse):
    print(f"{name}: {mse:.6f}")
\end{lstlisting}

\section{Python code for Generalized Score-Matching (GSM) method}
\begin{lstlisting}
\begin{lstlisting}
# Defining the p(Ni1l, Ni2l), Case- 1
def compute_pmf11(tau, zeta, a1, b1, a2, b2):
      product1 = 1
      sum2 = 0
      product3 = 1
      sum4 = 0
      sum5 = 0

      for l in range(1, 3):  # Iterate over l = 1, 2, 3
          term1 = a1 * (tau[l]**b1 - tau[l-1]**b1)
          term2 = a2 * (tau[l]**b2 - tau[l-1]**b2)
          sum4 += (term1 + term2)
          for n1l in range(0, 5):
              for n2l in range(0, 5):
                  product1 *= (term1**n1l) * (term2**n2l)
                  sum2 += scipy.special.factorial(n1l+n2l)
                  product3 *= (scipy.special.factorial(n1l))*(scipy.special.factorial(n2l))
                  sum5 += (n1l+n2l)

      r = sum2/product3
      f1 = ((zeta*product1)/(((zeta+sum4)**sum5)))*r
      return f1

def compute_pmf12(tau, ni1l, ni2l, zeta, a1, b1, a2, b2):
      product1 = 1
      sum2 = 0
      product3 = 1
      sum4 = 0
      sum5 = 0

      for l in range(1, 3):  # Iterate over l = 1, 2, 3
          term1 = a1 * (tau[l]**b1 - tau[l-1]**b1)
          term2 = a2 * (tau[l]**b2 - tau[l-1]**b2)
          product1 *= (term1** ni1l[l]) * (term2**ni2l[l])
          sum2 += scipy.special.factorial(ni1l[l]+ni2l[l])
          product3 *= (scipy.special.factorial(ni1l[l]))*(scipy.special.factorial(ni2l[l]))
          sum4 += (term1 + term2)
          sum5 += (ni1l[l]+ni2l[l])

      r = sum2/product3
      f2 = ((zeta*product1)/(((zeta+sum4)**sum5)))*r
      return f2

# Defining the p(Ni1l+, Ni2l), Case- 2
def compute_pmf21(tau, zeta, a1, b1, a2, b2):
      product1 = 1
      sum2 = 0
      product3 = 1
\end{lstlisting}
\begin{lstlisting}
      sum4 = 0
      sum5 = 0

      for l in range(1, 3):  # Iterate over l = 1, 2, 3
          term1 = a1 * (tau[l]**b1 - tau[l-1]**b1)
          term2 = a2 * (tau[l]**b2 - tau[l-1]**b2)
          sum4 += (term1 + term2)
          for n1l in range(0, 5):
            if n1l == 0:
                continue
            for n2l in range(0, 5):
                  product1 *= (term1**(n1l+1)) * (term2**n2l)
                  sum2 += scipy.special.factorial(n1l+n2l+1)
                  product3 *= (scipy.special.factorial(n1l+1))*(scipy.special.factorial(n2l))
                  sum5 += (n1l+n2l+1)

      r = sum2/product3
      f1 = ((zeta*product1)/(((zeta+sum4)**sum5)))*r
      return f1


def compute_pmf22(tau, ni1l, ni2l, zeta, a1, b1, a2, b2):
      product1 = 1
      sum2 = 0
      product3 = 1
      sum4 = 0
      sum5 = 0

      for l in range(1, 3):  # Iterate over l = 1, 2, 3
          term1 = a1 * (tau[l]**b1 - tau[l-1]**b1)
          term2 = a2 * (tau[l]**b2 - tau[l-1]**b2)
          if np.any(ni1l == 0):
            return 0

          product1 *= (term1**(ni1l[l]+1)) * (term2**ni2l[l])
          sum2 += scipy.special.factorial(ni1l[l]+ni2l[l]+1)
          product3 *= (scipy.special.factorial(ni1l[l]+1))*(scipy.special.factorial(ni2l[l]))
          sum4 += (term1 + term2)
          sum5 += (ni1l[l]+ni2l[l]+1)

      r = sum2/product3
      f2 = ((zeta*product1)/(((zeta+sum4)**sum5)))*r
      return f2


# Defining the p(Ni1l, Ni2l+), Case- 3
def compute_pmf31(tau, zeta, a1, b1, a2, b2):
      product1 = 1
      sum2 = 0
      product3 = 1
      sum4 = 0
      sum5 = 0

      for l in range(1, 3):  # Iterate over l = 1, 2, 3
          term1 = a1 * (tau[l]**b1 - tau[l-1]**b1)
          term2 = a2 * (tau[l]**b2 - tau[l-1]**b2)
          sum4 += (term1 + term2)
          for n1l in range(0, 5):
            for n2l in range(0, 5):
                if n2l == 0:
                    continue
                product1 *= (term1**n1l) * (term2**(n2l+1))
                sum2 += scipy.special.factorial(n1l+n2l+1)
                product3 *= (scipy.special.factorial(n1l))*(scipy.special.factorial(n2l+1))
                sum5 += (n1l+n2l)

      r = sum2/product3
      f1 = ((zeta*product1)/(((zeta+sum4)**sum5)))*r
      return f1


def compute_pmf32(tau, ni1l, ni2l, zeta, a1, b1, a2, b2):
      product1 = 1
      sum2 = 0
      product3 = 1
      sum4 = 0
      sum5 = 0
\end{lstlisting}
\begin{lstlisting}
      for l in range(1, 3):  # Iterate over l = 1, 2, 3
          term1 = a1 * (tau[l]**b1 - tau[l-1]**b1)
          term2 = a2 * (tau[l]**b2 - tau[l-1]**b2)
          if np.any(ni2l == 0):
            return 0

          product1 *= (term1** ni1l[l]) * (term2**(ni2l[l]+1))
          sum2 += scipy.special.factorial(ni1l[l]+ni2l[l]+1)
          product3 *= (scipy.special.factorial(ni1l[l]))*(scipy.special.factorial(ni2l[l]+1))
          sum4 += (term1 + term2)
          sum5 += (ni1l[l]+ni2l[l])

      r = sum2/product3
      f2 = ((zeta*product1)/(((zeta+sum4)**sum5)))*r
      return f2


# Defining the p(Ni1l-, Ni2l), Case- 4
def compute_pmf41(tau, zeta, a1, b1, a2, b2):
      product1 = 1
      sum2 = 0
      product3 = 1
      sum4 = 0
      sum5 = 0

      for l in range(1, 3):  # Iterate over l = 1, 2, 3
          term1 = a1 * (tau[l]**b1 - tau[l-1]**b1)
          term2 = a2 * (tau[l]**b2 - tau[l-1]**b2)
          sum4 += (term1 + term2)

          for n1l in range(0, 5):
            for n2l in range(0, 5):
                  adj_n1l = np.where(n1l > 0, n1l - 1, n1l)
                  product1 *= (term1**adj_n1l) * (term2**n2l)
                  sum2 += scipy.special.factorial(adj_n1l+n2l)
                  product3 *= (scipy.special.factorial(adj_n1l))*(scipy.special.factorial(n2l))
                  sum5 += (n1l+n2l+1)

      r = sum2/product3
      f1 = ((zeta*product1)/(((zeta+sum4)**sum5)))*r
      return f1


def compute_pmf42(tau, ni1l, ni2l, zeta, a1, b1, a2, b2):
      product1 = 1
      sum2 = 0
      product3 = 1
      sum4 = 0
      sum5 = 0

      adj_ni1l = np.where(ni1l > 0, ni1l - 1, ni1l)

      for l in range(1, 3):  # Iterate over l = 1, 2, 3
          term1 = a1 * (tau[l]**b1 - tau[l-1]**b1)
          term2 = a2 * (tau[l]**b2 - tau[l-1]**b2)
          product1 *= (term1**adj_ni1l[l]) * (term2**ni2l[l])
          sum2 += scipy.special.factorial(adj_ni1l[l]+ni2l[l])
          product3 *= (scipy.special.factorial(adj_ni1l[l]))*(scipy.special.factorial(ni2l[l]))
          sum4 += (term1 + term2)
          sum5 += (ni1l[l]+ni2l[l]-1)

      r = sum2/product3
      f2 = ((zeta*product1)/(((zeta+sum4)**sum5)))*r
      return f2


# Defining the p(Ni1l, Ni2l-), Case- 5
def compute_pmf51(tau, zeta, a1, b1, a2, b2):
      product1 = 1
      sum2 = 0
      product3 = 1
      sum4 = 0
      sum5 = 0

      for l in range(1, 3):  # Iterate over l = 1, 2, 3
          term1 = a1 * (tau[l]**b1 - tau[l-1]**b1)
          term2 = a2 * (tau[l]**b2 - tau[l-1]**b2)
\end{lstlisting}
\begin{lstlisting}
          sum4 += (term1 + term2)
          for n1l in range(0, 5):
            for n2l in range(0, 5):
                  adj_n2l = np.where(n2l > 0, n2l - 1, n2l)
                  product1 *= (term1**n1l) * (term2**adj_n2l)
                  sum2 += scipy.special.factorial(n1l+adj_n2l)
                  product3 *= (scipy.special.factorial(n1l))*(scipy.special.factorial(adj_n2l))
                  sum5 += (n1l+adj_n2l)

      r = sum2/product3
      f1 = ((zeta*product1)/(((zeta+sum4)**sum5)))*r
      return f1


def compute_pmf52(tau, ni1l, ni2l, zeta, a1, b1, a2, b2):
      product1 = 1
      sum2 = 0
      product3 = 1
      sum4 = 0
      sum5 = 0

      adj_ni2l = np.where(ni2l > 0, ni2l - 1, ni2l)

      for l in range(1, 3):  # Iterate over l = 1, 2, 3
          term1 = a1 * (tau[l]**b1 - tau[l-1]**b1)
          term2 = a2 * (tau[l]**b2 - tau[l-1]**b2)
          product1 *= (term1**ni1l[l]) * (term2**adj_ni2l[l])
          sum2 += scipy.special.factorial(ni1l[l]+adj_ni2l[l])
          product3 *= (scipy.special.factorial(ni1l[l]))*(scipy.special.factorial(adj_ni2l[l]))
          sum4 += (term1 + term2)
          sum5 += (ni1l[l]+adj_ni2l[l])

      r = sum2/product3
      f2 = ((zeta*product1)/(((zeta+sum4)**sum5)))*r
      return f2

# Define gamma values
gamma_values = np.arange(0.20, 0.61, 0.01)

# Placeholder to store results
results = []

# Perform optimization for each gamma
for gamma in gamma_values:
          def compute_pmf1(tau, zeta, a1, b1, a2, b2):
                product1 = 1
                sum2 = 0
                product3 = 1
                sum4 = 0
                sum5 = 0
    
                for l in range(1, 3):  # Iterate over l = 1, 2, 3
                    term1 = a1 * (tau[l]**b1 - tau[l-1]**b1)
                    term2 = a2 * (tau[l]**b2 - tau[l-1]**b2)
                    sum4 += (term1 + term2)
                    for n1l in range(0, 5):
                        for n2l in range(0, 5):
                            product1 *= (term1**n1l) * (term2**n2l)
                            sum2 += scipy.special.factorial(n1l+n2l)
                            product3 *= (scipy.special.factorial(n1l))* (scipy.special.factorial(n2l))
                            sum5 += (n1l+n2l)
                r = sum2/product3
                f1 = ((zeta*product1)/(((zeta+sum4)**sum5)))*r
                return f1

          def compute_pmf2(tau, ni1l, ni2l, zeta, a1, b1, a2, b2):
                product1 = 1
                sum2 = 0
                product3 = 1
                sum4 = 0
                sum5 = 0

                for l in range(1, 3):  # Iterate over l = 1, 2, 3
                    term1 = a1 * (tau[l]**b1 - tau[l-1]**b1)
                    term2 = a2 * (tau[l]**b2 - tau[l-1]**b2)
                    product1 *= (term1** ni1l[l]) * (term2**ni2l[l])
                    sum2 += scipy.special.factorial(ni1l[l]+ni2l[l])
\end{lstlisting}
\begin{lstlisting}
                    product3 *= (scipy.special.factorial(ni1l[l]))*(scipy.special.factorial(ni2l[l]))
                    sum4 += (term1 + term2)
                    sum5 += (ni1l[l]+ni2l[l])
                    
                r = sum2/product3
                f2 = ((zeta*product1)/(((zeta+sum4)**sum5)))*r
                return f2

          estimated_params = scp(theta_true, theta_init, rho_vec, k, gamma, tau, ni1l,                         ni2l, n, compute_pmf1, compute_pmf2, der_zeta1, der_zeta2,
                             der_a11, der_a12, der_b11, der_b12, der_a21, der_a22,
                             der_b21, der_b22)
          results.append((gamma, *estimated_params))

          # Print the gamma and corresponding estimated parameters
          # print(f"Gamma: {gamma:.2f}, Estimated Parameters: {estimated_params}")

# Simulation across D datasets and storing optimal theta values
D = 1000
results = []
initial_theta = estimated_params

# Set the seed for reproducibility
seed_value = 42
np.random.seed(seed_value)
random.seed(seed_value)

# Define the objective function
def compute_objective_function(zeta, gamma, a1, b1, a2, b2, ni1l, ni2l):
    cumulative_sum = 0

    for i in range(D):
        zi = np.random.exponential(zeta, n)
        tau_diff1 = [tau[l + 1]**b1 - tau[l]**b1 for l in range(3)]
        tau_diff2 = [tau[l + 1]**b2 - tau[l]**b2 for l in range(3)]
        ni1l = generate_poisson_samples(zi, a1, b1, tau_diff1)
        ni2l = generate_poisson_samples(zi, a2, b2, tau_diff2)

        k = len(tau) - 1

        for l in range(3):

            # Case- 1
            p11 = compute_pmf11(tau, zeta, a1, b1, a2, b2)
            p12 = compute_pmf12(tau, ni1l, ni2l, zeta, a1, b1, a2, b2)
            term1 = p11**(gamma+1)
            term2 = 1+(1/gamma)
            term3 = p12**gamma
            p1 = np.exp(term1 - term2*term3)

            # Case- 2
            p21 = compute_pmf21(tau, zeta, a1, b1, a2, b2)
            p22 = compute_pmf22(tau, ni1l, ni2l, zeta, a1, b1, a2, b2)
            term1 = p21**(gamma+1)
            term2 = 1+(1/gamma)
            term3 = p22**gamma
            p2 = np.exp(term1 - term2*term3)

            # Case- 3
            p31 = compute_pmf31(tau, zeta, a1, b1, a2, b2)
            p32 = compute_pmf32(tau, ni1l, ni2l, zeta, a1, b1, a2, b2)
            term1 = p31**(gamma+1)
            term2 = 1+(1/gamma)
            term3 = p32**gamma
            p3 = np.exp(term1 - term2*term3)

            # Case- 4
            p41 = compute_pmf41(tau, zeta, a1, b1, a2, b2)
            p42 = compute_pmf42(tau, ni1l, ni2l, zeta, a1, b1, a2, b2)
            term1 = p41**(gamma+1)
            term2 = 1+(1/gamma)
            term3 = p42**gamma
            p4 = np.exp(term1 - term2*term3)

            # Case- 5
            p51 = compute_pmf51(tau, zeta, a1, b1, a2, b2)
            p52 = compute_pmf52(tau, ni1l, ni2l, zeta, a1, b1, a2, b2)
\end{lstlisting}
\begin{lstlisting}
            term1 = p51**(gamma+1)
            term2 = 1+(1/gamma)
            term3 = p52**gamma
            p5 = np.exp(term1 - term2*term3)

            # Calculate normalized K for the individual
            cumulative_sum += ((p1 / (p1 + p2))**2 + (p1 / (p1 + p3))**2 + (p1 / (p1 + p4))**2 + (p1 / (p1 + p5))**2) - 2 * (p1 / (p1 + p2)) - 2 * (p1 / (p1 + p3))

    L = cumulative_sum / (n*k)
    return L

# Run the process for gamma values in the range [0.20, 0.61] with a step of 0.01
gamma_values = np.arange(0.20, 0.61, 0.01)

for gamma in gamma_values:
    # Estimate parameters using SCP
    estimated_params = scp(theta_true, theta_init, rho_vec, k, gamma, tau, ni1l, ni2l, n,
                                        compute_pmf1, compute_pmf2,
                                        der_zeta1, der_zeta2,
                                        der_a11, der_a12,
                                        der_b11, der_b12,
                                        der_a21, der_a22,
                                        der_b21, der_b22)
    # Extract the estimated parameters
    zeta, a1, b1, a2, b2 = estimated_params

    # Calculate the objective function value
    L = compute_objective_function(zeta, gamma, a1, b1, a2, b2, ni1l, ni2l)

    # Store the results
    results.append((gamma, L, *estimated_params))

# Convert results to a DataFrame for display
columns = ['gamma', 'L', 'zeta', 'a1', 'b1', 'a2', 'b2']
result_df = pd.DataFrame(results, columns=columns)

# Display the results
# print(result_df)

# Find the gamma with the minimum L
min_L_index = result_df['L'].idxmin() # Index of the minimum L value
min_gamma = result_df.loc[min_L_index, 'gamma'] # gamma corresponding to min L
min_L = result_df.loc[min_L_index, 'L'] # Minimum L value

# Display the results
print(f"Minimum L: {min_L:.6f} at gamma: {min_gamma:.2f}")

# Simulation across D datasets and storing optimal theta values
D = 1000
results = np.zeros((D, 5))
initial_theta = [4.6, 0.8, 0.4, 0.5, 0.1]
gamma = 0.55

# Set the seed for reproducibility
seed_value = 42
np.random.seed(seed_value)
random.seed(seed_value)

# Adding contamination in data, epsilon = 0.006, 0.044, 0.085, 0.153
# def generate_mixed_zi(zeta, n, mix_ratio=0.847):
    # num_exp = int(mix_ratio * n)
    # num_inv_gauss = n - num_exp

    # zi_exp = np.random.exponential(zeta, num_exp)
    # mu = zeta  # Set mean for inverse Gaussian distribution
    # zi_inv_gauss = invgauss.rvs(mu / zeta, scale=zeta, size=num_inv_gauss)

    # Combine the distributions
    # zi = np.concatenate([zi_exp, zi_inv_gauss])
    # np.random.shuffle(zi)  # Shuffle to mix the two distributions
    # return zi

# Contaminated data
# for i in range(D):
  # zi = generate_mixed_zi(zeta, n, mix_ratio=0.847)
\end{lstlisting}
\begin{lstlisting}
# Pure data, epsilon = 0.000
for i in range(D):
    zi = np.random.exponential(zeta, n)
    tau_diff1 = [tau[j + 1]**b1 - tau[j]**b1 for j in range(3)]
    tau_diff2 = [tau[j + 1]**b2 - tau[j]**b2 for j in range(3)]
    ni1l = generate_poisson_samples(zi, a1, b1, tau_diff1)
    ni2l = generate_poisson_samples(zi, a2, b2, tau_diff2)

    # Extract the best_theta from the returned tuple
    best_theta = scp(theta_true, theta_init, rho_vec, k, gamma, tau, ni1l, ni2l, n,
                                   compute_pmf1, compute_pmf2,
                                   der_zeta1, der_zeta2,
                                   der_a11, der_a12,
                                   der_b11, der_b12,
                                   der_a21, der_a22,
                                   der_b21, der_b22)
    results[i] = best_theta # Assign only the best_theta to results

# Store the results in a DataFrame
columns = ['zeta', 'a1', 'b1', 'a2', 'b2']
index = [f'D{i+1}' for i in range(D)]
result_df = pd.DataFrame(results, columns=columns, index=index)
# print(result_df)

true_theta = [4.5, 0.9, 0.5, 0.6, 0.2]

# Mean, Bias, MSE
column_means = np.sum(result_df, axis=0)/D
column_bias = column_means - true_theta
column_mse = np.sum((result_df - true_theta)**2, axis=0)/D

print("\nMean of each parameter:")
for name, mean in zip(columns, column_means):
    print(f"{name}: {mean:.6f}")

print("\nBias of each parameter:")
for name, bias in zip(columns, column_bias):
    print(f"{name}: {bias:.6f}")

print("\nMSE of each parameter:")
for name, mse in zip(columns, column_mse):
    print(f"{name}: {mse:.6f}")
\end{lstlisting}

\section{Python code for Iterative Warwick and Jones (IWJ) method}
\begin{lstlisting}
# Gradient of log(f_n) w.r.t. zeta
def der_log_zeta(tau, ni1l, ni2l, zeta, a1, b1, a2, b2):
      sum4 = 0
      sum5 = 0

      for l in range(1, 3):  # Iterate over l = 1, 2, 3
          term1 = a1 * (tau[l]**b1 - tau[l-1]**b1)
          term2 = a2 * (tau[l]**b2 - tau[l-1]**b2)
          sum4 += (term1 + term2)
          for n1l in range(0, 5):
              for n2l in range(0, 5):
                  sum5 += (n1l+n2l)

      dzeta = (1/zeta) + (sum5/(zeta+sum4))
      return dzeta

grad_log_zeta = der_log_zeta(tau, ni1l, ni2l, zeta, a1, b1, a2, b2)
# print("grad_log_zeta:", grad_log_zeta)

# Gradient of log(f_n) w.r.t. a1
def der_log_a1(tau, ni1l, ni2l, zeta, a1, b1, a2, b2):
      sum6 = 0
      sum7 = 0
      sum8 = 0
      sum9 = 0

      for l in range(1, 3):  # Iterate over l = 1, 2, 3
\end{lstlisting}
\begin{lstlisting}
          term1 = a1 * (tau[l]**b1 - tau[l-1]**b1)
          term2 = a2 * (tau[l]**b2 - tau[l-1]**b2)
          sum6 += (term1 + term2)
          for n1l in range(0, 5):
              for n2l in range(0, 5):
                  sum7 += (n1l+n2l)
                  sum8 += (n1l*(tau[l]**b1 - tau[l-1]**b1))
                  sum9 += (tau[l]**b1 - tau[l-1]**b1)

      da1 = sum8 - ((sum7*sum9)/(zeta+sum6))
      return da1

grad_log_a1 = der_log_a1(tau, ni1l, ni2l, zeta, a1, b1, a2, b2)
# print("grad_log_a1:", grad_log_a1)

# Gradient of log(f_n) w.r.t. a2
def der_log_a2(tau, ni1l, ni2l, zeta, a1, b1, a2, b2):
      sum6 = 0
      sum7 = 0
      sum8 = 0
      sum9 = 0

      for l in range(1, 3):  # Iterate over l = 1, 2, 3
          term1 = a1 * (tau[l]**b1 - tau[l-1]**b1)
          term2 = a2 * (tau[l]**b2 - tau[l-1]**b2)
          sum6 += (term1 + term2)
          for n1l in range(0, 5):
              for n2l in range(0, 5):
                  sum7 += (n1l+n2l)
                  sum8 += (n2l*(tau[l]**b2 - tau[l-1]**b2))
                  sum9 += (tau[l]**b2 - tau[l-1]**b2)

      da2 = sum8 - ((sum7*sum9)/(zeta+sum6))
      return da2

grad_log_a2 = der_log_a2(tau, ni1l, ni2l, zeta, a1, b1, a2, b2)
# print("grad_log_a2:", grad_log_a2)

# Gradient of log(f_n) w.r.t. b1
def der_log_b1(tau, ni1l, ni2l, zeta, a1, b1, a2, b2):
      sum10 = 0
      sum11 = 0
      sum12 = 0
      sum13 = 0

      for l in range(1, 3):  # Iterate over l = 1, 2, 3
          term1 = a1 * (tau[l]**b1 - tau[l-1]**b1)
          term2 = a2 * (tau[l]**b2 - tau[l-1]**b2)
          term3 = tau[l]**b1 - tau[l-1]**b1
          term4 = ((tau[l]**b1)*np.log(tau[l])) - ((tau[l-1]**b1)*np.log(tau[l-1]))
          sum10 += (term1 + term2)
          for n1l in range(0, 5):
              for n2l in range(0, 5):
                  sum11 += n1l*term3*term4
                  sum12 += (n1l+n2l)
                  sum13 += term3*term4

      db1 = (a1*sum11) - ((sum12*a1*sum13)/(zeta+sum10))
      return db1

grad_log_b1 = der_log_b1(tau, ni1l, ni2l, zeta, a1, b1, a2, b2)
# print("grad_log_b1:", grad_log_b1)

# Gradient of log(f_n) w.r.t. b2
def der_log_b2(tau, ni1l, ni2l, zeta, a1, b1, a2, b2):
      sum10 = 0
      sum11 = 0
      sum12 = 0
      sum13 = 0

      for l in range(1, 3):  # Iterate over l = 1, 2, 3
          term1 = a1 * (tau[l]**b1 - tau[l-1]**b1)
          term2 = a2 * (tau[l]**b2 - tau[l-1]**b2)
          term3 = tau[l]**b2 - tau[l-1]**b2
          term4 = ((tau[l]**b2)*np.log(tau[l])) - ((tau[l-1]**b2)*np.log(tau[l-1]))
          sum10 += (term1 + term2)
          for n1l in range(0, 5):
\end{lstlisting}
\begin{lstlisting}
              for n2l in range(0, 5):
                  sum11 += n1l*term3*term4
                  sum12 += (n1l+n2l)
                  sum13 += term3*term4

      db2 = (a2*sum11) - ((sum12*a2*sum13)/(zeta+sum10))
      return db2

grad_log_b2 = der_log_b2(tau, ni1l, ni2l, zeta, a1, b1, a2, b2)
# print("grad_log_b2:", grad_log_b2)

# Sequential Convex Programming
def scp(theta_true, theta_init, rho_vec, k, gamma, tau, ni1l, ni2l, n,
                                   compute_pmf1, compute_pmf2,
                                   der_zeta1, der_zeta2,
                                   der_a11, der_a12,
                                   der_b11, der_b12,
                                   der_a21, der_a22,
                                   der_b21, der_b22):
    theta = np.array(theta_init)

    def f(theta):
        zeta, a1, b1, a2, b2 = theta
        f_ni = compute_pmf1(tau, zeta, a1, b1, a2, b2)
        f_Ni = compute_pmf2(tau, ni1l, ni2l, zeta, a1, b1, a2, b2)
        term1 = f_ni ** (gamma + 1)
        term2 = 1 + (1 / gamma)
        term3 = f_Ni ** gamma
        return term1 - term2 * term3

    def grad_f(theta):
        zeta, a1, b1, a2, b2 = theta
        f_ni = compute_pmf1(tau, zeta, a1, b1, a2, b2)
        f_Ni = compute_pmf2(tau, ni1l, ni2l, zeta, a1, b1, a2, b2)
        term1 = f_ni ** (gamma + 1)
        term2 = 1 + (1 / gamma)
        term3 = f_Ni ** gamma

        dzeta1 = der_zeta1(tau, zeta, a1, b1, a2, b2)
        dzeta2 = der_zeta2(tau, ni1l, ni2l, zeta, a1, b1, a2, b2)
        grad_zeta = (((gamma + 1) * term1 * dzeta1) - (gamma * term2 * term3 * dzeta2)) / (n * (1 + gamma))

        da11 = der_a11(tau, zeta, a1, b1, a2, b2)
        da12 = der_a12(tau, ni1l, ni2l, zeta, a1, b1, a2, b2)
        grad_a1 = (((gamma + 1) * term1 * da11) - (gamma * term2 * term3 * da12)) / (n * (1 + gamma))

        db11 = der_b11(tau, zeta, a1, b1, a2, b2)
        db12 = der_b12(tau, ni1l, ni2l, zeta, a1, b1, a2, b2)
        grad_b1 = (((gamma + 1) * term1 * db11) - (gamma * term2 * term3 * db12)) / (n * (1 + gamma))

        da21 = der_a21(tau, zeta, a1, b1, a2, b2)
        da22 = der_a22(tau, ni1l, ni2l, zeta, a1, b1, a2, b2)
        grad_a2 = (((gamma + 1) * term1 * da21) - (gamma * term2 * term3 * da22)) / (n * (1 + gamma))

        db21 = der_b21(tau, zeta, a1, b1, a2, b2)
        db22 = der_b22(tau, ni1l, ni2l, zeta, a1, b1, a2, b2)
        grad_b2 = (((gamma + 1) * term1 * db21) - (gamma * term2 * term3 * db22)) / (n * (1 + gamma))

        return np.array([grad_zeta, grad_a1, grad_b1, grad_a2, grad_b2])

    for _ in range(k):
        grad = grad_f(theta)

        def f_cap(theta_var):
            return f(theta) + grad @ (theta_var - theta)

        constraints = [
            {"type": "ineq", "fun": lambda th, i=i: rho_vec[i] - abs(th[i] - theta_true[i])}
            for i in range(len(theta_true))
        ]

        bounds = [(1e-2, None)] * len(theta_true)

        res = minimize(f_cap, theta, method="SLSQP", bounds=bounds, constraints=constraints)

        if not res.success:
            print(f"Iteration failed: {res.message}")
\end{lstlisting}
\begin{lstlisting}
            break

        theta = res.x

    return theta

# Parameters
theta_true = np.array([4.5, 0.9, 0.5, 0.6, 0.2])
theta_init = np.array([4.6, 0.8, 0.4, 0.5, 0.1])
rho_vec = np.array([0.11464, 0.05746, 0.07415, 0.06524, 0.07670])
k = 10

# Pilot gamma
gamma_pilot = [0.01, 0.20, 0.40, 0.60, 0.80, 1.00]
pilot_matrix = []
results = [] # Initialize results as an empty list

# Perform optimization for each gamma
for gamma in gamma_pilot:
          def compute_pmf1(tau, zeta, a1, b1, a2, b2):
            product1 = 1
            sum2 = 0
            product3 = 1
            sum4 = 0
            sum5 = 0

            for l in range(1, 3):  # Iterate over l = 1, 2, 3
                term1 = a1 * (tau[l]**b1 - tau[l-1]**b1)
                term2 = a2 * (tau[l]**b2 - tau[l-1]**b2)
                sum4 += (term1 + term2)
                for n1l in range(0, 5):
                    for n2l in range(0, 5):
                        product1 *= (term1**n1l) * (term2**n2l)
                        sum2 += scipy.special.factorial(n1l+n2l)
                        product3 *= (scipy.special.factorial(n1l))*(scipy.special.factorial(n2l))
                        sum5 += (n1l+n2l)
            r = sum2/product3
            f1 = ((zeta*product1)/(((zeta+sum4)**sum5)))*r
            return f1

          def compute_pmf2(tau, ni1l, ni2l, zeta, a1, b1, a2, b2):
                product1 = 1
                sum2 = 0
                product3 = 1
                sum4 = 0
                sum5 = 0

                for l in range(1, 3):  # Iterate over l = 1, 2, 3
                    term1 = a1 * (tau[l]**b1 - tau[l-1]**b1)
                    term2 = a2 * (tau[l]**b2 - tau[l-1]**b2)
                    product1 *= (term1** ni1l[l]) * (term2**ni2l[l])
                    sum2 += scipy.special.factorial(ni1l[l]+ni2l[l])
                    product3 *= (scipy.special.factorial(ni1l[l]))*(scipy.special.factorial(ni2l[l]))
                    sum4 += (term1 + term2)
                    sum5 += (ni1l[l]+ni2l[l])

                r = sum2/product3
                f2 = ((zeta*product1)/(((zeta+sum4)**sum5)))*r
                return f2

          pilot_matrix_row = scp(theta_true, theta_init, rho_vec, k, gamma,
                              tau, ni1l, ni2l, n,
                              compute_pmf1, compute_pmf2,
                              der_zeta1, der_zeta2,
                              der_a11, der_a12,
                              der_b11, der_b12,
                              der_a21, der_a22,
                              der_b21, der_b22)

          # Add the gamma value and the results to the pilot_matrix
          pilot_row = [gamma, *pilot_matrix_row]
          pilot_matrix.append(pilot_row) # append results to pilot_matrix

          # print("gamma    zeta      a1        b1        a2        b2")
          # print(f"{pilot_row[0]:.2f}  {pilot_row[1]:.6f}  {pilot_row[2]:.6f}  {pilot_row[3]:.6f}  {pilot_row[4]:.6f}  {pilot_row[5]:.6f}") 
\end{lstlisting}
\begin{lstlisting}
# Proposed gamma
gamma_grid = np.round(np.linspace(0.20, 0.60, 41), 2)
proposed_matrix = []
results = [] # Initialize results as an empty list

# Perform optimization for each gamma
for gamma in gamma_grid:
          def compute_pmf1(tau, zeta, a1, b1, a2, b2):
            product1 = 1
            sum2 = 0
            product3 = 1
            sum4 = 0
            sum5 = 0

            for l in range(1, 3):  # Iterate over l = 1, 2, 3
                term1 = a1 * (tau[l]**b1 - tau[l-1]**b1)
                term2 = a2 * (tau[l]**b2 - tau[l-1]**b2)
                sum4 += (term1 + term2)
                for n1l in range(0, 5):
                    for n2l in range(0, 5):
                        product1 *= (term1**n1l) * (term2**n2l)
                        sum2 += scipy.special.factorial(n1l+n2l)
                        product3 *= (scipy.special.factorial(n1l))*(scipy.special.factorial(n2l))
                        sum5 += (n1l+n2l)

            r = sum2/product3
            f1 = ((zeta*product1)/(((zeta+sum4)**sum5)))*r
            return f1

          def compute_pmf2(tau, ni1l, ni2l, zeta, a1, b1, a2, b2):
                product1 = 1
                sum2 = 0
                product3 = 1
                sum4 = 0
                sum5 = 0

                for l in range(1, 3):  # Iterate over l = 1, 2, 3
                    term1 = a1 * (tau[l]**b1 - tau[l-1]**b1)
                    term2 = a2 * (tau[l]**b2 - tau[l-1]**b2)
                    product1 *= (term1** ni1l[l]) * (term2**ni2l[l])
                    sum2 += scipy.special.factorial(ni1l[l]+ni2l[l])
                    product3 *= (scipy.special.factorial(ni1l[l]))*(scipy.special.factorial(ni2l[l]))
                    sum4 += (term1 + term2)
                    sum5 += (ni1l[l]+ni2l[l])

                r = sum2/product3
                f2 = ((zeta*product1)/(((zeta+sum4)**sum5)))*r
                return f2

          proposed_matrix_row = scp(theta_true, theta_init, rho_vec, k, gamma,
                              tau, ni1l, ni2l, n,
                              compute_pmf1, compute_pmf2,
                              der_zeta1, der_zeta2,
                              der_a11, der_a12,
                              der_b11, der_b12,
                              der_a21, der_a22,
                              der_b21, der_b22)

          # Add the gamma value and the results to the proposed_matrix
          proposed_row = [gamma, *proposed_matrix_row]
          proposed_matrix.append(proposed_row) # append results to proposed_matrix

          # print("gamma    zeta      a1        b1        a2        b2")
          # print(f"{proposed_row[0]:.2f}  {proposed_row[1]:.6f}  {proposed_row[2]:.6f}  {proposed_row[3]:.6f}  {proposed_row[4]:.6f}  {proposed_row[5]:.6f}") 

# Function to compute MSE for a given gamma
def compute_mse(theta_grid, theta_pilot, gamma_value, n):  
    # Compute PMFs
    f_n = compute_pmf1(tau, zeta, a1, b1, a2, b2)

    # Gradient of logarithmic components
    u_zeta = der_log_zeta(tau, ni1l, ni2l, zeta, a1, b1, a2, b2)
    u_a1 = der_log_a1(tau, ni1l, ni2l, zeta, a1, b1, a2, b2)
    u_b1 = der_log_b1(tau, ni1l, ni2l, zeta, a1, b1, a2, b2)
    u_a2 = der_log_a2(tau, ni1l, ni2l, zeta, a1, b1, a2, b2)
    u_b2 = der_log_b2(tau, ni1l, ni2l, zeta, a1, b1, a2, b2)
\end{lstlisting}
\begin{lstlisting}
    u_n = np.array([u_zeta, u_a1, u_b1, u_a2, u_b2])

    # Compute matrices J, E, K
    J = np.outer(u_n, u_n) * (f_n ** (gamma + 1)) 
    E = u_n * (f_n ** (gamma + 1)) 
    K = (np.outer(u_n, u_n) * (f_n ** (2 * gamma + 1))) - np.outer(E, E)

    # Compute S
    J_inv = np.linalg.pinv(J)
    H = np.array([
        [0,  1,  0, -1,  0],
        [0,  0,  1,  0, -1]
    ])

    U = H.T   # (5,2)
    V = H     # (2,5)
    
    # Compute L
    U = H.T
    V = H
    P = -V @ J_inv @ U
    P_inv = np.linalg.pinv(P)
    L = -(J_inv + J_inv @ U @ P_inv @ V @ J_inv)

    # Compute Sigma
    Sigma = L.T @ K @ L

    # Calculate MSE(gamma)
    e1 = (theta_grid - theta_pilot) @ (theta_grid - theta_pilot).T
    sigma_trace = np.trace(Sigma)  
    e2 = sigma_trace / n
    M_gamma = e1 + e2

    # Ensure M_gamma is always a 2D array
    return np.array([[M_gamma]])  # Encapsulate the scalar value in a 2D array

# Gamma grid
gamma_grid = np.round(np.linspace(0.20, 0.60, 41), 2)

# Pilot gamma values
gamma_pilot_list = [0.01, 0.20, 0.40, 0.60, 0.80, 1.00]

# Set the seed for reproducibility
seed_value = 42
np.random.seed(seed_value)
random.seed(seed_value)

# Iterative algorithm to find gamma_opt
def optimal_gamma(proposed_matrix, theta_true, n, gamma_pilot, max_iterations=10, tolerance=1e-4):
    gamma_grid = np.round(np.linspace(0.20, 0.60, 41), 2)  # Gamma search grid
    theta_grid = np.array(proposed_matrix)[:, 1:]  # Drop gamma column
    gamma_history = []

    current_gamma = gamma_pilot
    prev_gamma = None

    for iteration in range(max_iterations):
        gamma_history.append(round(current_gamma, 2))
        # print(f"Iteration {iteration + 1}: Current gamma = {current_gamma:.2f}")

        # Pilot estimate at current gamma
        pilot_matrix_row = scp(theta_true, theta_init, rho_vec, k, current_gamma,
                               tau, ni1l, ni2l, n,
                               compute_pmf1, compute_pmf2,
                               der_zeta1, der_zeta2,
                               der_a11, der_a12,
                               der_b11, der_b12,
                               der_a21, der_a22,
                               der_b21, der_b22)
        theta_pilot = pilot_matrix_row

        # Compute MSE for each gamma in the grid
        mse_values = [
            compute_mse(theta_g, theta_pilot, gamma_val, n)
            for theta_g, gamma_val in zip(theta_grid, gamma_grid)
        ]
\end{lstlisting}
\begin{lstlisting}
        # Update gamma to the minimizer
        gamma_next = gamma_grid[np.argmin(mse_values)]

        # Check convergence
        if prev_gamma is not None and np.isclose(gamma_next, prev_gamma, atol=tolerance):
            # print("Convergence achieved.")
            break

        prev_gamma = current_gamma
        current_gamma = gamma_next

    return gamma_history

# Run for all pilot gamma values
gamma_iterations_matrix = {}
for gamma_pilot in gamma_pilot_list:
    gamma_path = optimal_gamma(
        proposed_matrix=proposed_matrix,
        theta_true=theta_true,
        n=n,
        gamma_pilot=gamma_pilot
    )
    gamma_iterations_matrix[round(gamma_pilot, 2)] = gamma_path

# Put results in a DataFrame
gamma_df = pd.DataFrame.from_dict(gamma_iterations_matrix, orient='index')
gamma_df.columns = [f"Iter {i+1}" for i in range(gamma_df.shape[1])]
gamma_df.index.name = "Pilot \gamma"
print("Gamma Iteration Matrix (n = 100):\n")
print(gamma_df.to_string())

# Converged Gamma via Majority Vote
from collections import Counter
def extract_consensus_gamma(gamma_df, k_last=3):
    last_k = gamma_df.iloc[:, -k_last:]
    flattened = last_k.values.flatten()
    most_common_gamma, freq = Counter(flattened).most_common(1)[0]

    return most_common_gamma, freq

# Extract consensus gamma
optimal_gamma, frequency = extract_consensus_gamma(gamma_df, k_last=3)

if frequency >= 5:
    print(f"Consensus optimal gamma (n = 100): {optimal_gamma}")
else:
    print("\n No strong consensus found in final gamma values.")
\end{lstlisting}

\bibliographystyle{elsarticle-num}
\bibliography{ref}

\begin{thebibliography}{10}
\expandafter\ifx\csname url\endcsname\relax
  \def\url#1{\texttt{#1}}\fi
\expandafter\ifx\csname urlprefix\endcsname\relax\def\urlprefix{URL }\fi
\expandafter\ifx\csname href\endcsname\relax
  \def\href#1#2{#2} \def\path#1{#1}\fi

\bibitem{wang2001analyzing}
M.-C. Wang, J.~Qin, C.-T. Chiang, Analyzing recurrent event data with informative censoring., Journal of the American Statistical Association 96 (2001) 1057--1065.
\newblock \href {https://doi.org/https://doi.org/10.1198/016214501753209031} {\path{doi:https://doi.org/10.1198/016214501753209031}}.

\bibitem{lin2000semiparametric}
D.~Y. Lin, L.-J. Wei, I.~Yang, Z.~Ying, Semiparametric regression for the mean and rate functions of recurrent events., Journal of the Royal Statistical Society: Series B (Statistical Methodology) 62 (2000) 711--730.
\newblock \href {https://doi.org/https://doi.org/10.1111/1467-9868.00259} {\path{doi:https://doi.org/10.1111/1467-9868.00259}}.

\bibitem{kalbfleisch1991methods}
J.~D. Kalbfleisch, J.~F. Lawless, J.~A. Robinson, Methods for the analysis and prediction of warranty claims., Technometrics 33 (1991) 273--285.
\newblock \href {https://doi.org/https://doi.org/10.2307/1268780} {\path{doi:https://doi.org/10.2307/1268780}}.

\bibitem{huang2006analysing}
C.-Y. Huang, M.-C. Wang, Y.~Zhang, Analysing panel count data with informative observation times., Biometrika 93 (2006) 763--775.
\newblock \href {https://doi.org/https://doi.org/10.1093/biomet/93.4.763} {\path{doi:https://doi.org/10.1093/biomet/93.4.763}}.

\bibitem{sun2013statistical}
J.~Sun, X.~Zhao, Statistical analysis of panel count data., Springer, 2013.

\bibitem{sun1995estimation}
J.~Sun, J.~Kalbfleisch, Estimation of the mean function of point processes based on panel count data., Statistica Sinica (1995) 279--289.

\bibitem{wellner2000two}
J.~A. Wellner, Y.~Zhang, Two estimators of the mean of a counting process with panel count data., The Annals of Statistics 28 (2000) 779--814.
\newblock \href {https://doi.org/10.1214/aos/1015951998} {\path{doi:10.1214/aos/1015951998}}.

\bibitem{hu2009generalized}
X.~J. Hu, S.~W. Lagakos, R.~A. Lockhart, Generalized least squares estimation of the mean function of a counting process based on panel counts., Statistica Sinica 19 (2009) 561.

\bibitem{tang2017bayesian}
Y.~Tang, J.~Fu, W.~Liu, A.~Xu, Bayesian analysis of repairable systems with modulated power law process, Applied Mathematical Modelling 44 (2017) 357--373.
\newblock \href {https://doi.org/https://doi.org/10.1016/j.apm.2017.01.067} {\path{doi:https://doi.org/10.1016/j.apm.2017.01.067}}.

\bibitem{li2017nhpp}
Q.~Li, H.~Pham, Nhpp software reliability model considering the uncertainty of operating environments with imperfect debugging and testing coverage, Applied Mathematical Modelling 51 (2017) 68--85.
\newblock \href {https://doi.org/https://doi.org/10.1016/j.apm.2017.06.034} {\path{doi:https://doi.org/10.1016/j.apm.2017.06.034}}.

\bibitem{slimacek2016nonhomogeneous}
V.~Slimacek, B.~H. Lindqvist, Nonhomogeneous poisson process with nonparametric frailty., Reliability Engineering \& System Safety 149 (2016) 14--23.
\newblock \href {https://doi.org/https://doi.org/10.1016/j.ress.2015.12.005} {\path{doi:https://doi.org/10.1016/j.ress.2015.12.005}}.

\bibitem{he2009semiparametric}
X.~He, X.~Tong, J.~Sun, Semiparametric analysis of panel count data with correlated observation and follow-up times., Lifetime data analysis 15 (2009) 177--196.
\newblock \href {https://doi.org/10.1007/s10985-008-9105-1} {\path{doi:10.1007/s10985-008-9105-1}}.

\bibitem{sun2000regression}
J.~Sun, L.~Wei, Regression analysis of panel count data with covariate-dependent observation and censoring times., Journal of the Royal Statistical Society Series B: Statistical Methodology 62 (2000) 293--302.
\newblock \href {https://doi.org/https://doi.org/10.1111/1467-9868.00232} {\path{doi:https://doi.org/10.1111/1467-9868.00232}}.

\bibitem{he2008regression}
X.~He, X.~Tong, J.~Sun, R.~J. Cook, Regression analysis of multivariate panel count data., Biostatistics 9 (2008) 234--248.
\newblock \href {https://doi.org/https://doi.org/10.1093/biostatistics/kxm025} {\path{doi:https://doi.org/10.1093/biostatistics/kxm025}}.

\bibitem{wellner2007two}
J.~A. Wellner, Y.~Zhang, Two likelihood-based semiparametric estimation methods for panel count data with covariates., The Annals of Statistics 35 (2007) 2106--2142.
\newblock \href {https://doi.org/10.1214/009053607000000181} {\path{doi:10.1214/009053607000000181}}.

\bibitem{zhu2018semiparametric}
L.~Zhu, Y.~Zhang, Y.~Li, J.~Sun, L.~L. Robison, A semiparametric likelihood-based method for regression analysis of mixed panel-count data., Biometrics 74 (2018) 488--497.
\newblock \href {https://doi.org/https://doi.org/10.1111/biom.12774} {\path{doi:https://doi.org/10.1111/biom.12774}}.

\bibitem{li2021regression}
Y.~Li, L.~Zhu, L.~Liu, L.~L. Robison, Regression analysis of mixed panel-count data with application to cancer studies., Statistics in biosciences 13 (2021) 178--195.
\newblock \href {https://doi.org/https://doi.org/10.1007/s12561-020-09291-2} {\path{doi:https://doi.org/10.1007/s12561-020-09291-2}}.

\bibitem{zeng2021maximum}
D.~Zeng, D.~Lin, Maximum likelihood estimation for semiparametric regression models with panel count data., Biometrika 108 (2021) 947--963.
\newblock \href {https://doi.org/https://doi.org/10.1093/biomet/asaa091} {\path{doi:https://doi.org/10.1093/biomet/asaa091}}.

\bibitem{basu1998robust}
A.~Basu, I.~R. Harris, N.~L. Hjort, M.~Jones, Robust and efficient estimation by minimising a density power divergence., Biometrika 85 (1998) 549--559.
\newblock \href {https://doi.org/https://doi.org/10.1093/biomet/85.3.549} {\path{doi:https://doi.org/10.1093/biomet/85.3.549}}.

\bibitem{ghosh2013robust}
A.~Ghosh, A.~Basu, Robust estimation for independent non-homogeneous observations using density power divergence with applications to linear regression., Electronic Journal of Statistics 7 (2013) 2420--2456.
\newblock \href {https://doi.org/10.1214/13-EJS847} {\path{doi:10.1214/13-EJS847}}.

\bibitem{basu2022robust}
A.~Basu, S.~Chakraborty, A.~Ghosh, L.~Pardo, Robust density power divergence based tests in multivariate analysis: A comparative overview of different approaches., Journal of Multivariate Analysis 188 (2022) 104846.
\newblock \href {https://doi.org/https://doi.org/10.1016/j.jmva.2021.104846} {\path{doi:https://doi.org/10.1016/j.jmva.2021.104846}}.

\bibitem{mandal2023robust}
A.~Mandal, B.~H. Beyaztas, S.~Bandyopadhyay, Robust density power divergence estimates for panel data models., Annals of the Institute of Statistical Mathematics 75 (2023) 773--798.
\newblock \href {https://doi.org/https://doi.org/10.1007/s10463-022-00862-2} {\path{doi:https://doi.org/10.1007/s10463-022-00862-2}}.

\bibitem{baghel2024analysis}
S.~Baghel, S.~Mondal, Analysis of one-shot device testing data under logistic-exponential lifetime distribution with an application to seer gallbladder cancer data, Applied Mathematical Modelling 126 (2024) 159--184.
\newblock \href {https://doi.org/https://doi.org/10.1016/j.apm.2023.10.037} {\path{doi:https://doi.org/10.1016/j.apm.2023.10.037}}.

\bibitem{balakrishnan2019robust}
N.~Balakrishnan, E.~Castilla, N.~Mart{\'\i}n, L.~Pardo, Robust estimators for one-shot device testing data under gamma lifetime model with an application to a tumor toxicological data., Metrika 82 (2019) 991--1019.
\newblock \href {https://doi.org/https://doi.org/10.1007/s00184-019-00718-5} {\path{doi:https://doi.org/10.1007/s00184-019-00718-5}}.

\bibitem{xiong2022minimum}
L.~Xiong, F.~Zhu, Minimum density power divergence estimator for negative binomial integer-valued garch models, Communications in Mathematics and Statistics 10~(2) (2022) 233--261.

\bibitem{basu2018testing}
A.~Basu, A.~Mandal, N.~Martin, L.~Pardo, Testing composite hypothesis based on the density power divergence., Sankhya B 80 (2018) 222--262.
\newblock \href {https://doi.org/https://doi.org/10.1007/s13571-017-0143-0} {\path{doi:https://doi.org/10.1007/s13571-017-0143-0}}.

\bibitem{felipe2023restricted}
{\'A}.~Felipe, M.~Jaenada, P.~Miranda, L.~Pardo, Restricted distance-type gaussian estimators based on density power divergence and their applications in hypothesis testing., Mathematics 11 (2023) 1480.
\newblock \href {https://doi.org/https://doi.org/10.3390/math11061480} {\path{doi:https://doi.org/10.3390/math11061480}}.

\bibitem{baghel2024robust}
S.~Baghel, S.~Mondal, Robust estimation of dependent competing risk model under interval monitoring and determining optimal inspection intervals., Applied Stochastic Models in Business and Industry 40 (2024) 926--944.
\newblock \href {https://doi.org/http://dx.doi.org/10.1002/asmb.2854} {\path{doi:http://dx.doi.org/10.1002/asmb.2854}}.

\bibitem{dinh2010local}
Q.~T. Dinh, M.~Diehl, Local convergence of sequential convex programming for nonconvex optimization, in: Recent Advances in Optimization and its Applications in Engineering: The 14th Belgian-French-German Conference on Optimization, Springer, 2010, pp. 93--102.

\bibitem{debrouwere2013time}
F.~Debrouwere, W.~Van~Loock, G.~Pipeleers, Q.~T. Dinh, M.~Diehl, J.~De~Schutter, J.~Swevers, Time-optimal path following for robots with convex--concave constraints using sequential convex programming, IEEE Transactions on Robotics 29 (2013) 1485--1495.
\newblock \href {https://doi.org/https://doi.org/10.1109/TRO.2013.2277565} {\path{doi:https://doi.org/10.1109/TRO.2013.2277565}}.

\bibitem{morgan2014model}
D.~Morgan, S.-J. Chung, F.~Y. Hadaegh, Model predictive control of swarms of spacecraft using sequential convex programming, Journal of Guidance, Control, and Dynamics 37 (2014) 1725--1740.
\newblock \href {https://doi.org/https://doi.org/10.2514/1.G000218} {\path{doi:https://doi.org/10.2514/1.G000218}}.

\bibitem{wang2020improved}
Z.~Wang, Y.~Lu, Improved sequential convex programming algorithms for entry trajectory optimization, Journal of Spacecraft and Rockets 57 (2020) 1373--1386.
\newblock \href {https://doi.org/https://doi.org/10.2514/1.A34640} {\path{doi:https://doi.org/10.2514/1.A34640}}.

\bibitem{warwick2005choosing}
J.~Warwick, M.~Jones, Choosing a robustness tuning parameter., Journal of Statistical Computation and Simulation 75 (2005) 581--588.
\newblock \href {https://doi.org/https://doi.org/10.1080/00949650412331299120} {\path{doi:https://doi.org/10.1080/00949650412331299120}}.

\bibitem{sugasawa2021selection}
S.~Sugasawa, S.~Yonekura, On selection criteria for the tuning parameter in robust divergence., Entropy 23 (2021) 1147.
\newblock \href {https://doi.org/https://doi.org/10.3390/e23091147} {\path{doi:https://doi.org/10.3390/e23091147}}.

\bibitem{yonekura2023adaptation}
S.~Yonekura, S.~Sugasawa, Adaptation of the tuning parameter in general bayesian inference with robust divergence., Statistics and Computing 33 (2023) 39.
\newblock \href {https://doi.org/https://doi.org/10.1007/s11222-023-10205-7} {\path{doi:https://doi.org/10.1007/s11222-023-10205-7}}.

\bibitem{hyvarinen2005estimation}
A.~Hyv{\"a}rinen, P.~Dayan, Estimation of non-normalized statistical models by score matching., Journal of Machine Learning Research 6 (2005).

\bibitem{hyvarinen2007some}
A.~Hyv{\"a}rinen, Some extensions of score matching., Computational statistics \& data analysis 51 (2007) 2499--2512.
\newblock \href {https://doi.org/https://doi.org/10.1016/j.csda.2006.09.003} {\path{doi:https://doi.org/10.1016/j.csda.2006.09.003}}.

\bibitem{lyu2012interpretation}
S.~Lyu, Interpretation and generalization of score matching., arXiv preprint arXiv:1205.2629 (2012).

\bibitem{xu2022generalized}
J.~Xu, J.~L. Scealy, A.~T. Wood, T.~Zou, Generalized score matching for regression., arXiv preprint arXiv:2203.09864 (2022).
\newblock \href {https://doi.org/https://doi.org/10.48550/arXiv.2203.09864} {\path{doi:https://doi.org/10.48550/arXiv.2203.09864}}.

\bibitem{duchi2018scp}
J.~Duchi, S.~Boyd, J.~Mattingley, Sequential convex programming, Lecture notes, notes for EE364b, Stanford University, Spring 2018 (2018).

\bibitem{basak2021optimal}
S.~Basak, A.~Basu, M.~Jones, On the ‘optimal’density power divergence tuning parameter., Journal of Applied Statistics 48 (2021) 536--556.

\bibitem{hanson1965inequality}
M.~Hanson, Inequality constrained maximum likelihood estimation, Annals of the Institute of Statistical Mathematics 17 (1965) 311--321.
\newblock \href {https://doi.org/https://doi.org/10.1007/BF02868175} {\path{doi:https://doi.org/10.1007/BF02868175}}.

\bibitem{aitchison1958maximum}
J.~Aitchison, S.~Silvey, Maximum-likelihood estimation of parameters subject to restraints, The annals of mathematical Statistics (1958) 813--828\href {https://doi.org/10.1214/aoms/1177706538} {\path{doi:10.1214/aoms/1177706538}}.

\bibitem{sen2010finite}
P.~K. Sen, J.~M. Singer, A.~C.~P. de~Lima, From finite sample to asymptotic methods in statistics, Cambridge University Press, 2010.
\newblock \href {https://doi.org/https://doi.org/10.1017/CBO9780511806957} {\path{doi:https://doi.org/10.1017/CBO9780511806957}}.

\end{thebibliography}

\end{document}